\documentstyle[fleqn]{article}
\begin{document}
\title{\bf Bondi mass in classical field theory}
\author{Jacek Jezierski\\
 Department of Mathematical Methods in Physics, \\ University of
 Warsaw, ul. Ho\.za 74, 00-682 Warsaw, Poland}
\maketitle

\newfont{\msa}{msam10 scaled\magstep1}

\newcommand{\kolo}[1]{\stackrel{\circ}{#1}}
\newcommand{\ve}{\varepsilon}
\newcommand{\ol}{\overline}
\newcommand{\scri}{\mathop {\cal S}\nolimits }
\newcommand{\ind}{\mathop {\rm ind}\nolimits }
\newcommand{\be}{\begin{equation}}
\newcommand{\ee}{\end{equation}} 
\newcommand{\ssr}[1]{{\, \strut ^{w}}\! {#1}}
\newcommand{\ssd}[1]{{\, \strut ^{d}}\! {#1}}
\newcommand{\ssm}[1]{{\, \strut ^{m}}\! {#1}}
\newcommand{\rd}{{\rm d}} 
\newcommand{\E}[1]{{\rm e}^{#1}} 

\begin{abstract}

We analyze three classical field theories based on the wave equation:
 scalar field, electrodynamics and linearized gravity.  We derive
 certain generating formula on a hyperboloid and on a null surface for
 them.  The linearized Einstein equations are analyzed around null
 infinity.  It is shown how the dynamics can be reduced to gauge
 invariant quanitities in a quasi-local way.  The quasi-local
 gauge-invariant ``density'' of the hamiltonian is derived on the
 hyperboloid and on the future null infinity ${\cal S}^+$.  The result
 gives a new interpretation of the Bondi mass loss formula
\cite{BVM}, \cite{Sachs}, \cite{Burg}, \cite{XIV}.
 We show also how to define angular momentum.\\ Starting from affine
 approach for Einstein equations we obtain variational formulae for
 Bondi-Sachs type metrics related with energy and angular momentum
 generators. The original van der Burg
\cite{Burg} asymptotic
 hierarchy is revisited and the relations between linearized and
 asymptotic nonlinear situations are established. We discuss also
 supertranslations, Newman-Penrose charges and Janis solutions.
\end{abstract}

\section{Introduction}

In the papers \cite{BVM},
\cite{Sachs}, \cite{Burg} from the series ``Gravitational waves in general
 relativity'' Bondi, van der Burg, Metzner and Sachs have analyzed
 asymptotic behaviour of the gravitational field at null infinity. The
 energy in this regime so called Bondi mass was defined and the main
 property -- loss of the energy was proved.  See also discussion on p.
 127 in the last paper \cite{XIV} in this series. The energy at null
 infinity was also proposed by Trautman
\cite{AT} and it will be called a Trautman-Bondi energy.

We interprete their result from symplectic point of view and it is
 shown that the concept of Trautman-Bondi energy arises not only in
 gravity but can be also defined for other fields.  In this case the
 TB energy can be treated formally as a ``hamiltonian'' and the loss
 of energy formula has a natural interpretation given by eq.
 (\ref{Hdot}). We apply similar technique to define angular momentum. 

We introduce here the language of {\em generating functions} which
 simplifies enormously our calculations.  This point of view on
 dynamics is due to W.~M.~Tulczyjew (see \cite{Tulcz2}).

We start from an example of a scalar field for which we define TB
 energy as a ``hamiltonian'' on a hyperboloid. The motivation for
 concerning hyperboloids in gravitation one can find in
\cite{HF}, \cite{AChF} and
\cite{ACh}. 

In section 3 we give an example from electrodynamics.

Next we prove analogous formulae for the linearized gravity. The
 result is formulated in a nice gauge-independent way. We show how the
 formula (\ref{Hdot}) can be related with original Bondi-Sachs result
 -- mass loss eqn (35) of \cite{BVM} (cf. also eqn (4.16) in
\cite{Sachs}, eqn (13) in \cite{Burg} and eqn
 (3.8) in \cite{XIV}).  Our result is an important gauge-independent
 generalization of this original mass loss equation. It shows the
 straightforward relation between the Weyl tensor on scri and the flux
 of the radiation energy through it.  We show how to define angular
 momentum from this point of view.

In section 8 we give covariant on a sphere formulation of the
 asymptotic equations from \cite{Burg}. We discuss several features of
 the theory like supertranslations, charges etc. and also the
 relations between linear and nonlinear theory.

\section{Scalar field}
 
Consider a scalar field theory derived from a Lagrangian $L =
 L(\varphi,
\varphi_\mu)$, where $\varphi_\mu := \partial_\mu \varphi$.  The entire
 information about field dynamics may be encoded in equation
\begin{equation}
\delta L(\varphi, \varphi_\mu) = \partial_\mu (p^\mu \delta \varphi) =
 (\partial_\mu p^\mu) \delta \varphi + p^\mu \delta \varphi_\mu \
.
\label{Lpole} \end{equation}

The above generating formula is equivalent to the system of equations:
\begin{eqnarray} \partial_\mu p^\mu = \frac {\partial L}{\partial \varphi}\ \ \
 ; \ \ \ \ \ \ \ \ \ p^\mu = \frac {\partial L}{\partial
\varphi_\mu} \ .
\label{E-Lpole} \end{eqnarray}

Hamiltonian description of the theory is based on a chronological
 analysis, i.~e.~on a (3+1)-foliation of space-time.  Treating
 separately time derivative and the space derivatives, we rewrite
 (\ref{Lpole}) as
\begin{equation}
\label{3+1} \delta L = {(\pi \delta \varphi)}\, \dot{} + \partial_k (p^k \delta
\varphi) \ , \end{equation} where we denoted $\pi := p^0$.  Integrating over a
 3-dimensional space--volume $V$ we obtain:
\begin{eqnarray} \delta \int_V
 L & = & \int_V \left( {\dot \pi} \delta \varphi + \pi \delta {\dot
\varphi} \right) +
\int_{\partial V} p^\perp \delta \varphi \label{dL-scalar}  =  \int_V
\left( {\dot \pi} \delta \varphi - {\dot \varphi} \delta \pi + \delta (\pi
 {\dot \varphi}) \right) + \int_{\partial V} p^\perp \delta
\varphi \ .
\end{eqnarray} 
 where by $p^\perp$ we denote normal part of the momentum $p^k$.
 Hence, Legendre transformation between $\pi$ and ${\dot
\varphi}$ gives
 us:
\begin{equation} - \delta {\cal H} (\varphi, \pi ) = \int_V \left( {\dot \pi}
\delta \varphi - {\dot \varphi} \delta \pi \right) + \int_{\partial V} p^\perp
\delta \varphi \ , \label{dH} \end{equation} where \begin{equation} {\cal H} =
\int_V \pi {\dot \varphi} - L \ .  \end{equation} Equation (\ref{dH}) is
 equivalent to Hamilton equations: \begin{eqnarray} {\dot \pi} = -
\frac {\delta {\cal H}}{\delta\varphi}\ \ \ ; \ \ \ \ \ \ \ \
\ {\dot \varphi} =
\frac {\delta {\cal H}}{\delta \pi} \label{rHamilt-pole} \end{eqnarray}
 provided no boundary terms remain when integration by parts is
 performed.  To kill these boundary terms we restrict ourselves to an
 infinitely dimensional functional space of initial data $(
\varphi , \pi )$, which
 are defined on $V$ and fulfill the Dirichlet boundary conditions
 $\varphi
\vert_{\partial V} \equiv f$ on its boundary.  Imposing these conditions,
 we kill the boundary integral in (\ref{dH}), because $\delta
\varphi
\equiv 0$ within the space of fields fulfilling boundary conditions.  This
 way formula (\ref{dH}) becomes an infinitely dimensional Hamiltonian
 formula.  Without any boundary conditions, the field dynamics in $V$
 {\em can not} be formulated in terms of any Hamiltonian system,
 because the evolution of initial data in $V$ may be influenced by the
 field outside of $V$.

Physically, a choice of boundary conditions corresponds to an
 insulation of a physical system composed of a portion of the field
 contained in $V$.  The choice of Dirichlet conditions is not unique.
 Performing e.~g.~Legendre transformation between $\varphi$ and
 $p^\perp$ in the boundary term of (\ref{dH}), we obtain:
\begin{equation} \int_{\partial V} p^\perp \delta \varphi = \delta
\int_{\partial V} p^\perp \varphi - \int_{\partial V} \varphi \delta
 p^\perp \ .
\end{equation} Hence, we have \begin{equation} - \delta \overline{\cal H}=
\int_V \left( {\dot \pi} \delta \varphi - {\dot \varphi} \delta \pi \right) -
\int_{\partial V} \varphi \delta p^\perp \ .  \label{dHbar} \end{equation} The
 new Hamiltonian \begin{equation} \overline{\cal H} = {\cal H} +
\int_{\partial V} p^\perp \varphi \label{freeH} \end{equation} generates
 formally the same partial differential equations governing the
 dynamics, but the evolution takes place in a {\em different phase
 space}.  Indeed, to derive Hamiltonian equations (\ref{rHamilt-pole})
 from (\ref{dHbar}) we have now to kill $\delta p^\perp$ at the
 boundary.  For this purpose we have to impose Neumann boundary
 condition $p^\perp \vert_{\partial V}=
\tilde{f}$.  The space of fields fulfilling this condition becomes now our
 infinite dimensional phase space, different from the previous one.

The difference between the above two dynamical systems is similar to
 the difference between the evolution of a thermodynamical system in
 two different regimes: in an adiabatic insulation and in a thermal
 bath (see
\cite{Thermo}).  As another example we may consider the dynamics of an
 elastic body: Dirichlet conditions mean controlling exactly the
 position of its surface, whereas Neumann conditions mean controlling
 only the forces applied to the surface.  We see that the same field
 dynamics may lead to different Hamiltonian systems according to the
 way we control the boundary behaviour of the field.  Without imposing
 boundary conditions field dynamics {\em can not} be formulated in
 terms of a Hamiltonian system.

\subsection{Coordinates in Minkowski space}
 We shall consider flat Minkowski metric of the following form in
 spherical coordinates:
\[ \eta_{\mu\nu}\rd y^\mu \rd y^\nu = -\rd t^2 +\rd r^2 + r^2 (\rd
\theta^2 +\sin^2\theta \rd \phi^2) \]
 Minkowski space $M$ has a natural structure of spherical foliation
 around null infinity, more precisely, the neighbourhood of $\scri^+$
 looks like $S^2 \times M_2$.

We shall use several coordinates on $M_2$: $s,t,r,\rho,\omega ,v,u$.
 They are defined as follows:
\[ r=\sinh \omega=\rho^{-1} \]
\[ t=s+\cosh \omega=s +\rho^{-1}\sqrt{1+\rho^2} \]
\[ u=t-r= s +{\rho\over1+\sqrt{1+\rho^2}} \]
\[ v=t+r=s +\rho^{-1}(\sqrt{1+\rho^2}+1) \, . \]

\subsection{Scalar field on a hyperboloid}
 We shall consider a scalar field $\varphi$ in a flat Minkowski space
 $M$ with the metric:
\be\label{etaonM} \eta_{\mu\nu}{\rm d}x^\mu{\rm d}x^\nu=
\rho^{-2}\left( -\rho^2{\rm d}s^2 +{2{\rm d}s{\rm d}\rho \over
\sqrt{1+\rho^2}}
 +{{\rm d}\rho^2\over 1+\rho^2} + {\rm d}\theta^2 +\sin^2\theta {\rm
 d}\phi^2 \right)
\ee
 Let us fix a coordinate chart $(x^\mu)$ on $M$ such that
 $x^1=\theta$, $x^2=\phi$ (spherical angles), $x^3=\rho$ and $x^0=s$
 and let us denote by $\kolo\gamma_{AB}$ a metric on a unit sphere
 ($\kolo\gamma_{AB}{\rm d}x^A{\rm d}x^B:={\rm d}\theta^2 +\sin^2\theta
 {\rm d}\phi^2$).

We shall consider an initial value problem on a hyperboloid $\Sigma$:
\[ \Sigma_s:= \left\{ x\in M \;| \; x^0=s=const. \right\} \] 
 for our scalar field $\varphi$ with density of the lagrangian
 (corresponding to the wave equation)
\[ L:= -\frac 12 \sqrt{-\det \eta_{\mu\nu}}
\eta^{\mu\nu}\varphi_\mu\varphi_\nu = 
 -\frac 12 \rho^{-2}\sin\theta\left[ \rho^2(\varphi_{3})^2
 -\frac{(\varphi_0)^2 }{1+\rho^2} +{2\varphi_{3} \varphi_{0}
\over \sqrt{1+\rho^2}}
 +\kolo\gamma^{AB}\varphi_{A} \varphi_{B} \right] \]
 
We use the following convention for indices: greek indices $\mu,
\nu,
\ldots$ run from 0 to 3; 
 $k,l, \ldots$ are coordinates on a hyperboloid $\Sigma_s$ and run
 from 1 to 3; $A,B,\ldots$ are coordinates on $S(s,\rho)$ and run from
 1 to 2, where $ S(s,\rho):= \left\{ x\in \Sigma_s \;|
\; x^3=\rho=const. \right\} $.

The formula (\ref{Lpole}) can be written as follows: \[ \delta
\int_V L=\int_V (p^0\delta\varphi)_{,0} + \int_{\partial V}
 p^3 \delta\varphi \] and \[ p^0=\frac {\partial L}{\partial
\varphi_0}=
\rho^{-2}\sin\theta\left(\frac{\varphi_0}{1+\rho^2}
 -\frac{\varphi_{3}}{\sqrt{1+\rho^2}}\right) \] \[ p^3=-\frac
 {\partial L}{\partial \varphi_3}=\rho^{-2}\sin\theta \left(\frac
 1{\sqrt{1+\rho^2}}\varphi_{0}+\rho^{2}\varphi_{3}\right) \] Let us
 observe that the integral $ \displaystyle\int_V L $ is in general not
 convergent on $\Sigma$ if we assume that $\varphi=O(
\rho)$ and
 $\varphi_{3}=O\left( 1 \right)$.  We can ``renormalize'' $L$ adding a
 full divergence: \[ {\overline L}:= -\frac 12
\sin\theta\left[
\rho^2(\psi_{3})^2 -\frac 1{1+\rho^2}(\psi_0)^2 +{2 \over
\sqrt{1+\rho^2}}\psi_{3} \psi_{0} 
 +\kolo\gamma^{AB}\psi_{A} \psi_{B} \right] =\] \be\label{Lren} =
 L+\frac 12 \partial_0 \left( \sin\theta
\frac{\rho^{-3}}{\sqrt{1+\rho^2}}\varphi^2 \right)
 -\frac 12 \partial_3 \left( \sin\theta \rho^{-1}\varphi^2
\right)
\ee where we have introduced a new field variable
 $\psi:=\rho^{-1}\varphi$ which is natural close to null infinity.
 The generating formula takes the following form:
\[\delta \int_V {\overline L} =
\int_V \left( \pi^0 \delta\psi \right)_{,0} + \int_{\partial V}
\pi^3 \delta\psi \]
 and \[ \pi^0=\frac {\partial \overline L}{\partial \psi_0}=
\sin\theta\left(\frac{\psi_0}{1+\rho^2}
 -\frac{\psi_{3}}{\sqrt{1+\rho^2}}\right) \] \[ \pi^3=\frac {\partial
\overline L}{\partial \psi_3}= -\sin\theta \left(\frac
 1{\sqrt{1+\rho^2}}\psi_{0} +\rho^{2}\psi_{3} \right) \] \[
\pi^A=\frac {\partial \overline L}{\partial \psi_A}=
 - \sin\theta\kolo\gamma^{AB}\psi_B \] \[ \frac {\partial
\overline L}{\partial \psi}=0
\] It is easy to check that all terms are finite at null
 infinity provided $\psi=O(1)$ and $\psi_{3}=O( 1)$.

From the above equations (in hamiltonian form) one can easily obtain
 wave equation: \be\label{waveq}
\overline{\mbox{\msa\symbol{3}}} \, \psi=0 \ee
 where the wave operator $\overline{\mbox{\msa\symbol{3}}}$ is defined
 with respect to the metric \be\label{etabar}
\overline\eta_{\mu\nu}{\rm d}x^\mu{\rm d}x^\nu:=
 -\rho^2{\rm d}s^2 +{2{\rm d}s{\rm d}\rho \over \sqrt{1+\rho^2}}
 +{{\rm d}\rho^2\over 1+\rho^2} + {\rm d}\theta^2 +\sin^2\theta {\rm
 d}\phi^2 = \rho^{2}\eta_{\mu\nu}{\rm d}x^\mu{\rm d}x^\nu
\ee
 which is conformally related to the original flat metric
 $\eta_{\mu\nu}$.

\underline{Remark} Let us observe that 
\[ \overline L = -\frac 12 \sqrt{-\det \overline\eta_{\mu\nu}}
\overline\eta^{\mu\nu}\psi_\mu\psi_\nu = L +\frac 12 \left(
\sqrt{-\det \overline\eta_{\mu\nu}}(\ln\rho)_{,\nu}
\overline\eta^{\mu\nu}\psi^2 \right)_{,\mu} \]
 so we are not surprised that (\ref{waveq}) holds.  It can be easily
 checked that equation (\ref{waveq}) is equivalent to the original
 wave equation: \be\label{waveq0} \mbox{\msa\symbol{3}}
\varphi=0 \ee 
 by the usual conformal transformation for the conformally invariant
 operator $ \overline{\mbox{\msa\symbol{3}}} +\frac 16
\overline R$    
 because the scalar curvature $\overline R$ of the metric
 $\overline\eta_{\mu\nu}$ vanishes.

If we want to have a closed hamiltonian system we have to assume that
 $\dot\psi\vert_{\partial V}=0$ and then the energy will be conserved
 in time. But we would like to describe the situation with any data on
 $\scri^+$. In this case we can define Bondi energy which is no longer
 conserved, formally we can treat it as a hamiltonian and formula
 (\ref{dH}) is useful as a definition of the Bondi energy together
 with its changes in time.  In our case boundary condition $f$ depends
 on time (see disscussion after formula (\ref{rHamilt-pole})) and an
 interesting case for us is to compare the data with different
 boundary conditions.  Although the energy defined on a hyperboloid is
 not a hamiltonian in a usual sense it plays an important role for the
 description of the radiation at null infinity.  The metod is useful
 to the construction of the other generators of the Poincare group and
 will be applied for the angular momentum.

Let us define ``hamiltonian density'': \[ H:= \frac 12
\sin\theta\left[ \left( \rho\psi_{,3}\right)^2
 + \frac 1{1+\rho^2}(\psi_0)^2 +\kolo\gamma^{AB}\psi_{A} \psi_{B}
\right] = 
\pi^0\psi_0 - {\overline L} \] and formally the following
 variational relation holds: \begin{equation}\label{Hpipsi} -
\delta \int_V H = \int_V \left( {\dot \pi}
\delta \psi - {\dot \psi} \delta \pi \right) + \int_{\partial V} \pi^3
\delta \psi \ , \label{pipsi} \end{equation}
 here $\pi:=\pi^0$.
 
\underline{Remark} Relation between (\ref{Hpipsi}) coming from
 $\overline L$ and (\ref{dH}) with respect to $L$ can be described by
 following observations \[ {\dot \pi}
\delta \psi - {\dot \psi} \delta \pi ={\dot p^0}
\delta \varphi - {\dot \varphi} \delta p^0 \]
\[ \pi^3 \delta \psi - p^3 \delta \varphi = \frac12 \delta
\sin\theta\rho\psi^2 \]
 So the formulae give the same hamiltonian because $\rho\psi^2$
 vanishes on $\scri^+$.
 
We should express our hamiltonian as a functional of $(\pi,\psi)$: \[
 H:= \frac 12 \sin\theta\left[ \left(
\rho\psi_{3}\right)^2
 + \left(\frac{\pi\sqrt{1+\rho^2}}{\sin\theta}+\psi_3\right)^2
 +\kolo\gamma^{AB}\psi_{A} \psi_{B} \right] \] and the equations
 (\ref{rHamilt-pole}) are the following: \be\label{eqm1} \dot\psi =
\frac{\pi}{\sin\theta}(1+\rho^2)+ \psi_3\sqrt{1+\rho^2} \ee
\be\label{eqm2} \dot\pi=(\pi\sqrt{1+\rho^2})_{,3}
 + \left[ (1+\rho^2)\sin\theta\psi_3 \right]_{,3}
 +(\sin\theta\kolo\gamma^{AB}\psi_B)_{,A} \ee and they correspond to
 the wave equation (\ref{waveq}).

Although formally the formula (\ref{Hpipsi}) looks similar to the
 usual hamiltonian system in our case there is no possibility to kill
 boundary term.  Our ``hamiltonian'' is not conserved in time:
\be\label{Hdot} -\partial_0\left(\int_\Sigma H \right) =
\int_{\partial \Sigma} \pi^3 \dot\psi =
\int_{S(s,0)} \sin\theta (\dot\psi)^2 \ee ($\partial
\Sigma$=$S(s,0)$ is odd oriented).

Nevertheless this formal calculation is very useful. For example we
 can easily define angular momentum along the $z$-axis as a generator
 for the vector field $\partial \over \partial\phi$:
\[ \int_\Sigma \left(  \pi_{,\phi} \delta \psi -
\psi_{,\phi} \delta \pi \right)=-\delta \int_\Sigma \pi
\psi_{,\phi}
 = -\delta J_z \] Using equations of motion we can check that the
 angular momentum is not conserved in time: \be\label{Jdot}
 -\partial_0 J_z = -\partial_0\left(\int_\Sigma \pi \psi_{,\phi}
\right) =
\int_{\partial \Sigma} \pi^3 \psi_{,\phi} \rightarrow
\int_{S(s,0)} \sin\theta \dot\psi \psi_{,\phi} \ee
 We will show in the sequel that the formulae (\ref{Hdot}),
 (\ref{Jdot}) can be written for linearized gravity and have also
 interpretation as the TB mass loss formula and angular momentum loss
 equation.
 
Equations (\ref{Hdot}) and (\ref{Jdot}) are examples of the general
 formula which has the following form for a vector field
 $X=X^\mu\partial_\mu$:
\be\label{dotHX} \partial_0 \left( \int_\Sigma X^0 H +\pi X^k\psi_k \right) =
\int_{\partial \Sigma} \pi X^3 \dot\psi -\pi^3 X^k\psi_k -\pi^3
 X^0 \dot\psi \ee
\underline{Remark}
 The formula (\ref{dotHX}) defines usual hamiltonian system if the
 field $X$ vanishes on the boundary. For example this situation take
 place for the so-called CMC (constant mean curvature) foliation which
 ``ends'' on the same sphere at $\scri$.

We can take a boost generator along $z$-axis restricted to the
 hyperboloid $\Sigma_0$ (with coordinate $s=0$) \[ X=
 -\rho\sqrt{1+\rho^2}\cos\theta\partial_3
 -\sqrt{1+\rho^2}\sin\theta\partial_{\theta} \] and the formula
 (\ref{dotHX}) takes the form: \be\label{boost0} \partial_0
\left( \int_\Sigma \pi X^k\psi_k \right) =
\int_{\partial \Sigma} \pi X^3 \dot\psi -\pi^3 X^k\psi_k =
\int_{S(0,0)} \sin^2\theta \dot\psi \psi_\theta
\ee Similarily the generator for linear momentum in $z$
 direction: \[ X=-\frac{\cos\theta}{\sqrt{1+\rho^2}}\partial_0
 -\rho^2\cos\theta\partial_3 -\rho\sin\theta\partial_{\theta} \] gives
 the loss formula: \be \label{pz0} \partial_0 P_z =
\int_{\partial \Sigma} \pi X^3 \dot\psi -\pi^3 X^k\psi_k
 -\pi^3 X^0 \dot\psi = -\int_{S(s,0)}\sin\theta\cos\theta (\dot\psi)^2
\ee where $\displaystyle P_z:= \int_\Sigma X^0 H
 +\pi X^k\psi_k$.

The equations (\ref{Hdot}), (\ref{Jdot}), (\ref{boost0}) and
 (\ref{pz0}) express {\it non-conservation} law of the
 ``hamiltonians'' defined at null infinity.
 
Let us formulate the following theorem:\\[1ex]
\underline{Theorem.} If the TB mass is conserved than angular momentum is
 conserved too.\\[1ex] This means that it is impossible to radiate
 away angular momentum without loss of mass. The proof is a simple
 consequance of (\ref{Hdot}) and (\ref{Jdot}).  If the TB mass is
 conserved than (from (\ref{Hdot})) $\dot\psi$ have to vanish on
 $\scri$ and from (\ref{Jdot}) we get that angular momentum is
 conserved.

We shall see in the sequel that this theorem also holds for
 Bondi-Sachs type metrics describing asymptotically flat solutions at
 null infinity for the full (nonlinear) Einstein equations.

\subsection{Scalar field on a null cone}

We shall consider an initial value problem on a null surface $N$:
\be \label{defN}
 N:= \left\{ x\in M \;| \; v=s+\rho^{-1}(1+\sqrt{1+\rho^2})={\rm
 const}
\right\} \ee  where we have introduced a null coordinate 
 $v:=s+\rho^{-1}(1+\sqrt{1+\rho^2}) $ which plays a role of time in
 our analysis. Formally $\scri^+$ corresponds to the surface $\rho=0$.
 Let us rewrite the Minkowski metric (\ref{etaonM}) using new
 coordinates $v,\overline u$ instead of $s,\rho$:
\[ \eta_{\mu\nu}{\rm d}x^\mu{\rm d}x^\nu=
\rho^{-2}\left( -\rho^2{\rm d}v^2 -\rho^2{\rm d}v{\rm
 d}\overline u + {\rm d}\theta^2 +\sin^2\theta {\rm d}\phi^2
\right)
\]
 The relation between coordinates $(v,\overline u)$ and $(x^0,x^3)$
 used in the previous subsection is the following:
\[ v=x^0+\rho^{-1}(1+\sqrt{1+\rho^2}) \, , \quad
\overline u = -2\rho^{-1} \, , \quad \rho=x^3 \] 
\[ \partial_0=\partial_v \, , \quad
\partial_3=2\rho^{-2}\partial_{\overline u} 
 -\rho^{-2}\left(1+\frac 1{\sqrt{1+\rho^2}}\right)\partial_v \]
\[ {\rm d}x^0= {\rm d}v 
 +\frac 12\left(1+\frac 1{\sqrt{1+\rho^2}}\right){\rm d}\overline u \,
 , \quad {\rm d} x^3 =\frac 12 \rho^2\rd\overline u \] The density of
 the lagrangian takes the following form:
\[ \overline L:= -\frac 12 \sqrt{-\det \overline\eta_{\mu\nu}}
\overline\eta^{\mu\nu}\psi_\mu\psi_\nu = 
\sin\theta\left[ \psi_{\overline u}\psi_v - \psi_{\overline u}^2
 -\frac 14 \rho^2 \kolo\gamma^{AB} \psi_A\psi_B \right] \]
 
The formula (\ref{Lpole}) can be written as follows: \[ \delta
\int_N \overline L=\int_N (\pi^v\delta\psi)_{,v} + \int_{\partial N}
\pi^{\overline u} \delta\psi \] and \[ \pi^v=\frac {\partial
 L}{\partial \psi_v}=
\sin\theta\psi_{\overline u}   \]
\[ \pi^{\overline u}=\frac {\partial L}{\partial \psi_{\overline
 u}} =\sin\theta \left( \psi_{v}-2\psi_{\overline u} \right) =
 -\sin\theta\left(\rho^2\psi_3+
\frac{\psi_0}{\sqrt{1+\rho^2}}\right) \]
 In this way we obtain density of the hamiltonian on a cone $N$
\[ H= \pi^v\psi_v - \overline L =
\sin\theta\left[ (\psi_{\overline u})^2 + \frac{\rho^2}{4}
\kolo\gamma^{AB}\psi_{A} \psi_{B} \right] \]
\be\label{HonS} \lim_{\rho\rightarrow 0^+} H=\sin\theta 
 (\psi_{\overline u})^2 = \sin\theta \dot\psi^2 \ee
\be\label{formonS} \int_N \pi\delta\psi=\int_N \sin\theta\psi_{\overline u}
\delta\psi \stackrel{\rho\rightarrow 0^+}{\longrightarrow}
\int_{\scri^+} \sin\theta\dot\psi\delta\psi
\rd{\overline u}\rd\theta\rd\phi \ee
 We will show in the sequel that the above formulae exist in
 electrodynamics and linearized gravity.  Similarily, the equation
\be\label{JonS}
\int_N \pi\psi_{,\phi}=\int_N \sin\theta\psi_{\overline
 u}\psi_{,\phi}
\stackrel{\rho\rightarrow 0^+}{\longrightarrow}
\int_{\scri^+} \sin\theta\dot\psi \psi_{,\phi}
\rd{\overline u}\rd\theta\rd\phi  \ee
 describes the flux of angular momentum through $\scri^+$.

\subsection{ADM mass}
 We have tried to treat separately hyperboloid and scri and we have
 learned that there is no possibility to get a nice hamiltonian
 system. Let $N$ denotes in this subsection a ``piece'' of $\scri^+$
 between $\Sigma$ and $i^0$.  If we take the surface $\Sigma\cup N$
 together:
\begin{equation} \label{HnaV+N} - \delta\left( \int_\Sigma H
 +\int_N H\right) = \int_\Sigma \left( {\dot \pi} \delta \psi - {\dot
\psi} \delta \pi \right) +\int_N \left( {\dot \pi} \delta
\psi - {\dot \psi} \delta \pi \right) +
\int_{\partial \Sigma} \pi^3 \delta \psi +\int_{\partial N}
\pi^{\overline u} \delta \psi   \end{equation}
 we will obtain hamiltonian system provided we can kill boundary term.
 This can be achieved assuming for example that
\[ \lim_{u\rightarrow-\infty} \dot\psi =0 \]
 which simply means that $\dot\psi$ is vanishing at spatial infinity.
 This usually hapens for initial data on Cauchy surface $t=$const with
 compact support or vanishing sufficiently fast at spatial infinity.
 The following relations confirm our theorem:
\[ \left. \pi^3\right|_{\partial \Sigma}=-\sin\theta\dot\psi =
\pi^{\overline u} \] $\partial \Sigma=S(s,0)$, $\partial N
 =S(s,0)\cup S(-\infty,0)$ \begin{equation} \label{MADM} - \delta
 m_{\rm ADM} = \int_{\Sigma\cup N} \left( {\dot \pi} \delta \psi -
 {\dot \psi} \delta
\pi \right) 
 + \int_{\partial (\Sigma\cup N)} \sin\theta\dot\psi \delta \psi
\end{equation} 
 where $m_{\rm ADM}:= \int_{\Sigma\cup N} H$.

\subsubsection{one-parameter family of hamiltonian systems and their limit}
\[
\Sigma_{\tau,\epsilon}:= \left\{ s=\tau, \;  \rho \geq \epsilon \right\}
\]  
\[
 N_{\tau,\epsilon}:= \left\{
 v=\tau+\frac{1+\sqrt{1+\epsilon^2}}{\epsilon}, \;
\frac{\epsilon}{1+\sqrt{1+\epsilon^2}} \leq \rho \leq \epsilon \right\}
\]  
\[
 I_{\tau,\epsilon}:= \left\{ t=\tau, \; 0< \rho \leq
\frac{\epsilon}{1+\sqrt{1+\epsilon^2}}  \right\}
\]  

\[ \lim_{\epsilon\rightarrow 0^+} \Sigma_{\tau,\epsilon}=\Sigma_{\tau} \]

\[ \lim_{\epsilon\rightarrow 0^+} N_{\tau,\epsilon}=N_{\tau}\subset \scri^+ \]

\[ \lim_{\epsilon\rightarrow 0^+} I_{\tau,\epsilon}=i^0 \]

\[
 N_{\tau,\epsilon}:= \left\{
 v=\tau+\frac{1+\sqrt{1+\epsilon^2}}{\epsilon}, \;
\frac{-2\sqrt{1+\epsilon^2}}{\epsilon}+\tau \leq u \leq \tau \right\}
\]  

\[
 I_{\tau,\epsilon}:= \left\{ t=\tau, \; r \geq
\frac{1+\sqrt{1+\epsilon^2}}{\epsilon}  \right\}
\]

$ \Sigma_{\tau,\epsilon}\cup N_{\tau,\epsilon}\cup I_{\tau,\epsilon} $
 is an explicit example of a one-parameter family of surfaces (with
 respect to $\tau$) and the hamiltonian related to this family is an
 ADM mass. On the other hand the hamiltonian system (\ref{HnaV+N}) is
 a limit of these systems with respect to the second parameter
 $\epsilon$ ($\epsilon\rightarrow 0^+$). This way we have certain
 ``finite'' procedure for the hamiltonian system (\ref{HnaV+N}) at
 infinity.

\subsection{Energy--momentum tensor}
 Let us consider energy--momentum tensor for the scalar field
 $\varphi$:
\[ T^\mu{_\nu} = \frac1{\sqrt{-\eta}}\left( p^\mu\varphi_\nu
 -\delta^\mu{_\nu} L \right) \] where $\eta:=\det\eta_{\mu\nu}$ and by
 $\delta^\mu{_\nu}$ we have denoted Kronecker delta.  In our case (for
 scalar field) symmetric and canonical energy momentum tensors are
 equal.

For a Killing vector field $X^\mu$ we can integrate the equation
\[ \partial_\mu \left( \sqrt{-\eta} T^\mu{_\nu}X^\nu \right)=0 \]
 and obtain as follows:
\[ \partial_0 \int_\Sigma \sqrt{-\eta} T^0{_\nu}X^\nu
 =-\int_{\partial\Sigma} \sqrt{-\eta} T^3{_\nu}X^\nu \]

It can be easily verified that for energy and angular momentum we have
 respectively
\[ \int_\Sigma \sqrt{-\eta} T^0{_0} = \int_\Sigma H \]
\[ \int_\Sigma\sqrt{-\eta} T^0{_\phi} =\int_\Sigma \pi \psi_{,\phi}  \]
 and the boundary terms can be expressed in terms of energy--momentum
 tensor
\[ -\int_{\partial\Sigma} T^3{_0}\rho^{-4}\sin\theta\rd\theta\rd\phi =
 -\frac 12 \int_{\partial\Sigma} T^{\overline u}{_v}\rho^{-2}
\sin\theta\rd\theta\rd\phi  \;\left(=\int_{\partial\Sigma\subset\scri^+}
\dot\psi^2\sin\theta\rd\theta\rd\phi \right)
\]
\[ \int_N T^v{_v}\rho^{-4}\rd\rho\sin\theta\rd\theta\rd\phi =
\int_N \rd\overline u \left( \frac 12 T^v{_v}\rho^{-2} \sin
\theta \rd
\theta\rd\phi \right) \quad \mbox{and} \quad 
\left.\frac 12\rho^{-2} T^v{_v}\right|_{\scri^+}=\dot\psi^2
\]
 Similarily for angular momentum:
\[ -\int_{\partial\Sigma} T^3{_\phi}\rho^{-4}\sin\theta\rd\theta\rd\phi =
 -\frac 12 \int_{\partial\Sigma} T^{\overline u}{_\phi}\rho^{-2}
\sin\theta\rd\theta\rd\phi \;\left(=\int_{\partial\Sigma\subset\scri^+}
\dot\psi \psi_{,\phi} \sin\theta\rd\theta\rd\phi \right)
\]
\[ \int_N T^v{_\phi}\rho^{-4}\rd\rho\sin\theta\rd\theta\rd\phi =
\int_N \rd\overline u \left( \frac 12 T^v{_\phi}\rho^{-2}
\sin \theta \rd \theta\rd\phi \right) \quad \mbox{and} \quad 
\left.\frac 12\rho^{-2} T^v{_\phi}\right|_{\scri^+}=\dot\psi
\psi_{,\phi}
\]
 Some of the equalities above hold for any energy-momentum tensor not
 only for the scalar field. Compare with (\ref{Hdot}--\ref{Jdot}) and
 (\ref{HonS}--\ref{JonS}) for scalar field.
 
This calculation shows that {\em quasi-local} density of the energy on
 $\scri^+$ has two different interpretations. It is a boundary term
 which describes non-conservation of the ``hamiltonian'' on a
 hyperboloid $\Sigma$ or a density of a ``hamiltonian'' on $\scri^+$
 as a limit of $N$.  More precisely, it is a density with respect to
 the parameter $\overline u$ but integrated over sphere. This is an
 example of an object which is local on $M_2$ but non-local on $S^2$.
 We call such objects {\em quasi-local}. It will be shown in the
 sequel that this concept of {\em quasi-locality} is useful in
 electrodynamics and gravitation.
\section{Electrodynamics}
 This section should convince the reader that the TB mass and angular
 momentum at null infinity can be described in classical
 electrodynamics in a similar way as the scalar field in previous
 section.

Field equations for linear electrodynamics may be written as follows:
\begin{equation}  \label{rsymp}
\delta L = \partial_{\mu}({\cal F}^{\nu \mu}\delta A_{\nu}) =
\partial_{\mu}({\cal F}^{\nu \mu})\delta A_{\nu} +
 {\cal F}^{\nu \mu}\delta A_{\nu \mu}\, ,
\end{equation}
 where $A_{\nu \mu} := \partial_{\mu}A_{\nu}$ and $L$ is the
 Lagrangian density of the theory. The above formula (see
\cite{Kij-Tulcz}) is a convenient way to write the Euler-Lagrange
 equations
\begin{equation}
\partial_{\mu}{\cal F}^{\nu \mu} = \frac{\delta L}{\delta A_{\nu}}
\end{equation}
 together with the relation between the electromagnetic field
 $f_{\mu\nu}=A_{\nu\mu}-A_{\mu\nu}$ and the electromagnetic induction
 density ${\cal F}^{\nu \mu}$ describing the momenta canonically
 conjugate to the potential:
\begin{equation}    \label{rel}
 {\cal F}^{\nu \mu} = \frac{\delta L}{\delta A_{\nu \mu}}\, .
\end{equation}
 For the linear Maxwell theory the Lagrangian density is given by the
 standard formula:
\begin{equation}
 L = -\frac{1}{4}\sqrt{-\eta}f^{\mu \nu}f_{\mu \nu}\, ,
\end{equation}
 and relation (\ref{rel}) reduces in this case to ${\cal F}^{\mu
\nu}
 := \sqrt{-\eta}\eta^{\mu \alpha}\eta^{\nu \beta}f_{\alpha
\beta}$.

Integrating (\ref{rsymp}) over $V$ we obtain:
\begin{eqnarray}    \label{VL1}
\delta\,\int_{V}L & = & \int_{V}\partial_{0}({\cal F}^{k 0}\delta
 A_{k}) + \int_{\partial V}{\cal F}^{\nu 3}\delta A_{\nu} =\nonumber\\
 & = &
\int_{V}\partial_{0}({\cal F}^{B 0}\delta A_{B} + {\cal F}^{3
 0}\delta A_{3}) + \int_{\partial V}({\cal F}^{B 3}\delta A_{B} +
 {\cal F}^{0 3}\delta A_{0})\, .
\end{eqnarray}

We assume that the charge $e$ defined by the surface integral
\begin{equation}   \label{e0}
 e:= \int_{S(s,\rho)}{\cal F}^{03}
\end{equation}
 vanishes. The situation with $e\neq 0$ can be described similarily as
 in
\cite{darek} but we are interested in ``wave'' degrees of freedom and we
 are going to show how the volume part of (\ref{VL1}) can be reduced
 to the gauge-invariant quantities.

Let $ {\bf a}:=\rho^{-2}\triangle$ where $\triangle$ denotes the
 2-dimensional Laplace-Beltrami operator on a sphere $S(s,\rho)$ and
 one can easily check that the operator $\bf a$ does not depend on
 $\rho$ and is equal to the Laplace-Beltrami operator on the unit
 sphere $S(1)$. Operator $\bf a$ is invertible on the space of
 monopole--free functions (functions with a vanishing mean value on
 each $S(s,\rho)$).

Let us denote by $\varepsilon^{AB}$ the Levi--Civita antisymmetric
 tensor on a sphere $S(s,\rho)$).  We can rewrite (\ref{VL1}) in the
 following way provided that the electric charge $e$ vanishes.
\begin{eqnarray}
\delta\,\int_{V}L & = & \int_{V}\partial_{0}\left[ {\cal F}^{0B},_{B}{\bf
 a}^{-1} \delta\rho^{-2} A^B{_{||B}} + {\cal F}^{3 0}\delta A_{3} +
 {\cal F}^{0B},_{C}\varepsilon_{B}^{\ C} {\bf a}^{-1}
\delta\rho^{-2} (\varepsilon^{AB}A_{A||B})\right] +\nonumber\\
 & + & \int_{\partial V}({\cal F}^{3B},_{B}{\bf a}^{-1} \delta
\rho^{-2} A^B{_{||B}} + {\cal F}^{03}\delta A_{0} 
 + {\cal F}^{3B},_{C}\varepsilon_{B}^{\ C} {\bf a}^{-1}
\delta \rho^{-2}(\varepsilon^{AB}A_{A||B}))
\end{eqnarray}
 Here, by ``$||$''we denote the 2-dimensional covariant derivative on
 each sphere $S(s,\rho)$.  Using identities $\partial_{B}{\cal F}^{B0}
 + \partial_{3}{\cal F}^{30} = 0$ and $\partial_{B}{\cal F}^{B3} +
\partial_{0}{\cal F}^{03} = 0$
 implied by the Maxwell equations and again integrating by parts we
 finally obtain:
\begin{eqnarray}   \label{a2}
\delta\,\int_{V}L & = & \int_{V}\partial_{0}\left[{\cal
 F}^{30} {\bf a}^{-1} \delta ({\bf a}A_{3} - (\rho^{-2}
 A^B{_{||B}})_{,3}) + ({\cal F}^{0B||C}\varepsilon_{BC}) {\bf a}^{-1}
\delta\rho^{-2}(\varepsilon^{AB}A_{A||B})\right] +\nonumber\\ & + &  
\int_{\partial V}\left[{\cal F}^{0 3}{\bf a}^{-1}\delta
 ({\bf a}A_{0} - \rho^{-2} A^B{_{||B}},_{0}) + ({\cal
 F}^{3B||C}\varepsilon_{BC}) {\bf a}^{-1} \delta
\rho^{-2}(\varepsilon^{AB}A_{A||B})\right]\, .  
\end{eqnarray}
 The quantities ${\bf a}A_{0} - \rho^{-2} A^B{_{||B,0}}$ and $({\bf a}
 A_{3} - (\rho^{-2} A^B{_{||B}})_{,3})$ are gauge invariant and it may
 be easily checked that
\[
\sin\theta \left( {\bf a}A_{0} - \rho^{-2}A^B{_{||B,0}}\right) =
\rho^2 {\cal F}^A{_{0||A}} \; (=\pi^3)
\]
 and
\[
\sin\theta \left[{\bf a}A_{3} - (\rho^{-2} A^B{_{||B}})_{,3}\right] =
\rho^2 {\cal F}^A{_{3||A}} \; (=\pi)
\]
 Let us introduce the following gauge invariants:
\[ \psi := {\cal F}^{30}/\sin\theta \]
\[ *\psi :=\rho^{-2}\varepsilon^{AB}A_{B||A}=- *{\cal F}^{30}/\sin\theta \]
\[ \pi:= -\rho^2 {\cal F}_3{^A}{_{||A}} \]
\[ *\pi:= {\cal F}^{0B||C}\varepsilon_{BC}=\rho^2*{\cal F}_3{^A}{_{||A}} \]
 where
\[ *{\cal F}^{\mu\nu}=\frac 12 \varepsilon^{\mu\nu\lambda\sigma}
 {\cal F}_{\lambda\sigma} \]

Now we will show how the vacuum Maxwell equations
\[ \partial_{\mu}{\cal F}^{\mu\nu}=0 \, , \quad 
\partial_{\mu}*{\cal F}^{\mu\nu}=0 \] allow to introduce
 equations for gauge-invariants and the result is analogous to
 (\ref{eqm1}) and (\ref{eqm2}) describing scalar field
\begin{equation}\label{el1} \dot\psi = \frac{\pi}{\sin\theta}(1+\rho^2)+
\psi_3\sqrt{1+\rho^2} \end{equation}
\begin{equation}\label{el2} \dot\pi=(\pi\sqrt{1+\rho^2})_{,3}
 + \left[ (1+\rho^2)\sin\theta\psi_3 \right]_{,3} +\sin\theta{\bf
 a}\psi \end{equation} \begin{equation}\label{el1*} *\!\dot\psi =
\frac{*\!\pi}{\sin\theta}(1+\rho^2)+ *\!\psi_3\sqrt{1+\rho^2}
\end{equation}
\begin{equation}\label{el2*} *\!\dot\pi=(*\!\pi\sqrt{1+\rho^2})_{,3}
 + \left[ (1+\rho^2)\sin\theta *\!\psi_3 \right]_{,3} +\sin\theta{\bf
 a}*\!\psi \end{equation} The proof of (\ref{el1}) is based on the
 following observations: \[
\sin\theta\psi_{,0}={\cal F}^{30}{_{,0}}=-{\cal F}^{3A}{_{,A}}=
\frac{\pi}{\sin\theta}(1+\rho^2)+ \psi_3\sqrt{1+\rho^2} \] \[
 -\pi=\rho^2{\cal F}_3{^A}{_{||A}}={ {\cal F}^{3A}{_{||A}}\over
 1+\rho^2} +{ {\cal F}^{0A}{_{||A}} \over \sqrt{1+\rho^2} } \] and \[
\sin\theta\psi_3={\cal F}^{30}{_{,3}}={\cal
 F}^{0A}{_{||A}} \] Similarily for (\ref{el2}) we have the following
 relations: \[ -\left( *{\cal
 F}^{0A}{_{||B}}\varepsilon_A{^B}\right)_{,0} - \left( *{\cal
 F}^{3A}{_{||B}}\varepsilon_A{^B}\right)_{,3} + (*{\cal
 F}^{AB}{_{||BC}}\varepsilon_A{^C})_{,0} =0 \] \[ *{\cal
 F}^{0A}=-\rho^2\varepsilon^{AB}{\cal F}_{3B}\] \[ *{\cal
 F}^{3A}=\rho^2\varepsilon^{AB}{\cal F}_{0B}\] \[ *{\cal F}^{AB}=
\rho^2\varepsilon^{AB}{\cal F}_{03}\]
\[ {\cal F}_{03}= -\rho^{-4} {\cal F}^{03} \] which allow to get
 an equation \[ \left({ {\cal F}^{3A}{_{||A}}\over 1+\rho^2} +{ {\cal
 F}^{0A}{_{||A}} \over \sqrt{1+\rho^2} } \right)_{,0} +
\left(\rho^2 {\cal F}^{0A}{_{||A}} -
 { {\cal F}^{3A}{_{||A}}\over \sqrt{1+\rho^2} } \right)_{,3} + {\bf a}
 {\cal F}^{30}=0 \] which is equivalent to (\ref{el2}).\\ For $(*\psi,
 *\pi)$ the proof is the same provided we apply the Hodge dual $*$ for
 the variables and equations: \[ (\pi,\psi)
\stackrel{-*}{\longrightarrow} (*\pi, *\psi)
\stackrel{*}{\longrightarrow} (\pi,\psi) \]
  
Now we will show how our variables appear in formula (\ref{a2}).  Let
 us perform the Legendre transformation in the volume $V$
\begin{eqnarray}
 -{\cal F}^{03}\delta \left[ A_3 -{\bf a}^{-1}(\rho^{-2}
 A^B{_{||B}})_{,3} \right] & = & -\delta\left[ {\cal F}^{03}\left( A_3
 -{\bf a}^{-1}(\rho^{-2} A^B{_{||B}})_{,3}
\right) \right] +
\nonumber \\ & & +\left[ A_3 -{\bf a}^{-1}(\rho^{-2}
 A^B{_{||B}})_{,3} \right] \delta {\cal F}^{03} \nonumber
\end{eqnarray}
 and on the boundary $\partial V$
\begin{eqnarray}
 {\cal F}^{03}\delta \left( A_{0} -{\bf a}^{-1}
\rho^{-2}A^B{_{||B,0}}\right)
 & = & \delta\left[ {\cal F}^{03}\left( A_{0} -{\bf a}^{-1}
\rho^{-2}A^B{_{||B,0}}\right) \right] +\nonumber \\
 & & - \left( A_{0} -{\bf a}^{-1}
\rho^{-2}A^B{_{||B,0}}\right)\delta{\cal
 F}^{03} \, . \nonumber
\end{eqnarray}
 This way the formula (\ref{a2}) may be written as follows: 
\begin{eqnarray}   \label{a3}
\lefteqn{\delta \int_{V}\left[ L -  \partial_{0}( \psi{\bf a}^{-1}\pi)
 - \partial_{3}(\psi{\bf a}^{-1}\rho^2 {\cal F}^A{_{0||A}} )\right] =
 - \int_{V} \partial_{0}\left[ \pi{\bf a}^{-1} \delta
\psi +
 *\pi {\bf a}^{-1} \delta *\psi
\right] + }\nonumber\\ & & + \int_{\partial V} \left[
\rho^2 {\cal F}^A{_{0||A}} {\bf a}^{-1} \delta \psi
 - {\cal F}^{3A||B}\varepsilon_{AB}\delta *\!\psi \right] \, .
\end{eqnarray}

Finally we have obtained the following variational principle:
\begin{equation}   \label{a5}
\delta \int_{V} {\overline L}  =
 -\int_{V}\partial_{0}(\pi{\bf a}^{-1}\delta \psi+ *\!\pi{\bf
 a}^{-1}\delta *\!\psi) + \int_{\partial V} \pi^{3} {\bf a}^{-1}
\delta \psi 
 + *\!\pi^{3} {\bf a}^{-1} \delta \! *\!\psi \, ,
\end{equation}
 where the lagrangian $\overline L$ is defined by
\begin{equation}   \label{Lel}
 {\overline L} = L - \partial_{0}\left(\psi{\bf a}^{-1}\pi
\right)-
\partial_{3}\left(\psi{\bf a}^{-1}\rho^2 {\cal F}^A{_{0||A}} \right)  \, .
\end{equation}
 and boundary momenta:
\[ \pi^3:=-\rho^2 {\cal F}_0{^A}{_{||A}}=\rho^2{\cal F}^{0A}{_{||A}}
 -\frac{{\cal F}^{3A}{_{||A}}}{\sqrt{1+\rho^2}}= \sin\theta
\left(
\frac{\dot\psi}{\sqrt{1+\rho^2}}+\rho^2\psi_3 \right) \]
\[ *\!\pi^3 :=  - {\cal F}^{3A||B}\varepsilon_{AB} = \rho^2 *\!{\cal
 F}_0{^A}{_{||A}} = \sin\theta \left( \frac{*\!
\dot\psi}{\sqrt{1+\rho^2}}+\rho^2*\!\psi_3 \right) \]

From lagrangian relation (\ref{a5}) we immediately obtain the
 hamiltonian one performing the Legendre transformation:
\begin{eqnarray}    \label{a6}
 -\,\delta \int_V H = - \int_{V} \dot{\pi}{\bf a}^{-1}\delta\psi
 -\dot{\psi}{\bf a}^{-1}\delta\pi+ *\!\dot{\pi}{\bf a}^{-1}\delta
 *\!\psi - *\!\dot{\psi}{\bf a}^{-1}\delta*\!\pi + \int_{\partial V}
\pi^{3}{\bf a}^{-1}\delta \psi + *\!\pi^{3}{\bf a}^{-1}\delta
 *\!\psi
\end{eqnarray}
 where
\begin{equation}    \label{a7}
 H := - \pi{\bf a}^{-1}\dot{\psi} - *\!\pi{\bf a}^{-1} *\!\dot{\psi} -
 {\overline L} \, , \end{equation} is the density of the hamiltonian
 of the electromagnetic field on a hyperboloid.

The value of $\int_V H$ is equal to the amount of electromagnetic
 energy contained in a volume $V$ and defined by the energy--momentum
 tensor.  \[ T^\mu{_\nu}= f^{\mu\lambda}f_{\lambda\nu} + \frac 14
\delta^\mu_\nu
 f^{\kappa\lambda} f_{\kappa\lambda} \]

We are not surprised that the quantity $H$ is related to $T^0{_0}$
\[ \int_{S(s,\rho)} H \rd\theta\rd\phi = \int_{S(s,\rho)} \sqrt{-\eta}
 T^0{_0} \rd\theta\rd\phi \] and to prove it we can use the following
 identity
\[ \rho^{-4}\sin\theta\left[ *\!\pi{\bf a}^{-1} *\!\dot{\psi} -
\dot\pi{\bf a}^{-1}{\psi}- \partial_{3}\left(\psi{\bf
 a}^{-1}\rho^2 {\cal F}^A{_{0||A}} \right) \right] =\] \[ = {\cal
 F}^{03}{\cal F}_{03} -{\cal F}^{0A}{_{||A}} {\bf
 a}^{-1}\rho^{-2}{\cal F}_0{^A}{_{||A}} -{\cal
 F}^{0A||B}\varepsilon_{AB}{\bf a}^{-1}{\cal F}_{0A||B}
\varepsilon^{AB} \]  

The non-conservation law for the energy we can denote as follows:
\be\label{HdotEM}
 -\partial_0 \int_\Sigma H = \int_{\partial\Sigma} \sin\theta
\left(
\dot\psi {\bf a}^{-1} \dot\psi 
 +{*\!\dot\psi} {\bf a}^{-1}{*\!\dot\psi} \right)=
 -\int_{S(s,0)}\left( \dot\psi {\bf a}^{-1} \dot\psi +{*\!\dot\psi}
 {\bf a}^{-1}{*\!\dot\psi} \right)
\sin\theta\rd\theta\rd\phi \ee

For angular momentum defined by
\[ J_z := - \int_{\Sigma} \pi{\bf a}^{-1}{\psi}_{,\phi} +
 *\!\pi{\bf a}^{-1}{*\!\psi}_{,\phi}
\]
 we have a similar formula
\be\label{JdotEM} -\partial_0 J_z=
\partial_0 \int_{\Sigma} \pi{\bf a}^{-1}{\psi}_{,\phi} +
 *\!\pi{\bf a}^{-1}{*\!\psi}_{,\phi} = -\int_{S(s,0)} \sin\theta
\left( \dot\psi{\bf a}^{-1}{\psi}_{,\phi} + *\!\dot\psi{\bf a}^{-1}
 {*\!\psi}_{,\phi} \right)\rd\theta\rd\phi \ee but the relation with
 symmetric energy-momentum tensor is not so obvious.
\[ {\tilde J}_z := \int_\Sigma \sqrt{-\eta} T^0{_\phi} =
\int_\Sigma {\cal F}^{03} f_{3\phi} + *{\cal F}^{03}\,
 *f_{3\phi} \] Using the relations
\[ \pi =-\rho^{-2}\sin\theta f_3{^A}_{||A} \quad
\psi_{,3}=-\rho^{-2}\varepsilon^{AB} f_{3A||B} \]
\[ *\!\pi = \rho^{-2}\sin\theta *\! f_3{^A}_{||A} \quad
 *\! \psi_{,3}=-\rho^{-2}\varepsilon^{AB} *\! f_{3A||B} \] we can
 express $\tilde J_z$ in terms of $(\pi ,\psi, *\!\pi ,*\!\psi)$ as
 follows: \[ \tilde J_z= \int_\Sigma \psi \left( {\bf
 a}^{-1}{\pi}_{,\phi} +\sin\theta\hat{\partial}_{\phi}{\bf
 a}^{-1}{*\!\psi}_{,3}\right) +
\int_\Sigma  *\!\psi \left( {\bf a}^{-1}{*\!\pi}_{,\phi} -\sin\theta
\hat{\partial}_{\phi}{\bf a}^{-1}{*\!\psi}_{,3}\right) =\]
\[
 = -\int_\Sigma \pi{\bf a}^{-1}{\psi}_{,\phi} + *\!\pi{\bf
 a}^{-1}{*\!\psi}_{,\phi} +\int_{\partial \Sigma} \sin\theta
\psi\hat{\partial}_{\phi}{\bf a}^{-1}{*\!\psi}  \]
 where $\hat{\partial}_A:=\varepsilon_A{^B}\partial_B$ and we have
 used the identity
\[ \int_{S^2}
\sin\theta \psi_{,3}  \hat{\partial}_{\phi} {\bf a}^{-1} *\!\psi =
 - \int_{S^2} \sin\theta {*\!\psi} \hat{\partial}_{\phi}{\bf
 a}^{-1}\psi_{,3} \]

The boundary term $ \int_{\partial \Sigma} \sin\theta
\psi \hat{\partial}_{\phi}{\bf a}^{-1}{*\!\psi} $ usually has to vanish if we
 want to interprete the integral $\int_\Sigma \sqrt{-\eta} T^0{_\phi}$
 as an angular momentum generator but in our case $\psi$, $*\!\psi$ do
 not vanish on $\scri^+$ and we obtain in general two different
 definitions of angular momenta $J_z$ and $\tilde J_z$. 

Let us observe that
\[ A_\phi= \left(\rho^{-2}A^B{_{||B}}\right)_{,\phi}
 -\hat{\partial}_{\phi}*\!\psi \] and
\[ \int_V \left({\cal F}^{\lambda 0} A_\phi\right)_{,\lambda} =
\int_{\partial V} \sin\theta \psi {\bf a}^{-1}
\rho^{-2}\left(A^B{_{||B}}\right)_{,\phi} - \int_{\partial V}
\sin\theta \psi  \hat{\partial}_{\phi} {\bf a}^{-1} *\!\psi 
\] so the angular momentum $J_z$ is related rather to the
 canonical energy-momentum tensor with gauge $A^B{_{||B}}=0$ than to
 the symmetric one. More precisely, the canonical energy-momentum
 density ${\cal T}^\mu{_\nu}$ is related with symmetric one as
 follows:
\[ {\cal T}^\mu{_\nu}={\cal F}^{\lambda\mu} A_{\lambda,\nu}-\frac 14
\delta^\mu{_\nu} L= \sqrt{-\eta} T^\mu{_\nu} +\left({\cal F}^{\lambda\mu}
 A_\nu\right)_{,\lambda} \] For angular momentum we obtain:
\[ \int_\Sigma  {\cal T}^0{_\phi} =
 -\int_\Sigma \pi{\bf a}^{-1}{\psi}_{,\phi} + *\!\pi{\bf
 a}^{-1}{*\!\psi}_{,\phi} -\int_{\partial \Sigma }
\sin\theta{\psi}_{,\phi}{\bf a}^{-1}\rho^{-2}
 A^B{_{||B}} = J_z +\]
\[ +\int_{S(s,0)} {\psi}_{,\phi}{\bf a}^{-1}\rho^{-2}
 A^B{_{||B}}\sin\theta\rd\theta\rd\phi \] Let us observe that if
 $\psi_{,\phi}$ and $*\!\psi_{,\phi}$ are vanishing on $\scri^+$ then
 $J_z$ is well defined in terms of the canonical energy-momentum
 tensor density ${\cal T}^0{_\phi}$ and conserved.

\subsection{Electrodynamics on a null surface}
 We will show now how the formula (\ref{formonS}) can be obtained in
 classical electrodynamics. Let us consider volume $V\subset N$ where
 $N$ has been already defined by (\ref{defN}).
\[ \int_V {\cal F}^{v\nu}\delta A_\nu \rd \rho\rd \theta \rd \phi =
\int_{\partial V} {\cal F}^{v\rho}{\bf a}^{-1} \delta \rho^{-2}
 A^B{_{||B}} \rd \theta \rd\phi + \] \[ + \int_V {\cal F}^{v\rho}
\delta \left[
 A_\rho -{\bf a}^{-1}(\rho^{-2}A^B{_{||B}})_{,\rho}\right]
\rd\rho\rd\theta\rd\phi
 -\int_V {\cal F}^{vA||B}\varepsilon_{AB} {\bf a}^{-1} \delta
\rho^{-2} A_{A||B}\varepsilon^{AB}
\rd\rho\rd\theta\rd\phi \] and similarily as on $\Sigma$ \[ {\bf
 a} A_\rho -(\rho^{-2}A^B{_{||B}})_{,\rho}=\frac 1{\sin\theta} {\cal
 F}^{vA}{_{||A}}=\psi_{,\rho}=2\rho^{-2}\psi_{,\overline u}
\]
 where $\overline u=-\frac 2\rho$ and
 $\partial_\rho=2\rho^{-2}\partial_{\overline u}$.  For dual degree of
 freedom the similar relation holds \[ {\cal
 F}^{vA||B}\varepsilon_{AB}=*{\cal F}^{vA}{_{||A}}=2\sin\theta
\rho^{-2} *\!\psi_{,\overline u} \]
 We obtain gauge--independent part $+$ boundary term $+$ full
 variation \[ \int_V {\cal F}^{v\nu}\delta A_\nu \rd \rho\rd
\theta \rd \phi =
\int_{\partial V} {\cal F}^{v\rho}{\bf a}^{-1} \delta \rho^{-2}
 A^B{_{||B}} \rd \theta \rd\phi -\delta\int_V \sin\theta\psi {\bf
 a}^{-1}
\psi_{,u} \rd u \rd\theta\rd\phi + \] 
\be + \int_V \sin\theta \left( \psi_{,u}{\bf a}^{-1}\delta\psi +
 *\!\psi_{,u}{\bf a}^{-1}\delta*\!\psi \right) \rd u
\rd\theta\rd\phi
\label{formonSEM} \ee This equality means that modulo boundary
 term and full variation we can reduce our form on $\scri^+$ and the
 final form is similar to (\ref{formonS}) and posseses {\em
 quasi-local} character.  ($u=v+\overline u$ so on the surface
 $v=const.$ we can use coordinate $u$ as well as $\overline u$ and
 this observation refers to the objects on $N$ but not on $M$).

Now we will show how the flux of energy through $\scri^+$ is related
 with energy-momentum tensor, similarily to section 2.5.

\[ T^v{_v} = \frac 12 \rho^{-4} (f^{v\rho})^2 +\frac 12 \rho^{-4}
 (*\! f^{v\rho})^2 + \frac 12 \eta_{AB} f^{vA}f^{vB} \]

\[ \int_{V} T^v{_v} \rho^{-4}\sin\theta \rd \rho\rd\theta\rd\phi =
 -\int_{V} \sin\theta \rd u \rd\theta\rd\phi \left( \psi_{,u}{\bf
 a}^{-1}
\psi_{,u} +  *\!\psi_{,u}{\bf a}^{-1} *\!\psi _{,u}  \right) +\]
\[
 + \frac 14 \int_V \rho^2 \sin\theta \rd u \rd\theta\rd\phi
\left( \psi^2 +
 *\!\psi^2 \right) \] the last term vanishes on scri \[ \rho^2
\left( \psi^2 + *\!\psi^2 \right) \stackrel{\rho\rightarrow
 0^+}{\longrightarrow} 0 \] so
\be \label{TvvEM}
\int_{\scri^+} \sqrt{-\eta} T^v{_v} =
 -\int_{\scri^+} \rd u \left[\sin\theta \rd\theta\rd\phi \left(
\psi_{,u}{\bf a}^{-1} \psi_{,u} +  *\!\psi_{,u}{\bf a}^{-1} *\!\psi _{,u}
\right) \right] \ee
 The integral on a sphere in quadratic brackets represents {\em
 quasi-local} density of the flux of the energy through $\scri^+$. The
 main difference comparing with scalar field is that here there is no
 possibility to work with local density because of the operator ${\bf
 a}^{-1}$ and only quasi-local object assigned to a sphere can be well
 defined.

\section{Linearized gravity on a hyperboloid}
 We start from ADM formulation of the initial value problem for
 Einstein equations \cite{ADM}.  In subsection 1 we introduce the
 hyperboloidal slicing and in subsection 2 we consider an initial
 value problem for the linearized Einstein equations on it.  In
 subsection 3 we discuss ``charges'' on the hyperboloid and in the
 next two sections we introduce invariants which describe reduced
 dynamics. In subsection 6 we derive ``hamiltonian'' in terms of gauge
 invariant quantities.

\subsection{Hyperboloidal conventions}
 The flat Minkowski metric of the following form in spherical
 coordinates:
\be \eta_{\mu\nu}\rd y^\mu \rd y^\nu = -\rd t^2 +\rd r^2 + r^2 (\rd
\theta^2 +\sin^2\theta \rd \phi^2) \ee
 with $ r=\sinh\omega$, $t=s+\cosh\omega $ already defined in section
 2, can be expressed in coordinates $s,\omega$ well adopted to
 ``hyperboloidal'' slicing of Minkowski spacetime $M$:
\be \eta_{\mu\nu}\rd y^\mu \rd y^\nu = -\rd s^2 - 2\sinh\omega \, \rd s \, \rd
\omega +\rd\omega^2 +\sinh^2\omega (\rd \theta^2 +\sin^2\theta \rd \phi^2) \ee

In this section we use different coordinate $\omega$ instead of $\rho$
 used previously but at the end we return to $\rho$ to compare the
 results for the scalar field and linearized gravity.  Let us fix a
 coordinate chart $(y^\mu)$ on $M$ such that $y^1=\theta$, $y^2=\phi$
 (spherical angles), $y^3=\omega$ and $y^0=s$.  So we have \be
\label{Hs} \Sigma_s:=\{ y\in M \; : \;
 y^0=s \} =\bigcup_{\omega \in [0,\infty[} S_s(\omega) \quad
\mbox{where} \; \; S_s(\omega):= \{ y \in \Sigma_s \; : \;
 y^3=\omega \} \ee and $\Sigma_s$ is a three--dimensional hyperboloid,
 $S_s(\omega)=S(s,\frac1{\sinh\omega})$ and $\partial
\Sigma_s=S_s(\infty)=S(s,0)$.  

We use the similar convention for indices (as for coordinates
 $(x^\mu)$), namely: greek indices $\mu, \nu, \ldots$ run from 0 to 3;
 $k,l, \ldots$ are coordinates on $\Sigma$ and run from 1 to 3;
 $A,B,\ldots$ are coordinates on $S(r)$ and run from 1 to 2.

Hyperboloid $\Sigma$ has a very simple geometry. The induced
 Riemannian metric $\eta_{kl}$ on $\Sigma$ in our coordinates takes
 the following form:
\be \label{eta3}
\eta_{kl}\rd y^k \rd y^l= \rd\omega^2 +\sinh^2\omega (\rd
\theta^2 +\sin^2\theta \rd \phi^2) 
\ee The hypersurface $\Sigma$ is a constant curvature space and
 the three--dimensional curvature tensor of $\Sigma$ can be expressed
 by the metric:
\be {^3}\! R_{ijkl}=\eta_{jk}\eta_{il} -\eta_{ik}\eta_{jl} \ee
 
\subsection{ADM formulation for linearized gravity on a hyperboloid} 

Let $(g_{kl}, {P}^{kl})$ be the Cauchy data for Einstein equations on
 a 3--dimensional hyperboloid $\Sigma$. This means that $g_{kl}$ is a
 riemanian metric on $\Sigma$ and $P^{kl}$ is a symmetric tensor
 density which we identify with the A.D.M.  momentum \cite{ADM}, i.e.
\[
 {P}^{kl} = \sqrt{\det g_{mn}} ({g}^{kl} Tr K - K^{kl})
\]
 where $K_{kl}$ is the second fundamental form (external curvature) of
 the imbedding of $\Sigma$ into a spacetime $M$ which is now curved.
\\ The 12 functions $(g_{kl}, {P}^{kl})$
 must fulfill 4 Gauss--Codazzi constraints:
\begin{equation} 
 P_i{^l}{_{| l}} = 8\pi\sqrt{\det g_{mn}}\, T_{i\mu}n^{\mu}
\label{ww}
\end{equation}
\begin{equation} 
 (\det g_{mn}){\cal R}- {P}^{kl}{P}_{kl} + \frac{1}{2}
 ({P}^{kl}{g}_{kl})^2 = 16\pi(\det g_{mn}) T_{\mu
\nu}n^{\mu}n^{\nu}
\label{ws}
\end{equation}
 where $T_{\mu\nu}$ is an energy momentum tensor of the matter, by
 $\cal R$ we denote the (three--dimensional) scalar curvature of
 $g_{kl}$, $n^\mu$ is a future timelike four--vector normal to the
 hypersurface $\Sigma$ and the calculations have been made with
 respect to the three--metric $g_{kl}$ ("$|$" denotes covariant
 derivative, indices are raised and lowered etc.).

Einstein equations and the definition of the metric connection imply
 the first order (in time) differential equations for $g_{kl}$ and
 $P^{kl}$ (see \cite{ADM} or \cite{MTW} p. 525) and contain the lapse
 function $N$ and the shift vector $N^k$ as parameters:

\be
\label{gdot}
\dot{g}_{kl}=\frac{2N}{\sqrt{g}}\left( P_{kl} -\frac 12 g_{kl} P \right) +
 N_{k|l} +N_{l|k}
\ee
 where $g:= \det g_{mn}$ and $P:= {P}^{kl}{g}_{kl}$
\[
\dot{P}^{kl} = -N\sqrt{g} \left( {\cal R}^{kl} -\frac 12 g^{kl}{\cal R}
\right)  - \frac{2N}{\sqrt{g}}\left( {P}^{km}{P}_{m}{^l}-
\frac{1}{2} P P^{kl} \right) + \left( P^{kl} N^m \right)_{|m} + \]
\[
 + \frac{N}{2\sqrt{g}}g^{kl} \left( {P}^{kl}{P}_{kl}-
\frac{1}{2}{P}^2 \right)
 -N^k{_{|m}} P^{ml}-N^l{_{|m}} P^{mk} + \sqrt{g} \left( N^{|kl}
 -g^{kl} N^{|m}{_{|m}} \right)+ \]
\be\label{Pdot}
 + 8\pi N\sqrt{g} T_{mn}g^{km}g^{ln}
\ee

We want to consider an initial value problem for the linearized
 Einstein equations on the hyperboloidal slicing introduced in the
 previous section.  For this purpose let us check first that on this
 slicing the ADM momentum $P^{kl}$ for background flat Minkowski
 spacetime on each hyperboloid $\Sigma_s$ is no longer trivial:
\be \label{Ptr}
 P^{kl}=-2\sqrt{g} g^{kl} \ee and
\be \label{gtr} g_{kl}\rd y^k \rd y^l = \eta_{kl}\rd y^k \rd y^l=
\rd\omega^2 +\sinh^2\omega (\rd \theta^2 +\sin^2\theta \rd
\phi^2) \ee
 where $g^{kl}$ is three--dimensional inverse of $g_{kl}$.

Let us define the linearized variations $(h_{kl},\varpi^{kl} )$ of the
 full nonlinear Cauchy data $(g_{kl}, P^{kl} )$ around background data
 (\ref{Ptr}), (\ref{gtr}):
\be  
h_{kl}:=g_{kl}-\eta_{kl}\, ,\quad \varpi^{kl}:=P^{kl}+2\Lambda\eta^{kl}  
\ee
 where $\Lambda:=\sqrt{\det \eta_{kl}}$ ($=\sinh^2\omega
\sin\theta$).

We should now rewrite equations (\ref{ww}--\ref{Pdot}) in a linearized
 form in terms of $(h_{kl},\varpi^{kl} )$.  Let us denote by
 $\varpi:=\eta_{kl}\varpi^{kl}$ and by $h:=\eta^{kl}h_{kl}$.  The
 vector constraint (\ref{ww}) can be linearized as follows:
\begin{equation} \label{ww1}
 P_i{^l}{_{| l}} \approx \varpi_i{^l}{_{| l}} -2\Lambda h_i{^k}{_{|k}}
 +\Lambda h_{|i}
\end{equation}
 Let us stress that the symbol ``$|$'' has a different meaning on the
 left-hand side and on the right-hand side of the above formula.  It
 denotes the covariant derivative with respect to the full nonlinear
 metric $g_{kl}$ when applied to the $P^{kl}$ but on the right hand
 side it means covariant derivative with respect to the background
 metric $\eta_{kl}$.  The scalar constraint (\ref{ws}) after
 linearization takes the form
\begin{equation}   \label{ws1}
\sqrt{g}{\cal R}-\frac 1{\sqrt{g}}\left( {P}^{kl}{P}_{kl}
 - \frac{1}{2} ({P}^{kl}{g}_{kl})^2 \right) \approx \Lambda
\left( h^{kl}{_{|l}}- h^{|k} \right)_{|k} - 2\varpi
\end{equation}
 Linearized constraints for vacuum ($T_{\mu\nu}=0$) have the following
 form:
\begin{equation} \label{wwl}
\varpi_l{^k}{_{| k}} -2\Lambda h_l{^k}{_{|k}}+\Lambda h_{|l} =0
 (=8\pi\Lambda T_{l\mu}n^{\mu})
\end{equation}
\begin{equation}   \label{wsl}
\Lambda \left( h^{kl}{_{|l}}- h^{|k} \right)_{|k} - 2\varpi =0
 (=16\pi\Lambda T_{\mu\nu}n^{\mu}n^{\nu})
\end{equation}
 The linearization of (\ref{gdot}) leads to the equation:
\begin{eqnarray}
\dot{h}_{kl} & = & \frac{2N}{\Lambda} \left( \varpi_{kl} -\frac 12
\eta_{kl} \varpi 
\right) + h_{0k|l} +h_{0l|k} + 2N\eta_{kl}(n+\frac 12 h) -2Nh_{kl} +
\nonumber \\
\label{hdot} & & -N^m (h_{mk|l} +h_{ml|k}-h_{kl|m}) 
\end{eqnarray}
 where $N:=\frac 1{\sqrt{-\eta^{00}}}=\cosh\omega$,
 $N_3=\eta_{03}=-\sinh\omega$, $N_A=\eta_{0A}=0$ are lapse and shift
 for the background and $n:=\frac{h^{00}}{2\eta^{00}}$ is the
 linearized lapse.
\begin{eqnarray}
\dot{\varpi}^{kl} &=& -N\Lambda  h^{kl} 
 + N \varpi^{kl} + N^m \varpi^{kl}{_{|m}} + 2\Lambda \left(
 h_0{^{k|l}} + h_0{^{l|k}} -\eta^{kl} h_0{^m}{_{|m}} \right) +
\nonumber \\
 & & + \Lambda \left[ (Nn)^{|kl} - \eta^{kl} (Nn)^{|m}{_{|m}}
\right]
 -\frac{\Lambda}2 N \left( h^{mk|l}{_m} +h^{ml|k}{_m} - h^{kl|m}{_m}
 -h^{|kl} \right) + \nonumber \\
\label{pidot} & &
 -\frac{\Lambda}2 N_m \left[ h^{kl|m} +3h^{ml|k} +3h^{mk|l} -
\eta^{kl}
 (h^{|m} +2h^{mn}{_{|n}}) \right]
\end{eqnarray}
 It is well known (see for example \cite{JJspin2}) that linearized
 Einstein equations are invariant with respect to the ``gauge''
 transformation:
\begin{equation}
 h_{\mu \nu} \rightarrow h_{\mu \nu} + \xi _{\mu ; \nu} +
\xi _{\nu ; \mu}   \label{gauge4}	 
\end{equation}
 where $\xi _{\mu}$ is a covector field, pseudoriemannian metric
 $g_{\mu \nu} = \eta _{\mu \nu} + h_{\mu \nu}$ and ``$;$'' denotes
 four--dimensional covariant derivative with respect to the flat
 Minkowski metric $\eta_{\mu \nu}$.  There is no (3+1) splitting of
 the gauge for hyperboloidal slicing similar to the situation
 described in \cite{JJspin2}. The (3+1) decomposition of the gauge
 acts on Cauchy data in the following way:
\begin{eqnarray}
\Lambda ^{-1} \varpi^{kl} & \rightarrow & \Lambda ^{-1} \varpi^{kl}
 +N \xi^{0 |kl} - N\eta^{kl} \xi{^{0| m}}{_m} - 2N\xi^0 \eta^{kl}
 -N^k\xi^{0|l}-N^l\xi^{0|k} + \nonumber \\ \label{pixi} & &
 +2\eta^{kl} N_m \xi^{0|m} +2\xi^{k|l}+2\xi^{l|k}-2\eta^{kl}
\xi^m{_{|m}}   \\
\label{hxi} h_{kl}& \rightarrow & h_{kl} + \xi_{l | k}+ \xi_{k |
 l} + 2N\eta_{kl}\xi^0
\end{eqnarray}
 It can be easily checked that scalar constraint (\ref{wsl}) and
 vector constraint (\ref{wwl}) are invariant with respect to the gauge
 transformations (\ref{pixi}) and (\ref{hxi}).  The Cauchy data
 ($h_{kl}$, $\varpi^{kl}$) and ($\overline{h}_{kl}$,
 $\overline{\varpi}^{kl}$) on $\Sigma$ are equivalent to each other if
 they can be related by the gauge transformation $\xi_{\mu}$.  The
 evolution of canonical variables $\varpi^{kl}$ and $h_{kl}$ given by
 equations (\ref{hdot}), (\ref{pidot}) is not unique unless the lapse
 function $n$ and the shift vector $h^0{_k}$ are specified.

Let us define the ``new momentum'' $p^{kl}$:
\[ p^{kl}:=\varpi^{kl} -\Lambda\left(2h^{kl}-\eta^{kl}h\right) \quad \quad
 (p:=\varpi+\Lambda h )\] This object can be also introduced in full
 nonlinear theory as $P^{kl}+2\sqrt{g}g^{kl}$ and after linearization
 we obtain $p^{kl}$:
\[ P^{kl}+2\sqrt{g}g^{kl} \approx p^{kl} \]
 Let us first observe that the new momentum is trivial for flat
 Minkowski data.  Secondly the symplectic structure is preserved:
\[ \rd P^{kl} \wedge \rd g_{kl} - \rd \left( P^{kl}+2\sqrt{g}g^{kl}\right)
\wedge \rd g_{kl} = -4\rd^2 \sqrt{g} =0 \]
 Moreover, the gauge transformation for $p^{kl}$ is simpler than for
 $\varpi^{kl}$:
\begin{eqnarray}  \label{pxi0} 
\Lambda ^{-1} p^{kl} & \rightarrow & \Lambda ^{-1} p^{kl}
 +N \xi^{0 |kl} - N\eta^{kl} \xi{^{0| m}}{_m}
 -N^k\xi^{0|l}-N^l\xi^{0|k} +2\eta^{kl} N_m \xi^{0|m}
\end{eqnarray}
 and the vector constraint has a familiar form:
\be \label{wwp} p_k{^l}{_{| l}} =0 (=8\pi\Lambda T_{l\mu}n^{\mu}) \ee
 We can also rewrite the dynamical equation (\ref{pidot}) in terms of
 the new momentum:
\begin{eqnarray}
\dot{p}^{kl} &=&  N^m p^{kl}_{|m} + N \left(\eta^{kl}p -3p^{kl}\right)+
\Lambda \left[ (Nn)^{|kl} - \eta^{kl} (Nn)^{|m}{_{|m}}
 +2Nn\eta^{kl} \right] + \nonumber \\ & & + N\Lambda \left(
\eta^{kl} h -3h^{kl}  \right)
 -\frac{\Lambda}2 N \left( h^{mk|l}{_m} +h^{ml|k}{_m} - h^{kl|m}{_m}
 -h^{|kl} \right) + \nonumber \\
\label{pdot} & &
 +\frac{\Lambda}2 N_m \left[ h^{ml|k} +h^{mk|l}-h^{kl|m} +
\eta^{kl}(
 h^{|m} -2h^{mn}{_{|n}}) \right]
\end{eqnarray}

We will show in the sequel that it is possible to define reduced
 dynamics in terms of invariants which is no longer sensitive on gauge
 conditions.  The construction is analogous to the analysis given in
\cite{GRG}.

\subsection{``Charges'' on a hyperboloid}

The equation (\ref{wwp}) allows to introduce ``charges'' related to
 the symmetries of the hyperboloid. There are six generators of the
 Lorentz group which are simultaneously Killing vectors on the
 hyperboloid $\Sigma$.  Let us denote this Killing field by $X^k$. It
 fulfills the following equation:
\be\label{Xk} X_{k|l} + X_{l|k} = 0 \ee
 Let $V\subset \Sigma$ be a compact region in $\Sigma$. For example
 $\displaystyle V:=\bigcup_{r\in [r_0,r_1]} S_s(r)$ and $\partial
 V=S_s(r_0) \cup S_s(r_1)$.  From (\ref{wwp}) and (\ref{Xk}) we get:
\be\label{Lc}
 (8\pi\int_V\Lambda T_{l\mu}n^{\mu})=0=
\int_V p^{kl}{_{|l}} X_k= \int_V (p^{kl}X_k){_{|l}} =
\int_{\partial V} p^{3k}X_k
\ee
 The equation (\ref{Lc}) expresses the ``Gauss'' law for the charge
 ``measured'' by the flux integral.

In particular for angular momentum when $X=\partial /\partial\phi$ we
 can show the relation of this charge with dipole part of invariant
 $\bf y$ which we introduce in the sequel (subsection 4.4). 
\[
 16\pi s^z:= 16\pi j^{xy} = -2\int_{\partial V} p^3{_\phi } =
 -2\int_{\partial V} p^3{_A} (r^2\varepsilon^{AB}\cos\theta)_{||B} =\]
\begin{equation} \label{sz}
 = 2\int_{\partial V} r^2 p^{3}{_{A||B}}\varepsilon^{AB}
\cos\theta
 =\hspace{0.5cm} \int_{\partial V} \Lambda {\bf y} \cos\theta
\end{equation}

The time translation defines a mass charge as follows: \[
 (16\pi\int_V\Lambda T_{0\mu}n^{\mu})=0=\int_V N\left[ \Lambda
\left( h^{kl}{_{|l}}- h^{|k} \right)_{|k} - 2\pi
\right] +2N_k p^{kl}{_{|l}}= \] \[ =\int_V \left[ 2N_k p^{kl}
 +N\Lambda \left( h^{lk}{_{|k}}- h^{|l} \right) +\Lambda \left( N_k
 h^{kl} -N^l h\right) \right]_{|l} = \] \begin{equation}
\label{BM}
 = \int_{\partial V} 2N_k p^{k3} + \Lambda \left( N h^{3k}{_{|k}}- N
 h^{|3} + N_k h^{k3} -N^3 h \right)
\end{equation}
 and it can be related to monopole part of an invariant $\bf x$
 (subsection 4.4).

\[
 16\pi p^0=\int_{\partial V} 2N_k p^{k3} + \Lambda \left( N
 h^{3k}{_{|k}}- N h^{|3} + N_k h^{k3} -N^3 h
\right)=\] \[
 =\int_{\partial V} \frac{\Lambda}{\sinh\omega} \left( 2\cosh^2\omega
 h^{33} -\cosh\omega\sinh\omega H,{_3} - H
 -\frac{2\sinh^2\omega}{\Lambda}p^{33} \right) = \]
\begin{equation} \label{p0}
 = \hspace*{0.5cm} \int_{\partial V} \frac{\Lambda}r {\bf x}
\end{equation} 
 
\underline{Remark} Traceless part of $h_{kl}$ and $p^{kl}$ has nice
 properties, gauge splits into 0-component (transversal to $\Sigma$)
 which acts on $p^{kl}$ and space components (tangent to $\Sigma$)
 which act on $h_{kl}-\frac 13 \eta_{kl} h$. But $h$ and $\varpi$
 remain nontrivial unless we impose gauge conditions. The most popular
 gauge condition, which allows to obtain scalar constraint as a full
 divergence (see below (\ref{Bm})), is to assume that $\varpi=0$.
 Assuming this gauge we can define different ``mass'' charge as a
 surface integral coming from the scalar constraint (\ref{wsl}) (but
 we obtain totally nonlocal object).  More precisely, one can analyze
 the scalar constraint (\ref{wsl}) in the same way as (\ref{Lc})
\begin{equation}   \label{Bm}
 2\int_V\varpi = \int_V \Lambda \left( h^{kl}{_{|l}}- h^{|k}
\right)_{|k} =
\int_{\partial V} \Lambda \left( h^{3l}{_{|l}}- h^{|3} \right)
\end{equation}
 and there is no ``Gauss'' law for the ``mass''defined by the surface
 integral of the right-hand side of (\ref{Bm}) unless we impose gauge
 condition $\varpi=0$. This means that such definition of the mass
 charge measured by the flux integral at null infinity is {\it not}
 gauge invariant like ADM mass at spatial infinity.

\subsection{2+1 decomposition and reduction}
 Now we introduce reduced gauge invariant data on $\Sigma$ for the
 gravitational field similar to the invariants introduced in
\cite{GRG}. For this purpose
 we use spherical foliation of $\Sigma$ (see equations (\ref{Hs}) and
 (\ref{eta3})).

In this section we present mainly results without detailed proofs as
 in the section about electrodynamics.

Let $\kappa:=\coth\omega$.  The gauge (\ref{hxi}) splits in the
 following way:
\begin{eqnarray}
\label{hxi33} h_{33}& \rightarrow & h_{33} + 2\xi_{3,3}+ 2N\xi^0
\\ \label{hxi3A}
 h_{3A}& \rightarrow & h_{3A} + \xi_{3,A}+ \xi_{A,3} -2 \kappa
\xi_A
\\ \label{hxiAB} h_{AB}& \rightarrow & h_{AB} + \xi_{A|| B}+
\xi_{B|| A} +2 \kappa
\eta_{AB} \xi_3 +  2N\eta_{AB}\xi^0
\end{eqnarray}
 where by ``$||$'' we denote covariant derivative with respect to the
 two--metric $\eta_{AB}$ on $S(r)$.  Similarily the gauge (\ref{pxi0})
 can be splitted as follows:
\begin{eqnarray}
\Lambda ^{-1} p^{33} & \rightarrow & \Lambda ^{-1} p^{33}
 +N \xi^{0 |33} - N \xi{^{0| m}}{_m} = \Lambda ^{-1} p^{33}
 -N\xi^{0||A}{_A} -2N\kappa \xi^{0,3} \label{pxi33} \\
\label{pxi3A} 
\Lambda ^{-1} p^{3A} & \rightarrow & \Lambda ^{-1} p^{3A}
 +N \xi^{0 |3A} -N^3\xi^{0|A} = \Lambda ^{-1} p^{3A}
 +N\xi^{0,3A}-\frac 1{\sinh\omega} \xi^{0,A} \\ \label{pxiAB}
\Lambda ^{-1} p^{AB} & \rightarrow & \Lambda ^{-1} p^{AB}
 +N \xi^{0 |AB} - N\eta^{AB} \xi{^{0| m}}{_m} +2\eta^{AB} N_m
\xi^{0|m} =
\Lambda ^{-1} p^{AB} + \\ & & +N\xi^{0||AB} -N\eta_{AB}\left(
\xi^{0,3}{_3} +\xi^{0||C}{_C} +
 (\kappa - 2N^3)\xi^0,{_3} \right) \nonumber \end{eqnarray} It is also
 quite easy to rewrite (2+1) decomposition of (\ref{hdot}):
\begin{eqnarray}
\dot{h}_{33} & = & \frac{2N}{\Lambda} \left( p_{33} -\frac 12 p
\right) + 2h_{03|3}  + 2N(n +h_{33})  -N^3 h_{33|3} =
\label{hdot33} \\ & = & \frac{2N}{\Lambda} \left( p_{33} -\frac
 12 p \right) + 2h_{03,3} + 2Nn +2Nh_{33} -N^3 h_{33,3} \nonumber
\\
\dot{h}_{3A} & = & \frac{2N}{\Lambda}  p_{3A} + h_{03|A} +h_{0A|3}
 +2Nh_{3A} -N^3 h_{33|A} = \label{hdot3A} \\ & = &
\frac{2N}{\Lambda}  p_{3A} + h_{03,A} +h_{0A,3}
 -2\kappa h_{0A} -N^3 h_{33,A} \nonumber \\
\dot{h}_{AB} & = & \frac{2N}{\Lambda} \left( p_{AB} -\frac 12 \eta_{AB} p
\right) + h_{0A|B} +h_{0B|A} + 2N\eta_{AB}n +2Nh_{AB} +\nonumber \\
\label{hdotAB} & & -N^3 (h_{3A|B} +h_{3B|A}-h_{AB|3}) =\frac{2N}{\Lambda}
\left( p_{AB} -\frac 12 \eta_{AB} p \right) + h_{0A||B} + \\
 & & + h_{0B||A} +2\kappa\eta_{AB}h_{03}+ 2N\eta_{AB}(n+h_{33})
 +2Nh_{AB}+\nonumber \\ & & -N^3 (h_{3A||B} +h_{3B||A}-h_{AB,3})
\nonumber
\end{eqnarray}
 The vector constraint (\ref{wwp}) can be splitted in the similar way:
\be  p_3{^k}{_{|k}}=p_3{^3}{,{_3}}+p_3{^A}{_{||A}} -\kappa
 p^{AB}\eta_{AB}=0 \ee \be
 p_A{^k}{_{|k}}=p_A{^3}{,{_3}}+p_A{^B}{_{||B}} = p_{3A,3}
 +S_A{^B}{_{||B}} +\frac 12 S_{||A} =0 \ee where $S:=p^{AB}\eta_{AB}$
 and $S^{AB}:=p^{AB} -\frac 12 \eta^{AB} S$.  Similarily let us denote
 $H:=h_{AB}\eta^{AB}$ and $\chi_{AB}:=h_{AB}-\frac12\eta_{AB}H$.  The
 invariants are defined as follows:
\[
 {\bf x} = 2\cosh^2\omega h^{33}+2\cosh\omega\sinh\omega
 h^{3C}{_{||C}}+ \sinh^2\omega
\chi^{AB}{_{||AB}} -\cosh\omega\sinh\omega H,{_3} + \]
\begin{equation}
 -\frac{1}{2}({\bf a}+2) H -
\frac{2\sinh^2\omega}{\Lambda}   p^{33}
\label{x}
\end{equation}
\begin{equation}
 {\bf X} = 2\sinh^2\omega S^{AB}{_{||AB}} + 2\cosh\omega\sinh\omega
 p^{3A}{_{||A}} +{\bf a}p^{33} \label{X}
\end{equation}
\begin{equation}
 {\bf y}=2\Lambda^{-1}\sinh^2\omega p^{3A||B}\varepsilon_{AB}
\label{y}
\end{equation}
\begin{equation}
 {\bf Y}= \Lambda({\bf a}+2)h^{3A||B}\varepsilon_{AB} -
\sinh^2\omega(\Lambda \chi^C{_{A||CB}}\varepsilon^{AB}),{_3} \label{Y}
\end{equation}
 (2+1) decomposition of the scalar constraint (\ref{wsl}) can be
 written in the form: \[
\Lambda \left( h ^{|l}{_l} - h^{kl}{_{|lk}} \right) +2\varpi =
\left[ \Lambda (H,{_3}-2h^{3A}{_{||A}}-2\kappa h^{33} +\kappa H)
\right],{_3} -2\Lambda(h^{33}+H)  +\]
\begin{equation}
 +2(p^{33}+S)-\Lambda (\chi^{AB}{_{||BA}}+2\kappa h^{3A}{_{||A}})
 +\Lambda \left( h^{33 ||A}{_A}+\frac{1}{2}H^{||C}{_C} \right) = 0
\label{w2+1}
\end{equation}
  	
The dynamical equations (\ref{pdot}) take the following $(2+1)$ form:
\[
\Lambda^{-1}\dot{p}^{33}= \Lambda^{-1}N_3(p^{33},{_3}-\kappa S)
 - (Nn)^{||A}{_A}-2\kappa N n,{_3} + \] \[ + \frac N2 \left[ h^{33
 ||A}{_A} +H,{_{33}}+2\kappa H,{_{3}} - 2\kappa(2h^{3A}{_{||A}} +
 h^{33},{_3} ) -(2h^{3A}{_{||A}}),{_{3}} \right]+ \] \begin{equation}
 +\frac 12 N_3 \left[ H,{_{3}} - 2h^{3A}{_{||A}} \right] \label{p33}
\end{equation}
\[
\Lambda^{-1}\dot{p}^{3A}=\Lambda^{-1}N^3(p^{3A},{_3}+2\kappa p^{3A})+
 [(Nn),{_3} -\kappa Nn]^{ ||A} +
\frac{N}{2}\left[ H,{_3}^{||A} -\kappa h^{33 ||A}  + \right.\]
\begin{equation} \left. +\frac 2{\sinh^2\omega}h^{3A} - h^{AB}{_{||B}},{_3}
 -2\kappa h^{AB}{_{||B}}-h^{3B||A}{_{||B}} +h^{3A}{_{||B}}^{||B}
\right]
 + \frac12 N_3 h^{33 ||A} \label{p3A}
\end{equation}
\[
\Lambda^{-1}\dot{p}_{AB} = \Lambda^{-1} N^3 p_{AB},{_3}+ \Lambda^{-1} N
\left[ \eta_{AB} (p^{33} + S) +p_{AB} \right] + \]
\[ +
 N(n_{||AB} - \eta_{AB}n^{||C}{_C}) -N\eta_{AB}(n,{_{33}}+\kappa
 n,{_{3}}) + 2\eta_{AB}N^3 n,{_3} + \]
\[ +
\frac{N^3}2 \left[ h^3{_{A||B}}+h^3{_{B||A}}-
\eta_{AB}h^{3C}{_{||C}} +\eta_{AB} ( \frac{1}{2} H,{_3}- h^{3A}{_{||A}} 
 - h^{33},{_3}) -\chi_A{^{C}},{_3}\eta_{CB} \right] +
\]
\[ +
\frac{N}2 \left[ (\chi^C{_B},{_3}\eta_{CA}),{_3}
 + \chi_{AB}{^{||C}}_{||C} - \chi^C{_{A||BC}} - \chi^C{_{B||AC}} +
 h^{33}{_{||AB}} +\frac{1}{2}\eta_{AB}H_{||C}{^C} + \right. \]
\[ + \eta_{AB}\frac 1{\sinh^2\omega}(h^{33}+H) +
\eta_{AB}(\kappa H,{_3}- 2\kappa h^{3A}{_{||A}} - 2\kappa^2 h^{33})
 + \frac 2{\sinh^2\omega}\chi_{AB} + \]
\begin{equation} \left.
 - \left( \eta_{AB} (\kappa h^{33}-\frac{1}{2} H,{_3})
 +h^3{_{A||B}}+h^3{_{B||A}} \right)\! ,{_3} \right] \label{pAB}
\end{equation}
 We can check the reduced field equations for our invariants:
\be\label{px} \dot{\bf x}=\frac{N}\Lambda {\bf X} +(N^3{\bf x}),{_3} \ee
\be\label{pX} \dot{\bf X}= N^3{\bf X},{_3} +
\Lambda N \triangle_{\scriptscriptstyle \Sigma} {\bf x} -\Lambda
 N^3({\bf x},{_3} +2\kappa {\bf x}) \ee \be\label{py} \dot{\bf
 y}=\frac{N}\Lambda {\bf Y} +\frac{N^3}{\Lambda}(\Lambda{\bf y}),{_3}
\ee
\be\label{pY} \dot{\bf Y}= \Lambda (N^3\Lambda^{-1}{\bf Y}),{_3} +
\Lambda N \triangle_{\scriptscriptstyle \Sigma} {\bf y} -\Lambda
 N^3({\bf y},{_3} +2\kappa {\bf y}) \ee where
 $\triangle_{\scriptscriptstyle \Sigma}$ is a Laplacian on a
 hyperboloid $\Sigma$.

It can be easily verified that the invariants $\bf x$ and $\bf y$
 fulfill usual d'Alambert equation (as a consequence of the above
 dynamical equations):
\[ \mbox{\msa\symbol{3}} \, {\bf x}=0 \]
\[ \mbox{\msa\symbol{3}} \, {\bf y}=0 \]

Let us notice that $\bf x$ and $\bf y$ are scalars on each sphere
 $S_s(r)$ with respect to the coordinates $y^A$.\\ For the scalar $f$
 on a sphere we can define ``monopole'' part mon$(f)$ and a ``dipole''
 part dip$(f)$ as a corresponding component with respect to spherical
 harmonics on $S^2$. Similarily ``dipole'' part of a vector $v^A$
 corresponds to the dipole harmonics for the scalars $v^A{_{||A}}$ and
 $\varepsilon^{AB}v_{A||B}$. Let us denote by $\underline f$ mono--,
 dipole-- free part of $f$.  According to this decomposition we have:

\[ {\bf x}=\mbox{mon}({\bf x}) +\mbox{dip}({\bf x}) +\underline{\bf x}
\]
\[ {\bf y}=\mbox{mon}({\bf y}) +\mbox{dip}({\bf y}) +\underline{\bf y}
\]
 Then mono-dipole part of the each scalar can be solved explicitly
 from the equations (\ref{px}-\ref{pY}) and the solution has the form:
\[ {\bf x} - \underline{\bf x}= \frac{4{\bf m}}{\sinh\omega}+
\frac{12{\bf k}}{\sinh^2\omega}
\]
\[ {\bf y}-\underline{\bf y}=\frac{12{\bf s}}{\sinh^2\omega}
\]
 Let ${\bf p}:= \dot{\bf k}$ than
\[ \dot{\bf m}=\dot{\bf p}=\dot{\bf s}=0 \] and
\[ {\bf k}={\bf p}(s+\cosh\omega) + {\bf k}_0 \]
 Moreover ${\bf a m}=0$, $({\bf a}+2){\bf p}=({\bf a}+2){\bf
 k}_0=({\bf a}+2) {\bf s}=0$ which simply means that $\bf m$ is a
 monopole and ${\bf k}_0$, $\bf p$, $\bf s$ are dipoles and they are
 constant on $M_2$.  They correspond to the charges introduced in
\cite{JJspin2}.  Let us rewrite the solution in
 coordinates $u,r$ which will be more useful in the sequel:
\be\label{cx}
 {\bf x} = \underline{\bf x}+ \frac{4{\bf m}+12{\bf p}}{r}+
\frac{12({\bf k}_0+{\bf p}u)}{r^2}
\ee
\be\label{cy}
 {\bf y} = \underline{\bf y} + \frac{12{\bf s}}{r^2}
\ee
 Let us remind the relation between spatial constant three-vectors in
 cartesian coordinates and dipole harmonics
\[ 
 {\bf k}_0=\frac{j^{l0}z_l}{r} \, , \quad {\bf p} =\frac{p^l z_l}{r}
\, , \quad {\bf s} =\frac{s^l z_l}{r} \] where $z_l$ are
 cartesian coordinates and $j^{l0}$, $p^l$, $s^l$ are corresponding
 three-vectors representing our charges (see
\cite{JJspin2}).

\subsection{Reduction of symplectic form on a hyperboloid}

We want to show the relation between the symplectic structure and the
 invariants introduced in the previous subsection. Let
 $(p^{kl},h_{kl})$ and $(s^{kl},q_{kl})$ denote two pairs of a Cauchy
 data on a hyperboloid.  The $(2+1)$-- splitting of the tensor
 $q_{kl}$ gives the following components on a sphere:
 $\stackrel{2}{q}:=\eta^{AB}q_{AB}$, $q_{33}$ -- scalars on $S^2$,
 $q_{3A}$ -- vector and $\stackrel{\circ}{q}\! {_{AB}}:=
 q_{AB}-\frac12\eta_{AB}\stackrel{2}{q}$ -- symmetric traceless tensor
 on $S^2$. Similarily we can decompose tensor density $p^{kl}$.  The
 quadratic form $\int_{V}{p}^{kl}{q}_{kl}$ can be decomposed into
 monopole part, dipole part and the remainder in a natural way.

The ``monodipole'' part we write separately:

\[ \mbox{mon}( \int_{V}{p}^{kl}{q}_{kl})= \int_{V}
\frac1{2\cosh^2\omega}{p}^{33}  \mbox{mon}(\xi ) 
 + \int_{V} \frac{\tanh^2\omega}{\Lambda} p^{33}
\mbox{mon}(s^{33})+
\]
\be
 +\frac{1}{2} \int_{\partial V} \tanh\omega p^{33}
\mbox{mon}(\stackrel{2}{q})
\ee  
\[ \mbox{dip}(\int_{V}{p}^{kl}{q}_{kl})=
\int_{V}\frac1{2\cosh^2\omega} {p}^{33} \mbox{dip}(\xi ) - 
 2 \int_{V}\mbox{dip} (\sinh^2\omega{p}^{3A|| B}\varepsilon_{AB}){\bf
 a}^{-1} ({q}_{3A|| B}\varepsilon^{AB}) +\]
\[
 + \int_{V} \frac{\tanh^2\omega}{\Lambda}p^{33}
\mbox{dip}(s^{33})
 +\frac{1}{2} \int_{\partial V} \tanh\omega p^{33}
\mbox{dip}(\stackrel{2}{q}) +\]
\be
 + \int_{V} \mbox{dip} \left( \frac{{p}^{33}}{2\cosh^2\omega} +
\tanh\omega {\bf a}^{-1} {p}^{3A}{_{|| A}} \right)  
\left(\frac{1}{2}{\bf a}\stackrel{2}{q} -2\sinh\omega
\cosh\omega{q}^{3A}{_{|| A}} \right)
\ee where invariant $\xi$ is defined as follows: \[ \xi:=
 2\cosh^2\omega q^{33} + 2\cosh\omega \sinh\omega {q}^{3A}{_{|| A}} +
\sinh^2\omega\stackrel{\circ}{q}\! {^{AB}}{_{|| AB}}
 -\cosh\omega \sinh\omega \stackrel{2}{q},{_3} + \]
\[ -\frac12 ({\bf a}+2)
\stackrel{2}{q} - 2\frac{\sinh^2\omega}\Lambda s^{33} \]

From vector constraints: \be
\sinh\omega {p}^{33},{_3} +\sinh\omega {p}^{3A}{_{|| A}} - \cosh\omega
\stackrel{2}{p} = 0 \ee 
\be
 (\sinh^2\omega {p}^{3A}{_{|| A}}),{_3} + (\sinh^2\omega
\stackrel{\circ}{p}\!
 {^{AB}}{_{|| AB}}) + \frac{1}{2}{\bf a} \stackrel{2}{p} = 0 \ee
\be
 (\sinh^2\omega {p}^{3A|| B}\varepsilon_{AB}),{_3} + (\sinh^2\omega
\varepsilon^{AC}\stackrel{\circ}{p}\!
 {_A{^B}}{_{|| BC}}) = 0 \ee we can partially reduce our form:
  \[ \int_{V}{p}^{kl}{q}_{kl} = \int_{V} {p}^{33}{q}_{33} +
 2{p}^{3A}{q}_{3A} + \frac{1}{2} \stackrel{2}{p} \stackrel{2}{q} +
\stackrel{\circ}{p}\! {^{AB}}\stackrel{\circ}{q}\! {_{AB}}  = 
\]
\[ 
 = \int_{V} {p}^{33}{q}_{33} - 2(\sinh\omega {p}^{3A}{_{|| A}}) {\bf
 a}^{-1} (\sinh\omega {q}^{3A}{_{|| A}}) - 2(\sinh\omega {p}^{3A||
 B}\varepsilon_{AB}) {\bf a}^{-1} (\sinh\omega {q}_{3A||
 B}\varepsilon^{AB}) +
\]
\[
 + \int_{V}
\frac{1}{2} \stackrel{2}{p} \stackrel{2}{q}
 + 2 (\sinh^2\omega\varepsilon^{AC}\stackrel{\circ}{p}\!
 {_A{^B}}{_{|| BC}}) {\bf a}^{-1}({\bf a}+2)^{-1}
 (\sinh^2\omega\varepsilon^{AC}\stackrel{\circ}{q}\!  {_A{^B}}{_{||
 BC}}) + \]
\[
 + 2 \int_{V} (\sinh^2\omega\stackrel{\circ}{p}\! {^{AB}}{_{|| AB}})
 {\bf a}^{-1}({\bf a}+2)^{-1} (\sinh^2\omega\stackrel{\circ}{q}\!
 {^{AB}}{_{|| AB}}) = \] \[ =
\int_{V} {p}^{33}{q}_{33} -
 2(\sinh\omega {p}^{3A}{_{|| A}}) {\bf a}^{-1} (\sinh\omega
 {q}^{3A}{_{|| A}}) - 2(\sinh\omega {p}^{3A|| B}\varepsilon_{AB}) {\bf
 a}^{-1} (\sinh\omega {q}_{3A|| B}\varepsilon^{AB}) + \]
\[ 
 + \int_{V} \frac{1}{2}(\tanh\omega {p}^{33},{_3} + \tanh\omega
 {p}^{3A}{_{|| A}})
\stackrel{2}{q} -
 2 \int_{V} (\sinh^2\omega{p}^{3A|| B}\varepsilon_{AB}),{_3} {\bf
 a}^{-1}({\bf a}+2)^{-1}
 (\sinh^2\omega\varepsilon^{AC}\stackrel{\circ}{q}\!  {_A{^B}}{_{||
 BC}}) +
\]
\[
 -2 \int_{V} \left[(\sinh^2\omega{p}^{3A}{_{|| A}}),{_3} +
\frac{1}{2}{\bf a}
 (\tanh\omega {p}^{33},{_3} + \tanh\omega {p}^{3A}{_{|| A}})
\right]
 {\bf a}^{-1}({\bf a}+2)^{-1} (\sinh^2\omega\stackrel{\circ}{q}\!
 {^{AB}}{_{|| AB}}) = \] \[ = \int_{\partial V} \tanh\omega
 {p}^{33}\left[ \frac{1}{2}\stackrel{2}{q} - ({\bf a}+2)^{-1}
 (\sinh^2\omega\stackrel{\circ}{q}\! {^{AB}}{_{|| AB}})
\right] + \]
\[
 -2 \int_{\partial V} (\sinh^2\omega{p}^{3A}{_{|| A}}) {\bf
 a}^{-1}({\bf a}+2)^{-1} (\sinh^2\omega\stackrel{\circ}{q}\!
 {^{AB}}{_{|| AB}}) + \] \[ -2 \int_{\partial V}
 (\sinh^2\omega{p}^{3A|| B}\varepsilon_{AB}) {\bf a}^{-1}({\bf
 a}+2)^{-1} (\sinh^2\omega\varepsilon^{AC}\stackrel{\circ}{q}\!
 {_A{^B}}{_{|| BC}}) + \] \[ -2\int_{V} \fbox{$\displaystyle
 (\sinh^2\omega{p}^{3A|| B}\varepsilon_{AB}) {\bf a}^{-1}\left[
 {q}_{3A|| B}\varepsilon^{AB} - ({\bf a}+2)^{-1}
 (\sinh^2\omega\varepsilon^{AC}\stackrel{\circ}{q}\!  {_A{^B}}{_{||
 BC}}),{_3} \right]$} + \]
\[
 + \int_{V}{p}^{33} \left[ {q}_{33} + ({\bf a}+2)^{-1}
 (\frac{\sinh^3\omega}{\cosh\omega} \stackrel{\circ}{q}\!
 {^{AB}}{_{||AB}}),{_3} - \frac{1}{2}(\tanh\omega
\stackrel{2}{q}),{_3} \right] + \]
\[ + \int_{V} \tanh\omega {p}^{3A}{_{|| A}} \left[
\frac{1}{2}\stackrel{2}{q} + 
 2\sinh\omega \cosh\omega {\bf a}^{-1} ({\bf a}+2)^{-1}
 (\sinh^2\omega\stackrel{\circ}{q}\!{^{AB}}{_{|| AB}}),{_3} +\right.
\] \[ \left.  - ({\bf a}+2)^{-1}
 (\sinh^2\omega\stackrel{\circ}{q}\! {^{AB}}{_{|| AB}}) - 2 {\bf
 a}^{-1} (\sinh\omega \cosh\omega {q}^{3A}{_{|| A}}) \right] \] The
 volume term in the framebox is mono-- dipole-- free and corresponds
 to the invariants $\bf y$, $\bf Y$. The last two terms we can proceed
 further but let us first write a scalar constraint in two equivalent
 forms: \[ (\frac{\sinh^3\omega}{\cosh\omega}\stackrel{\circ}{q}\!
 {^{AB}}{_{|| AB}}),{_3} + ({\bf a}+2) \left[ {q}_{33} -
\frac{1}{2}(\tanh
 r\stackrel{2}{q}),{_3} \right] = (\tanh\omega\xi ),{_3} + \xi +
\]
\be +\frac 2\Lambda\left( \tanh^2\omega s^{33} -\tanh\omega
\sinh^2\omega s^{3A}{_{||A}} \right)  \ee
\[ \frac{1}{2}{\bf a}({\bf a}+2) \stackrel{2}{q} +
 2\sinh\omega\cosh\omega(\sinh^2\omega\stackrel{\circ}{q}\!
 {^{AB}}{_{|| AB}}),{_3} - {\bf a}\sinh^2\omega
\stackrel{\circ}{q}\! {^{AB}}{_{|| AB}} +
\] \[ - 2({\bf a}+2) (\cosh\omega \sinh\omega{q}^{3A}{_{|| A}})
 = \] \be = 2 \cosh\omega(\sinh\omega\xi ),{_3} -{\bf a} \xi
 -\frac{2\sinh^2\omega}{\Lambda} \left( {\bf a} s^{33} +2\sinh\omega
\cosh\omega s^{3A}{_{||A}} \right) \ee
 
In ``radiation'' part we get the following result: \[
\int_{V}{p}^{33} \left[ {q}_{33}
 + ({\bf a}+2)^{-1}
 (\frac{\sinh^3\omega}{\cosh\omega}\stackrel{\circ}{q}\!
 {^{AB}}{_{||AB}}),{_3} -
\frac{1}{2}(\tanh\omega\stackrel{2}{q}),{_3} \right] \; + \]
\[ + \int_{V} \tanh\omega {p}^{3A}{_{|| A}} \left[
\frac{1}{2}\stackrel{2}{q} + 2\sinh\omega \cosh\omega {\bf a}^{-1}
 ({\bf a}+2)^{-1} (\sinh^2\omega\stackrel{\circ}{q}\!{^{AB}}{_{||
 AB}}),{_3} + \right. \] \[ \left.  - ({\bf a}+2)^{-1}
 (\sinh^2\omega\stackrel{\circ}{q}\! {^{AB}}{_{|| AB}}) - 2 {\bf
 a}^{-1} (\sinh\omega \cosh\omega {q}^{3A}{_{|| A}}) \right] = \]
\[
 = \int_{V}{p}^{33}({\bf a}+2)^{-1} \left[ (\tanh\omega \xi ),{_3} +
\xi +\frac 2\Lambda (\tanh^2\omega s^{33}
 -\tanh\omega\sinh^2\omega s^{3A}{_{|| A}})
\right] + \] 
\[
 +\int_{V} \tanh\omega{p}^{3A}{_{|| A}}{\bf a}^{-1} ({\bf a}+2)^{-1}
\left[ 2\cosh\omega (\sinh\omega\xi ),{_3} -{\bf a}
\xi -
\frac{2\sinh^2\omega}\Lambda ({\bf a} s^{33} +2\sinh\omega
\cosh\omega s^{3A}{_{|| A}}) \right] = \]
\[
 =\int_{\partial V}\left[ \tanh\omega {p}^{33} + 2\sinh^2\omega {\bf
 a}^{-1}{p}^{3A}{_{|| A}} \right] ({\bf a}+2)^{-1}\xi + \]
\[ 
 + \int_{V} \left[{\bf a}{p}^{33} - {\bf a}\tanh\omega{p}^{33},{_3}
 -2\sinh\omega(\sinh\omega{p}^{3A}{_{|| A}}),{_3}-{\bf a}
\tanh\omega{p}^{3A}{_{|| A}}\right] 
 {\bf a}^{-1} ({\bf a}+2)^{-1} \xi + \]
\[ + \int_{V} \frac2\Lambda \left[
\tanh\omega p^{33} ({\bf a}+2)^{-1} (\tanh\omega s^{33} -\sinh^2\omega
 s^{3A}{_{|| A}}) + \right. \]
\[ \left.
 -\sinh^2\omega {p}^{3A}{_{|| A}}{\bf a}^{-1} ({\bf a}+2)^{-1}
 (\tanh\omega{\bf a} s^{33} +2\sinh^2\omega s^{3A}{_{|| A}})
\right] = \]
\[
 =\int_{\partial V}[\tanh\omega {\bf a}{p}^{33} +
 2\sinh^2\omega{p}^{3A}{_{|| A}}] {\bf a}^{-1}({\bf a}+2)^{-1}\xi +\]
\[
 + \int_{V} \fbox{$ \displaystyle \left[ {\bf a}{p}^{33} +2
\sinh^2\omega\stackrel{\circ}{p}\!
 {^{AB}}{_{|| AB}} + 2\sinh\omega\cosh\omega{p}^{3A}{_{|| A}}
\right]
 {\bf a}^{-1} ({\bf a}+2)^{-1} \xi $} + \]
\[ + \int_{V} \frac2\Lambda \left[
\tanh^2\omega p^{33}({\bf a}+2)^{-1} s^{33} -
 2\sinh^4\omega{p}^{3A}{_{|| A}}{\bf a}^{-1} ({\bf a}+2)^{-1}
 s^{3A}{_{|| A}} \right] +\]
\[    
 -\int_{V} \frac2\Lambda \tanh\omega\sinh^2\omega \left( p^{33}({\bf
 a}+2)^{-1} s^{3A}{_{|| A}}+ p^{3A}{_{|| A}} ({\bf a}+2)^{-1} s^{33}
\right) \] and again framebox corresponds to
 the invariants but now $\bf x$ and $\bf X$.
  
Finally we get in volume integrand the gauge invariant part: \[
\int_{V}{p}^{kl}{q}_{kl} = \mbox{monodipole part in $V$} +
\fbox{``radiation'' part in $V$} + \] \[ + \int_{\partial
 V}[\tanh\omega {\bf a}{p}^{33} + 2\sinh^2\omega{p}^{3A}{_{|| A}}]{\bf
 a}^{-1}({\bf a}+2)^{-1}\xi +\] \[ -2 \int_{\partial V}
 (\sinh^2\omega{p}^{3A|| B}\varepsilon_{AB}) {\bf a}^{-1}({\bf
 a}+2)^{-1}(\sinh^2\omega\varepsilon^{AC}
\stackrel{\circ}{q}\! {_A{^B}}{_{|| BC}}) + \] 
\[ + \int_{\partial V} \tanh\omega{p}^{33}\left[
\frac{1}{2}\stackrel{2}{q} -
 ({\bf a}+2)^{-1} (\sinh^2\omega\stackrel{\circ}{q}\!  {^{AB}}{_{||
 AB}})
\right] +\]
\[ -2 \int_{\partial V} (\sinh^2\omega{p}^{3A}{_{|| A}}) {\bf a}^{-1}
 ({\bf a}+2)^{-1} (\sinh^2\omega\stackrel{\circ}{q}\!  {^{AB}}{_{||
 AB}})
\] 
 where
\[ \mbox{monodipole part in V}= \int_{V}
\frac1{2\cosh^2\omega}{p}^{33}  \mbox{mon}(\xi ) 
 + \int_{V} \frac{\tanh^2\omega}{\Lambda} p^{33}
\mbox{mon}(s^{33})+
\] \[ + 
\int_{V}\frac1{2\cosh^2\omega} {p}^{33} \mbox{dip}(\xi ) - 
 2 \int_{V}\mbox{dip} (\sinh^2\omega{p}^{3A|| B}\varepsilon_{AB}){\bf
 a}^{-1} ({q}_{3A|| B}\varepsilon^{AB}) +\]
\[
 + \int_{V} \frac{\tanh^2\omega}{\Lambda}p^{33}
\mbox{dip}(s^{33})
 + \int_{V} \mbox{dip} \left( \frac{{p}^{33}}{2\cosh^2\omega} +
\tanh\omega {\bf a}^{-1} {p}^{3A}{_{|| A}} \right)  
\left(\frac{1}{2}{\bf a}\stackrel{2}{q} -2\sinh\omega
\cosh\omega{q}^{3A}{_{|| A}} \right)
\]
\[   \fbox{``radiation'' part in $V$}= \int_{V} 
\left[ {\bf a}{p}^{33} +2 \sinh^2\omega\stackrel{\circ}{p}\!
 {^{AB}}{_{|| AB}} + 2\sinh\omega\cosh\omega{p}^{3A}{_{|| A}}
\right]
 {\bf a}^{-1} ({\bf a}+2)^{-1} \xi + \]
\[  -2\int_{V}    (\sinh^2\omega{p}^{3A|| B}\varepsilon_{AB}) {\bf
 a}^{-1}\left[ {q}_{3A|| B}\varepsilon^{AB} - ({\bf a}+2)^{-1}
 (\sinh^2\omega\varepsilon^{AC}\stackrel{\circ}{q}\!  {_A{^B}}{_{||
 BC}}),{_3} \right] \]

\subsection{Hyperboloid around ${\cal S}^+$} 
 Let us return to the coordinate $\rho:=\frac 1{\sinh\omega}$.  The
 metric on $M$ takes the starting form (\ref{etaonM}).  It is
 convenient to introduce new canonical field variables similar to the
 variables for the scalar field and electrodynamics:
\[ \Psi_x := \rho^{-1} \underline{\bf x} \, , \quad \Psi_y := \rho^{-1}
\underline{\bf y} \]  
\[ \Pi_x:={\underline{\bf X} \over \sqrt{1+\rho^2}} \, , \quad
\Pi_y:={\underline{\bf Y} \over \sqrt{1+\rho^2}} \]
\noindent
 Equations of motion are the same for both degrees of freedom:
\[ \frac{1}{\sqrt{1+\rho^2}} \dot{\Psi}_{\Upsilon} -
\Psi_\Upsilon ,_{\rho} =
\frac{\Pi_\Upsilon\sqrt{1+\rho^2}}{\sin\theta} \quad\quad \Upsilon=x,y \]
\[ \dot\Pi_\Upsilon -(\Pi_\Upsilon \sqrt{1+\rho^2}) ,_{\rho} =
 {\sin\theta}\left[ {\bf a}\Psi_\Upsilon + ((1+\rho^2)\Psi_\Upsilon
 ,_{\rho}),{_{\rho}} \right] \]
\noindent
 and they are similar to (\ref{eqm1}), (\ref{eqm2}) for the scalar
 field and (\ref{el1}), (\ref{el2}) for electrodynamics.

The reduction of symplectic form from the previous section allows to
 formulate the hamiltonian relation in terms of new canonical
 variables: \[ \sum_{\Upsilon=x,y} \int_V
\dot{\Psi}_\Upsilon{\bf a}^{-1}({\bf a}+2)^{-1}
\delta \Pi_\Upsilon - \dot\Pi_\Upsilon {\bf a}^{-1}({\bf
 a}+2)^{-1} \delta {\Psi}_\Upsilon = \delta {\cal H} + \] \be
 -\sum_{\Upsilon=x,y} \int_{\partial V}
\left[{\Pi_\Upsilon\sqrt{1+\rho^2}}
 + \sin\theta (1+\rho^2)\Psi_\Upsilon ,_{\rho} \right]{\bf
 a}^{-1}({\bf a}+2)^{-1} \delta {\Psi}_\Upsilon \ee where
\[ {\cal H}:= \frac1{2}\sum_{\Upsilon=x,y} \int_V \left( 
 {{\Pi_\Upsilon\sqrt{1+\rho^2}}\over \sin\theta} + \Psi_\Upsilon
 ,_{\rho} \right) {\bf a}^{-1}({\bf a}+2)^{-1} \left(
 {{\Pi_\Upsilon\sqrt{1+\rho^2}}\over \sin\theta} + \Psi_\Upsilon
 ,_{\rho} \right) +
\]
\be + \frac1{2}\sum_{\Upsilon=x,y} \int_V \rho^2 \sin\theta 
\Psi_\Upsilon ,_{\rho}{\bf a}^{-1}({\bf a}+2)^{-1} \Psi_\Upsilon
 ,_{\rho} - \sin\theta \Psi_\Upsilon ({\bf a}+2)^{-1}
\Psi_\Upsilon
\ee
 Similarily for angular momentum we propose the following expression:
\be J_z=\sum_{\Upsilon=x,y} 
\int_\Sigma \Pi_\Upsilon {\bf a}^{-1}({\bf a}+2)^{-1} 
 {\Psi}_{\Upsilon ,\phi}
\ee
 The non-conservation laws for the energy and angular momentum
\[ -\partial_0 {\cal H} = \sum_{\Upsilon=x,y}
\int_{S(0)} \sin\theta \dot\Psi_\Upsilon
 {\bf a}^{-1}({\bf a}+2)^{-1} \dot\Psi_\Upsilon
\]
\[ -\partial_0 J_z = \sum_{\Upsilon=x,y}
\int_{S(0)} \sin\theta \dot\Psi_\Upsilon
 {\bf a}^{-1}({\bf a}+2)^{-1} \Psi_\Upsilon ,_\phi
\]
 are similar to (\ref{Hdot}), (\ref{Jdot}) and (\ref{HdotEM}),
 (\ref{JdotEM}).  It should be also possible to formulate linear
 momentum $P_z$ in a similar way as (\ref{pz0})
\[ -\partial_0 P_z = \sum_{\Upsilon=x,y}
\int_{S(0)} \sin\theta\cos\theta \dot\Psi_\Upsilon
 {\bf a}^{-1}({\bf a}+2)^{-1} \dot\Psi_\Upsilon
\]
 but this will be analyzed in a separate paper.  It is obvious that
 all these formulae are {\em quasi-local}.

\subsection{Appendix -- explicit formulae on a hyperboloid}
 
\[  \Lambda=\sinh^2\omega \sin\theta \quad N_{|m}=-N_m \quad
 N_{|kl}=-N_{k|l}=N\eta_{kl} \quad N^{|k}{_k}=-N^k_{|k}=3N \]
\[N=\cosh\omega \quad N^3=N_3=-\sinh\omega \quad N^A=0 \quad
\eta_{AB,3}=2\kappa 
\eta_{AB} \] 
\[ \eta^{AB},{_3}=-2\kappa \eta^{AB} \quad
\varepsilon_{AB,3}=2\kappa \varepsilon_{AB}
\quad \varepsilon^{AB},{_3}=-2\kappa \varepsilon^{AB} \quad
\kappa=\coth\omega \] 
\[ \Lambda,{_3}=2\kappa\Lambda \quad \kappa N,{_3}=-\kappa N_3 =N \quad
\kappa,{_3} =-\frac 1{\sinh^2\omega} \]
\[ \Gamma^3{_{AB}}=-\kappa \eta_{AB} \quad \Gamma^A{_{B3}}=\kappa
\delta^A_B \quad \Gamma^A{_{BC}},{_3} =0 \]
\[ {^2}\! R_{ABCD}=\frac 1{\sinh^2\omega} \left(\eta_{AC}\eta_{BD} -
\eta_{AD}\eta_{BC}\right) \quad {^2}\! R_{AB}=\frac 1{\sinh^2\omega} \eta_{AB}
\] 
\[ \xi^{|33}=\xi^{,33} \quad \xi^{|3A}=\xi^{,3A}-\kappa \xi^{,A} \quad 
\xi^{|AB}=\xi^{||AB}+\kappa \eta^{AB} \xi^{,3} \]
\[ \xi_{3|3}=\xi_{3,3} \quad \xi_{3|A}=\xi_{3||A}-\kappa \xi_{A} \quad 
\xi_{A|3}=\xi_{A,3}-\kappa \xi_{A} \]
\[
\xi_{A|B}=\xi_{A||B}+\kappa \eta^{AB} \xi_{3} \]
\[ h_{33|3}=h_{33,3} \quad h_{33|A}=h_{33,A}-2\kappa h_{3A}  \]
\[ h^{AB}{_{|3}}= h^{AB},{_3} +2\kappa h^{AB} \quad
 h_{AB|3}= h_{AB},{_3} -2\kappa h_{AB} \quad h^{A}{_{B|3}}=
 h^{A}{_B},{_3} \] \[ h^{3A}{_{|3}}= h^{3A},{_3} +\kappa h^{3A}
\quad
 h_{3A|3}= h_{3A},{_3} -\kappa h_{3A} \]
\[ h_{3A|B}=h_{3A||B}-\kappa h_{AB}+\kappa \eta_{AB}h_{33} \]
\[
 h_{AB|C}=h_{AB||C}+\kappa\eta_{AC} h_{3B}+\kappa\eta_{BC} h_{3A}
\]

\section{Linearized gravity in null coordinates}
 We are going to follow the idea from subsection 2.2 and apply it to
 linearized gravity.
\subsection{Minkowski metric in null coordinates}
 Let us introduce null coordinates: $u:=t-r$, $v:=r+t$ together with
 the index $a$ corresponding to the coordinates $(u,v)$. The spherical
 foliation is the same as previously and the coordinates on a sphere
 $(x^A),(A=1,2)$, $(x^1=\theta, x^2=\phi)$ are the same.

For convenience we need also some more denotations:
 $\rho:=r^{-1}=\frac2{v-u}$, $\rho_{,a}=\rho^2 \epsilon_a$ where
 $\epsilon_u:=\frac12$, $\epsilon_v:=-\frac12$,
 $\eta^{ab}\epsilon_a\epsilon_b=1$. We will also need
 $\epsilon^a:=\eta^{ab}\epsilon_b$ and we can check that
 $\epsilon^u=1$, $\epsilon^v=-1$, $\eta_{ab}\epsilon^a\epsilon^b=1$.

The explicit formulae for the components of Minkowski metric can be
 denoted as follows:

\[ \eta_{AB}=\rho^{-2}\kolo\gamma_{AB} \, , \quad
\eta_{ab}=-\frac12 |E_{ab}| \, , \quad \eta_{aA}=0 \] where
 $E_{uu}=0=E_{vv}$ and $E_{uv}=1=-E_{vu}$.  Similarily the inverse
 metric has the following components
\[ \eta^{AB}=\rho^{2}\kolo\gamma^{AB} \, , \quad \eta^{ab}=-2 |E^{ab}| \,
 , \quad \eta^{aA}=0 \] where $E^{uu}=0=E^{vv}$ and
 $E^{uv}=1=-E^{vu}$.  We shall also need derivatives
\[ \eta^{AB}{_{,a}}=2\rho\epsilon_a\eta^{AB} \, , \quad
\eta_{AB}{_{,a}}=-2\rho\epsilon_a\eta_{AB} \] and finally the
 nonvanishing Christoffel symbols except $\Gamma^A{_{BC}}$ are the
 following:
\[ \Gamma^a{_{AB}}=\rho\epsilon^a\eta_{AB} \, , \quad
\Gamma^A{_{aB}}=-\rho\epsilon_a\delta^A{_B} \]

\subsection{ Riemann tensor in null coordinates}
 We need to derive linearized Riemann tensor in null coordinates:
\[ 
 2R_{abcd}= h{_{ad,bc}} -h{_{bd,ac}} + h_{bc,ad} - h{_{ac,bd}} \]
\[ 
 2R_{abcD}= h{_{aD,bc}} -h{_{bD,ac}} + h_{bc,aD} - h{_{ac,bD}} +
\] \[ 
 +\rho\epsilon_b \left( h_{aD,c} + h_{cD,a}- h_{ac,D} \right) -
\rho \epsilon_a \left( h_{bD,c} + h_{cD,b}- h_{bc,D} \right)
\]
\[ 
 2R_{AbCd}= h{_{dA||C,b}} +h{_{bC||A,d}} - h_{bd||AC} - h{_{AC,bd}} +
\] \[ +\rho\epsilon_b \left( h_{dA||C} -
 h_{dC||A}- h_{AC,d} \right) - \rho \epsilon_d \left( h_{bC||A} -
 h_{bA||C}- h_{AC,b} \right)+
\]
\[ +\rho\eta_{AC}\epsilon^a \left( h_{bd,a}-h_{ad,b}-h_{ab,d}\right)
 -2\rho^2\epsilon_b\epsilon_d h_{AC}
\]
\[ 
 2R_{ABCd}= h{_{dA||BC}} +h{_{BC||A,d}} - h_{Bd||AC} - h{_{AC||B,d}}
 +2\rho \epsilon_d \left( h_{bC||A} - h_{AC||B}
\right)+ \]
\[ +\rho\eta_{BC} \epsilon^a \left( h_{aA,d} - h_{dA,a}+
 h_{ad,A} +2\rho\epsilon_d h_{aA} \right) -\rho\eta_{AC}\epsilon^a
\left( h_{aB,d}-h_{dB,a}+h_{ad,B}
 +2\rho\epsilon_d h_{aB} \right) \]

\[ 
 2R_{abCD}= h{_{aD||C,b}} -h{_{bD||C,a}} + h_{bC||D,a} - h{_{aC||D,b}}
 + \] \[ +2\rho\epsilon_b \left( h_{aD||C} - h_{aC||D} \right) +2 \rho
\epsilon_a \left( h_{bC||D} -
 h_{bD||C} \right)
\]

\[ 
 2R_{ABCD}= h{_{AD||BC}} +h{_{BC||AD}} - h_{BD||AC} - h{_{AC||BD}} +\]
\[ +\rho\eta_{AC}\epsilon^a \left(
 h_{BD,a}-h_{aB||D}-h_{aD||B} \right) +\rho\eta_{BD}\epsilon^a
\left( h_{AC,a}-h_{aC||A}-h_{aA||C} \right)
 + \] \[ -\rho\eta_{BC} \epsilon^a \left( h_{AD,a} - h_{aA||D} -
 h_{aD||A} \right) -\rho\eta_{AD}\epsilon^a \left(
 h_{BC,a}-h_{aB||C}-h_{aC||B} \right) +\] \[ +\rho^2 \left( h_{BD}
\eta_{AC} +h_{AC}\eta_{BD} -h_{AD}\eta_{BC}
 -h_{BC}\eta_{AD} \right) +2\rho^2\epsilon^a\epsilon^b h_{ab}
\left( \eta_{AC}\eta_{BD}
 -\eta_{BC}\eta_{AD} \right)
\]

\subsection{ Ricci tensor in null coordinates}
 The Ricci tensor takes the following form:
\[ 
 2R_{ab}= h^c{_{b,ac}}+h_a{^c}{_{,cb}} - h_{ab}{^{,c}}{_c} -
 h^c{_{c,ab}} + h_{aA,b}{^{||A}} + h_{bA,a}{^{||A}} -
 h_{ab}{^{||A}}{_A} -H_{,ab}+ \] \[ +\rho\epsilon_a H_{,b}
 +\rho\epsilon_b H_{,a} + 2\rho \epsilon^c \left( h_{ab,c} -h_{ac,b}-
 h_{bc,a} \right)
\]
\[
 2R_{aB}= h^b{_{B,ab}} -h_{aB}{^{,c}}{_c} +h_a{^c}{_{,cB}} -
 h^c{_{c,aB}} +h_{a}{^{A}}{_{||BA}} - h_{aB}{^{||A}}{_A}
 +\chi_{B}{^{A}}{_{||A,a}} -\frac12 H_{||B,a} + \] \[ +\rho\epsilon_a
\left( 2 h^b{_{B,b}} - h^b{_{b,B}} \right)
 -2\rho\epsilon^b h_{bB,a} -2\rho^2\epsilon_a\epsilon^b h_{bB}
\]
\[
 2R_{AB}= \left( h^a{_{A||B}}+h^a{_{B||A}} \right)_{,a} -
 h^a{_{a||AB}} - \chi_{AB}{^{,a}}{_{a}}-2\rho\epsilon^a
\chi_{AB,a}
 +\chi_A{^C}{_{||CB}} +\chi_B{^C}{_{||CA}} + \]
\[ -\chi_{AB}{^{||C}}{_C} 
 +\eta_{AB} \left[ -\frac12 ( H^{||C}{_C} + H^{,a}{_a} )
 +2\rho\epsilon^a (H_{,a} -h_a{^A}{_{||A}}) +\rho^2 (2\epsilon^a
\epsilon^b h_{ab} -H )   \right]  
\]

\subsection{Gauge in null coordinates}
 The gauge transformation $\xi_\mu$
\[ h_{\mu\nu}\rightarrow h_{\mu\nu}+\xi_{(\mu;\nu)}\]
 splits in the following way:
\[ h_{ab} \longrightarrow h_{ab}+\xi_{a,b}+\xi_{b,a}\]

\[h_{aA}  \longrightarrow  h_{aA}+\xi_{a,A}+\xi_{A,a}+2\rho \epsilon_a \xi_A\]

\[ h_{AB} \longrightarrow h_{AB}+\xi_{A||B}+\xi_{B||A}-2\rho
\eta_{AB}\epsilon^a\xi_a\] 
 and it would be also useful the following formulae
\[\chi_{AB} \longrightarrow
\chi_{AB}+\xi_{A||B}+\xi_{B||A}-\eta_{AB}\xi^C{_{||C}}\] 

\[\frac12 H  \longrightarrow \frac12 H +\xi^A{_{||A}}-2\rho\epsilon^a\xi_a\]

\[ h_a{^A}  \longrightarrow h_a{^A} +\xi_a{^{||A}}+\xi^A{_{,a}}\]
 which are straightforward consequences of the previous one.

\subsection{Invariants}
 Let us introduce the following gauge invariant quantities:

\begin{eqnarray}\label{ya}
 {\bf y}_a :=& ({\bf a}+2)h_{aA||B}\epsilon^{AB}
 -(\rho^{-2}\chi_A{^C}{_{||CB}} \epsilon^{AB})_{,a}
\end{eqnarray} 
\begin{eqnarray} {\bf y} :=& 2\rho^{-2}(h_{bB||A}\epsilon^{AB})_{,a}E^{ab} 
\end{eqnarray} 
\begin{eqnarray}
 {\bf x} :=& \rho^{-2}\chi^{AB}{_{||BA}}-\frac12 {\bf
 a}H+\rho^{-1}\epsilon^a H_{,a}-H+2\epsilon^a
\epsilon^bh_{ab}-2\rho^{-1}\epsilon^ah_a{^A}{_{||A}}
\end{eqnarray} 
\begin{eqnarray} {\bf x}_{ab} :=& {\bf a}({\bf a}+2) h_{ab} -({\bf a}+2)
\left[(\rho^{-2}h_a{^A}{_{||A}})_{,b} +(\rho^{-2}h_b{^A}{_{||A}})_{,a}
\right] +\left[\rho^{-2}(\rho^{-2}\chi^{AB}{_{||AB}})_{,a} \right]_{,b}+
\nonumber \\  &
 +\left[\rho^{-2}(\rho^{-2}\chi^{AB}{_{||AB}})_{,b} \right]_{,a}
\label{xab}
\end{eqnarray} 

They fullfill the following equations:
\[ (\rho^{-2}{\bf y}^a)_{,a}=0 \quad 2\rho^{-2}{\bf y}_{b,a}E^{ab}
 =(a+2){\bf y}\]

\[ 2E^{ab}(\rho^{-2}{\bf y})_{,b} +\rho^{-2}{\bf y}^a=0 \]

\[ [\rho^{-4}({\bf y}_{a,b}-{\bf y}_{b,a})]^{,b}+\rho^{-2}(a+2) {\bf y}_a
 =0 \]

\[ (\rho^{-1}{\bf y})^{,a}{_a}+\rho^{-1}a{\bf y}=0\]

\[ (\rho^{-1} {\bf x})^{,a}{_a}=-\rho^{-1}a{\bf x}\]

\[ \rho^{-2} {\bf x}^{ab}{_{,ab}} ={\bf a}({\bf a}+2) {\bf x} \]

\be\label{etaxab} \eta^{ab} {\bf x}_{ab}=0 \ee
 
\[ {\bf x}_{ab}=2(\rho^{-2}{\bf x})_{,ab}
 -\eta_{ab}(\rho^{-2}{\bf x})^{,c}{_c} \] if we assume vacuum
 equations $R_{\mu\nu}=0$.

\subsection{Reduction of symplectic form on $\scri^+$}
 Now we will show how linearized symplectic form on $N$, which was
 introduced by Kijowski in the full nonlinear theory, can be reduced
 to the invariants in ``wave'' part similarily to the hyperboloid
 case.

We shall calculate this form in a convenient gauge but final result
 will be gauge-invariant. This way we shall prove that modulo boundary
 terms depending on the gauge the gauge-invariant part of the
 symplectic form can be obtained in the demanded shape in ``wave''
 part.

\subsubsection{gauge conditions}

We would like to work with the following gauge conditions:
\[ \chi_{AB}=0 \, , \quad h_a{^A}{_{||A}}=0 \]
 It is easy to verify that they are compatible for ``wave'' part:
\[ \rho^{-2}\chi^{AB}{_{||AB}} \longrightarrow \rho^{-2}\chi^{AB}{_{||AB}}
 + ({\bf a} +2)\xi^A{_{||A}} \] \[
\rho^{-2}\chi_A{^C}{_{||CB}}\varepsilon^{AB} \longrightarrow
\rho^{-2}\chi_A{^C}{_{||CB}}\varepsilon^{AB} + ({\bf a}
 +2)\xi_{A||B}\varepsilon^{AB} \] \[ h_a{^A}{_{||A}}
\longrightarrow h_a{^A}{_{||A}}
 +\rho^{-2}{\bf a}\xi_a +(\xi^A{_{||A}})_{,a} \] More precisely,
 mono-dipole free part of $\xi_a$ and $\xi_A$ is uniquely defined
 under these gauge conditions.
  
\subsubsection{partial reduction to extract gauge invariant part}
 The linearized $\pi^{\mu\nu}$ has the following form
\[ \pi^{\mu\nu}=-\Lambda h^{\mu\nu}+\frac 12 \eta^{\mu\nu}(h^a_a+H)\Lambda \]
 and it can be simplified in our gauge. Let us observe that invariants
 (\ref{ya}) and (\ref{xab}) have simple form in our gauge in terms of
 $h_{\mu\nu}$. From (\ref{xab}) and (\ref{etaxab}) we obtain
 $\underline h^a_a=-4\underline h_{uv}=0$.  Moreover $\underline
\pi^{AB}=\frac
 12\Lambda\eta^{AB}\underline h^a_a -\Lambda\chi^{AB}$ vanishes.
 Similarily $\pi^{Ab}{_{||A}}=0$ because $\pi^{Ab}=-\Lambda h^{Ab}$.
 Finally from the above considerations we obtain:
\[ \int_{S(s,\rho)} \underline\pi^{\mu\nu}\delta\underline A^a_{\mu\nu}
 =\int_{S(s,\rho)} \underline\pi^{cd}\delta \underline A^a_{cd}
 -2\underline{\pi{^{bA||B}}\varepsilon_{AB}}\rho^{-2}{\bf a}^{-1}
\delta \underline{A^a{_{bA||B}}\varepsilon^{AB}}  \]
 One can show the following relation:
\be\label{dxab} \int_V 
\underline \pi^{cd}\delta \underline A^v_{cd} \sim \int_V
\Lambda\rho^2 (\rho^{-1}{\bf x}^{ab})_{,u}
 {\bf a}^{-2}({\bf a} +2)^{-2}\delta(\rho^{-1}{\bf x}_{ab}) \ee where
 $\sim$ denotes equality modulo boundary terms and full variation.
 
Similarily one can prove \be\label{dyb} \int_V
\underline{\pi{^{bA||B}}\varepsilon_{AB}}\rho^{-2}{\bf a}^{-1}
\delta \underline{A^v{_{bA||B}}\varepsilon^{AB}} \sim \int_V
\Lambda\rho^2 {\bf y}^b
 {\bf a}^{-1}({\bf a} +2)^{-2}\delta\left[ (\rho^{-4}{\bf y}^v)_{,b} -
 (\rho^{-2}{\bf y}_b)^{,v} \right] \ee
 
\subsubsection{full reduction to x,y}
 We would like to obtain similar formula to (\ref{formonS}).  The
 ``curl'' part (\ref{dyb}) reduces easily to the demanded form
\[ \int_V \Lambda\rho^2 {\bf y}^b
 {\bf a}^{-1}({\bf a} +2)^{-2}\delta\left[ (\rho^{-4}{\bf y}^v)_{,b} -
 (\rho^{-2}{\bf y}_b)^{,v} \right] \sim \int_V 2\Lambda\rho^2
 (\rho^{-1} \underline{\bf y})_{,u} {\bf a}^{-1}({\bf a}+2)^{-1}
\delta (\rho^{-1} \underline{\bf y}) \] 
 On the other hand the second part (\ref{dxab}) can be rewritten in
 the following way
\[ \int_V 
\Lambda\rho^2 (\rho^{-1}{\bf x}^{ab})_{,u}
 {\bf a}^{-2}({\bf a} +2)^{-2}\delta(\rho^{-1}{\bf x}_{ab}) \sim
\int_V 
 2\Lambda\rho^2 (\rho^{-1} \underline{\bf x})_{,u} {\bf a}^{-1}({\bf
 a}+2)^{-1}
\delta (\rho^{-1} \underline{\bf x}) +\]
\[ +\int_V  16\Lambda\rho^2\left[\rho^{-1} (\rho^{-1} \underline{\bf x})_{,vv}
 +\frac12 {\bf a}(\rho^{-1} \underline{\bf x})_{,v} \right] {\bf
 a}^{-2}({\bf a}+2)^{-2} \delta (\rho^{-1} \underline{\bf x}) \] Let
 us observe that the last term vanishes on $\scri^+$, more precisely
 $(\rho^{-1}{\bf x})_{,v}=O(\rho^2)$.  Presented calculations should
 convince the reader that the following formula
\[ \int_N \underline\pi^{\mu\nu}\delta \underline A^v_{\mu\nu} \sim \int_N 
 2\Lambda\rho^2\left[ (\Psi_{ x})_{,u} {\bf a}^{-1}({\bf a}+2)^{-1}
\delta (\Psi_{ x}) +
 (\Psi_{ y})_{,u} {\bf a}^{-1}({\bf a}+2)^{-1}
\delta (\Psi_{ y}) \right]  \] 
 holds and this is {\em quasi-local} form which is similar to
 (\ref{formonS}) and (\ref{formonSEM}).

\section{Generating formula for Einstein equations}
\label{generating}

There are different variational principles which may be used to derive
 Einstein equations.  They may be classified as belonging to three
 basic approaches:
\begin{enumerate} \item The purely metric approach, where the variation is
 performed with respect to the metric tensor.  As a Lagrangian one can
 use the second order Hilbert Lagrangian or the first order (gauge
 dependent) Lagrangian, quadratic with respect to the Christoffel
 symbols.  \item The metric--affine approach, based on the Palatini
 variational principle, where the variation is performed independently
 with respect to the metric tensor and to the connection.
\item The purely affine approach where the variation is performed with respect
 to the connection.  The metric tensor arises as a momentum
 canonically conjugate to the connection -- see \cite{affine}.
\end{enumerate}

Each of these variational principles leads to the same Hamiltonian
 description of the theory.  In this paper we use the Hilbert
 variational principle.  At the end of this section we will show how
 the different variational principles converge to the same generating
 formula.  Hence, the canonical structure, derived from this formula,
 does not depend upon the variational principle we begin with.

The variation of the Hilbert Lagrangian \begin{equation} L =
\frac 1{16
\pi}
\sqrt{|g|} \ R \label{Hilbert} \end{equation} may be calculated as follows:
\begin{equation} \delta L = \delta \left( \frac 1{16 \pi} \sqrt{|g|} \
 g^{\mu\nu} \ R_{\mu\nu} \right) = - \frac 1{16 \pi} {\cal G}^{\mu\nu}
\delta g_{\mu\nu} + \frac 1{16 \pi} \sqrt{|g|} \ g^{\mu\nu} \delta
 R_{\mu\nu}
\label{deltaR} \end{equation} where \begin{equation} {\cal G}^{\mu\nu} :=
\sqrt{|g|} \ (R^{\mu\nu} - \frac 12 g^{\mu\nu} R) \ .  \end{equation} We are
 going to prove that the last term in (\ref{deltaR}) is a boundary
 term (a complete divergence).  For this purpose we denote:
\begin{equation}
\label{defpi} {\pi}^{\mu\nu} := \frac 1{16 \pi} \sqrt{|g|} \ g^{\mu\nu} \
 , \end{equation} and
\begin{equation}\label{defA}
 A^{\lambda}_{\mu\nu} := {\Gamma}^{\lambda}_{\mu\nu} -
 {\delta}^{\lambda}_{(\mu} {\Gamma}^{\kappa}_{\nu ) \kappa}
\end{equation}
 (do not try to attribute any sophisticated geometric interpretation
 to $A^{\lambda}_{\mu\nu}$; it is merely a frequently appearing
 combination of the connection coefficients, which we introduce in
 order to simplify the derivation of the final formula).  We have:
\begin{eqnarray}
\partial_\lambda A^{\lambda}_{\mu\nu} & = & \partial_\lambda
 {\Gamma}^{\lambda}_{\mu\nu} -
\partial_{(\mu} {\Gamma}^{\lambda}_{\nu ) \lambda} = R_{\mu\nu} -
 {\Gamma}^{\lambda}_{\sigma\lambda} {\Gamma}^{\sigma}_{\mu\nu} +
 {\Gamma}^{\lambda}_{\mu\sigma} {\Gamma}^{\sigma}_{\nu\lambda} \\
\nonumber
 & = & R_{\mu\nu} + A^{\lambda}_{\mu\sigma} A^{\sigma}_{\nu\lambda} -
\frac 13 A^{\lambda}_{\mu\lambda}
 A^{\sigma}_{\nu\sigma} \ .  \end{eqnarray} Hence, we obtain an
 identity \begin{eqnarray} \partial_\lambda \left( {\pi}^{\mu\nu}
\delta A^{\lambda}_{\mu\nu} \right) & = & {\pi}^{\mu\nu} \delta R_{\mu\nu} +
 {\pi}^{\mu\nu} \delta \left( A^{\lambda}_{\mu\sigma}
 A^{\sigma}_{\nu\lambda} -
\frac 13 A^{\lambda}_{\mu\lambda} A^{\sigma}_{\nu\sigma} \right) + \left(
\partial_\lambda {\pi}^{\mu\nu} \right) \delta A^{\lambda}_{\mu\nu}
\\ \nonumber 
 & = & {\pi}^{\mu\nu} \delta R_{\mu\nu} + \left( \nabla_\lambda
 {\pi}^{\mu\nu}
\right) \delta A^{\lambda}_{\mu\nu} \ .  \end{eqnarray} Due to metricity of
 $\Gamma$ we have $\nabla_\lambda {\pi}^{\mu\nu} = 0$.  This way we
 obtain
\begin{equation} \label{pidR} {\pi}^{\mu\nu} \delta R_{\mu\nu} =
\partial_\lambda \left( {\pi}^{\mu\nu} \delta A^{\lambda}_{\mu\nu} \right) =
\partial_\kappa \left( {\pi}_{\lambda}^{\ \mu\nu\kappa} \delta
 {\Gamma}^{\lambda}_{\mu\nu} \right) \ , \end{equation} where we
 denote
\begin{equation} {\pi}_{\lambda}^{\ \mu\nu\kappa} := {\pi}^{\mu\nu}
\delta^\kappa_\lambda - {\pi}^{\kappa ( \nu} \delta^{\mu )}_\lambda \ .
\end{equation} Inserting (\ref{pidR}) into (\ref{deltaR}) we have:
\begin{equation} \label{deltaL-grav} \delta L = - \frac 1{16 \pi} {\cal
 G}^{\mu\nu} \delta g_{\mu\nu} + \partial_\lambda \left(
 {\pi}^{\mu\nu}
\delta A^{\lambda}_{\mu\nu} \right) \ .  \end{equation} We conclude that
 Euler-Lagrange equations ${\cal G}^{\mu\nu} = 0$ are equivalent to
 the following generating formula, analogous to (\ref{Lpole}) in field
 theory: \begin{equation} \delta L =
\partial_\lambda \left( {\pi}^{\mu\nu} \delta A^{\lambda}_{\mu\nu} \right)
\label{dL=pidA} \end{equation} or, equivalently, \begin{equation} \delta L =
\partial_\kappa \left( {\pi}_{\lambda}^{\ \mu\nu\kappa} \delta
 {\Gamma}^{\lambda}_{\mu\nu} \right) \label{dL=pidgamma} \ .
\end{equation} This formula is a starting point for our derivation of
 canonical gravity.  Let us observe, that it is valid not only in the
 present, purely metric, context but also in any variational
 formulation of General Relativity.  For this purpose let us rewrite
 (\ref{deltaL-grav}) without using {\em a priori} the metricity
 condition $\nabla_\lambda {\pi}^{\mu\nu} = 0$.  This way we obtain
 the following, universal formula:
\begin{equation} \delta L = - \frac 1{16 \pi} {\cal G}^{\mu\nu} \delta
 g_{\mu\nu} - \left( \nabla_\kappa {\pi}_{\lambda}^{\
\mu\nu\kappa} \right) \delta {\Gamma}^{\lambda}_{\mu\nu} + \partial_\kappa
\left( {\pi}_{\lambda}^{\ \mu\nu\kappa} \delta {\Gamma}^{\lambda}_{\mu\nu}
\right) \end{equation} It may be proved that in this form, the formula remains
 valid also in the metric-affine approach and in the purely-affine
 one.  In metric--affine formulation, the vanishing of $\nabla_\lambda
 {\pi}^{\mu\nu}$ is not automatic: it is a part of field equations.
 We see that, again, the entire field dynamics is equivalent to
 (\ref{dL=pidgamma}).  Finally, in the purely affine formulation of
 General Relativity the Einstein equations are satisfied ``from the
 very beginning'' whereas the metricity condition for the connection
 becomes the dynamical equation.  We conclude that also in this case
 the entire information about the field dynamics is contained in
 generating formula (\ref{dL=pidgamma}).

This formula, compared with (\ref{Lpole}) suggests that the role of
 field potentials in General Relativity should be rather played by the
 connection $\Gamma$, whereas the metric $g$ should rather remain on
 the side of canonical momenta.  This observation was the origin of
 the purely affine formulation of the theory.  Also in the
 multisymplectic formulation (i.~e.~formulation in terms of
 Poincar\'e-Cartan form -- see
\cite{Kij-Szczyr}) the connection appears on
 the side of field configurations.  We stress, however, that the
 results presented in this paper do not depend upon the choice of a
 variational formulation.

\section{Metrics of Bondi--Sachs type}

In this section we shall consider the initial value problem for the
 curved space-time $M$ with a metric of the form:

\be\label{gB} 
 g_{\mu\nu}{\rm d}x^\mu{\rm d}x^\nu= -\frac Vr \E{2\beta}{\rm d}u^2
 -2\E{2\beta}{\rm d}u{\rm d}r + r^2 \gamma_{AB}({\rm d}x^A -U^A {\rm
 d}u)({\rm d}x^B -U^B {\rm d}u)
\ee
 on the null cone $C= \left\{ x\in M \;| \; x^0=u=\mbox{const}
\right\}$ (see
\cite{XIV}, \cite{BVM}, \cite{Burg}) and boundary at null infinity
 $\partial C=S(u,0)$.  We have the following non-vanishing components
 of the inverse metric $g^{\mu\nu}$:
\[ g^{33}=\frac Vr \E{-2\beta} \]
\[  g^{03}=-\E{-2\beta} \]
\[ g^{3A}= - \E{-2\beta} U^A \]
\[  g^{AB}=\frac 1{r^2}\gamma^{AB} \]
 where $\gamma^{AB}$ is the inverse metric to $\gamma_{AB}$.

Let us define the ``covector'' $U_B$ as follows:
\[ U_B:=g_{BA}U^A=r^2 \gamma_{BA}U^A \]
 We have in our coordinate system the following non-vanishing
 components of the metric $g_{\mu\nu}$:
\[ g_{00}=-\frac Vr \E{2\beta} + U_A U^A \]
\[  g_{03}=-\E{2\beta} \]
\[ g_{0A}= - U_A \]
\[  g_{AB}= {r^2}\gamma_{AB} \]
 We also assume that
\[ \sqrt{\det \gamma_{AB}}=\sin\theta \]

The metric (\ref{gB}) implies the following expressions for $16\pi
\pi^{\mu\nu}=\sqrt{-g}g^{\mu\nu} $ and $ A^\lambda{_{\mu\nu}}=
\Gamma^\lambda{_{\mu\nu}}-\delta^\lambda_{(\mu}
\Gamma^\sigma{_{\nu)\sigma}} $ defined by (\ref{defpi}) and (\ref{defA}).
 
\[ \sqrt{-g}= \E{2\beta}r^2\sin\theta \]
\[ 16\pi\pi^{03}=-r^2\sin\theta \] \[ 16\pi\pi^{AB}=
\E{2\beta}\sin\theta\gamma^{AB} \]

\[ 16\pi\pi^{33}= rV\sin\theta \] \[ 16\pi \pi^{3A}= -r^2
\sin\theta U^{A} \] \[
 16\pi\pi^{03}=-r^2\sin\theta \] \[ 16\pi\pi^{AB}=
\E{2\beta}\sin\theta\gamma^{AB} \]

\[ A^0{_{33}} =A^0{_{3A}}=0 \]
\[ A^0{_{03}} = -\beta _{\! ,3} -\frac 1r \] \[ A^0{_{AB}}=\frac
 12
\E{-2\beta}\left( r^2 \gamma_{AB}\right)_{\! ,3} \]

\[ A^3{_{33}} = -\frac 2r \] \[ A^3{_{3A}}=\frac 12 \E{-2\beta}
 U^B{_{,3}}g_{BA} -\frac 12 \left(\ln\sin\theta\right)_{,A} \] \[
 A^3{_{03}} = \frac Vr \beta_{,3} +\left(\frac{V}{2r}\right)_{,3} -
 U^B\beta_{,B} -\dot{\beta}-\frac 12 \E{-2\beta}U_A U^A{_{,3}}
\] \[
 A^3{_{AB}}=\frac 12 \E{-2\beta}\left( \dot{g}_{AB} -\frac Vr
 {g_{AB}}{_{,3}} +U_{A||B}+U_{B||A} \right) \]

The following expression was proposed by Bondi to call the mass:
\be\label{mB}
 m_{\mbox{\tiny TB}}:=\frac 1{8\pi} \int_{\partial C} r-V \ee

Choose a (3+1)-foliation of space-time and integrate (\ref{dL=pidA})
 over a 3-dimensional null-volume $V \subset C=\{ x^0 = \mbox{const}
\}$:
\begin{equation} 
\delta \int_V L = \int_V \left( {\pi}^{\mu\nu} \delta
 A^{0}_{\mu\nu} \right)\dot{} + \int_{\partial V} {\pi}^{\mu\nu}
\delta
 A^{3}_{\mu\nu} \ .  \label{dLV}
\end{equation} 
 Similarily as in the case of electrodynamics, we use here adapted
 coordinates; this means that coordinate $x^3$ is constant on the
 boundary $\partial V$.  Adapted coordinates simplify considerably
 derivation of the final formula.  We stress, however, that all our
 results have an independent, geometric meaning.  To rewrite them in a
 coordinate-independent form it is sufficient to replace ``dots'' by
 Lie derivatives ${\cal L}_X$, where $X$ is the vector field
 generating our one-parameter group of transformations which we are
 describing.  In adapted coordinates $X:= \frac {\partial}{\partial
 x^0}$.  Moreover, the upper index ``3'' has to be replaced everywhere
 by the sign $\perp$, denoting the transversal component with respect
 to the world tube.  This way our results have a
 coordinate-independent meaning as relations between well defined
 geometric objects and not just their specific components.

Because the translation between these two notations is so simple, we
 have decided to use much simpler language, based on adapted
 coordinates.  The volume part of the formula (\ref{dLV}) can be
 simplified (or reduced) as follows:

\begin{eqnarray} 16\pi{\pi}^{\mu\nu} \delta A^{0}_{\mu\nu} & = &
 16\pi{\pi}^{kl} \delta A^{0}_{kl} + 32\pi{\pi}^{0k} \delta A^{0}_{0k}
 + 16\pi{\pi}^{00} \delta A^{0}_{00} \nonumber \\ & = &
 32\pi\pi^{03}\delta A^0{_{03}} +16\pi\pi^{AB} \delta A^0{_{AB}}
\nonumber \\ & = & -\frac 12 \sin\theta \left( r
\gamma_{AB}\right)_{\! ,3} \delta \left( r
\gamma^{AB}\right)+\nonumber \\ & &   + \delta\left[ 2r^4\sin\theta
\left(\frac{\beta}{r^2}\right)_{,3}\right]  \label{piA0}
\end{eqnarray}
 The last term in the above formula is a full variation of the
 quantity which logarithmically diverges when we try to integrate it,
 $\beta=O(r^{-2})$ and $2r^4\sin\theta
\left(\frac{\beta}{r^2}\right)_{,3}=O(r^{-1})$. Removing of this term
 (-- the renormalization of the symplectic form) corresponds to the
 renormalization of the lagrangian for scalar field (\ref{Lren}).

On the other hand the boundary part in (\ref{dLV}) can be rewritten as
 follows:
\[ 16\pi \pi^{\mu\nu} \delta  A^3{_{\mu\nu}} = 16\pi\pi^{33}\delta A^3{_{33}}+
 32\pi\pi^{03}\delta A^3{_{03}} +32\pi\pi^{3A}\delta A^3{_{3A}}
 +16\pi\pi^{AB} \delta A^3{_{AB}}= \]
\[ = 2\sin\theta\left(2V- r^2 U^B{_{||B}}\right)\delta\beta
 +\sin\theta\gamma^{AB}\delta U_{A||B} -\frac 12 \sin\theta
\left(
\dot{g}_{AB} -\frac Vr {g_{AB}}{_{,3}}  \right)\delta \gamma^{AB} + \] \be
 +r^2\sin\theta \E{-2\beta}U^B{_{,3}}g_{BA}\delta U^A -\delta
\left[ 2
 r^2\sin\theta \left(\frac V{r^2}+ \left(\frac{V}{2r}\right)_{,3}
 +\frac Vr \beta _{,3} -\dot{\beta} -U^B\beta_{,B} \right)
\right] \label{piA3} \ee

where we have denoted by ``$||$'' a covariant derivative with respect
 to the two-metric $g_{AB}$ on $\partial V$.

Inserting these results into (\ref{dLV}) we obtain:
\begin{eqnarray} 16\pi
\delta \int_V L & = & -\int_V \frac 12\sin\theta\left[ \left( r
\gamma_{AB}\right)_{\! ,3} \delta \left( r \gamma^{AB}\right)\right]_{,0}+
\int_{\partial V} r^2\sin\theta \E{-2\beta}U^B{_{,3}}g_{BA}\delta U^A +
\nonumber \\ & &
 + \int_{\partial V} 2\sin\theta\left( 2V- r^2
 U^B{_{||B}}\right)\delta
\beta +\frac 12 \sin\theta \left(rV
 {\gamma_{AB}}{_{,3}-r^2\dot{\gamma}_{AB} -2 U_{A||B}}
\right)\delta
\gamma^{AB} +\nonumber \\ & & +\delta \int_V 4r^2\sin\theta
\dot{\beta}_{,3} -\delta \int_{\partial V}  r^2\sin\theta\left[
\frac {2V}{r^2}+
\left(\frac{V}{r}\right)_{,3} +2\frac Vr \beta _{,3}  -2U^B\beta_{,B}
\right] \end{eqnarray} 
 because 2-dimensional divergencies ``$\partial_A f^A$'' vanish when
 integrated over the boundary $\partial V$.

It is convenient to introduce the following asymptotic variables
 $(\Pi_{AB},\psi^{AB})$ related to asymptotic degrees of freedom:
\[ \psi^{AB}:=r\gamma^{AB} \]
\[ \Pi_{AB}:= -\frac 12 \sin\theta\left( r \gamma_{AB}\right)_{\! ,3} \]
 From (\ref{piA0}) we get the following relation:

\begin{eqnarray} 16\pi{\pi}^{\mu\nu} \dot{A}^{0}_{\mu\nu} & = &
 -\frac 12 \sin\theta \left( r \gamma_{AB}\right)_{\! ,3} \left( r
\dot\gamma^{AB}\right)   +  2r^4\sin\theta
\left(\frac{\dot\beta}{r^2}\right)_{,3} \label{piA00}
\end{eqnarray}

On the other hand from \cite{Kij-Tulcz} we know that:
\[ 16\pi \int_V {\pi}^{\mu\nu} {\cal L}_X {A}^{0}_{\mu\nu} =
\int_{\partial V} \sqrt{-g}\left(\nabla^3 X^0 -\nabla^0 X^3 \right) =
\]
\be \label{komar}
 = \int_{\partial V} r^2\sin\theta\left[
\left(\frac{V}{r}\right)_{,3} +2\frac Vr \beta _{,3}  -2U^B\beta_{,B}
 -2\dot\beta -\E{-2\beta} U_A U^A_{,3} \right] \; (=2r^2\sin\theta
 A^3_{03}) \ee where the last equality can be checked directly for the
 metric (\ref{gB}) and $X^\mu=\delta^\mu_0$.  From (\ref{piA00}) and
 (\ref{komar}) we obtain final formula: \begin{eqnarray} 16\pi
\delta \int_V L & = & \int_V \frac 12\sin\theta\left[
\left( r\dot\gamma^{AB}\right) \delta \left( r
\gamma_{AB}\right)_{,3}-
\left( r\dot\gamma_{AB}\right)_{\! ,3} \delta \left( r
\gamma^{AB}\right)\right] 
 -\delta \int_{\partial V}2V\sin\theta+ \nonumber \\ & & +\frac 12
\int_{\partial V}\sin\theta \left(rV
 {\gamma_{AB}}{_{,3}-r^2\dot{\gamma}_{AB} -2 U_{A||B}}
 +r^2\E{-2\beta}U^C_{,3}\gamma_{CA}U_B \right)\delta
\gamma^{AB} + \nonumber \\ & & + \int_{\partial V} 2r^2\sin\theta
\left(\frac{2V}{r^2} -U^B{_{||B}} +U_A U^A_{,3}  \right) \delta\beta 
 - r^2\sin\theta \E{-2\beta} U_A \delta U^A_{,3} \label{0nab}
\end{eqnarray} 
\underline{Remark} It seems to me that more natural ``control mode'' in
 the above formula corresponds to the control of the term
 $(r^2U^A)_{,3}$ than $U^A_{,3}$ and it can be achieved by the
 following manipulation:
\[   - r^2\sin\theta \E{-2\beta} U_A \delta U^A_{,3}  = 
 - \sin\theta \E{-2\beta} U_A \delta (r^2 U^A)_{,3} +\delta
\left(
 r\sin\theta \E{-2\beta} U_A U^A \right) +\] \[ + 2r\sin\theta
\E{-2\beta} U_A U^A\delta\beta +\frac 1r \sin\theta
\E{-2\beta} U_A U^A \delta \gamma^{AB} \]
 
If we pass to the limit the formula (\ref{0nab}) takes the following
 form:
\be\label{Mham}
 -16\pi\delta m_{\mbox{\tiny TB}}=-\delta\int_{\partial C}
 4M\sin\theta= \int_C \dot \Pi_{AB} \delta \psi^{AB} -
\dot\psi^{AB} \delta \Pi_{AB}
 -\frac 12 \int_{\partial C}\sin\theta r^2\dot{\gamma}_{AB}
\delta
\gamma^{AB} \ee
 where $V=r-2M+O(r^{-1})$ and the asymptotic conditions are given in
\cite{Burg} and will be summarized in the next section. We can denote
 non-conservation law for the TB mass:
\be\label{M0}
 -16\pi\partial_0 m_{\mbox{\tiny TB}}= -\frac 12 \int_{\partial
 C}\sin\theta r^2\dot{\gamma}_{AB}
\dot\gamma^{AB} \; \left(=\frac 12 \int_{\partial C}\sin\theta
\kolo\chi_{AB,u} \kolo\chi^{AB}_{,u} \right)
\ee 
 where the last form in the brackets becomes clear when we learn about
 asymptotics presented in the next section.\\ Similarily for angular
 momentum we get the answer from the superpotential proposed by Komar
\cite{Komar}:
\[ 16\pi \int_V {\pi}^{\mu\nu} {\cal L}_X {A}^{0}_{\mu\nu} =
\int_{\partial V} \sqrt{-g}\left(\nabla^3 X^0 -\nabla^0 X^3 \right) 
\]
 where now $X=\partial/\partial\phi$.

The right-hand side can be expressed in terms of the Bondi-Sachs type
 metric:
\[ \int_{\partial V} \sqrt{-g}\left(\nabla^3 X^0 -\nabla^0 X^3 \right) =
\int_{\partial V} r^4\sin\theta \E{-2\beta}\gamma_{\phi A}
 U^A_{,3} \longrightarrow 16\pi J_z \] the limit is taken on $\scri^+$
 and according to the asymptotics presented in the next section we
 obtain
\[ 16\pi J_z=- \int_{\partial C}(6N_\phi +\frac 12
\kolo{\chi}{_{\phi B}}{\kolo{\chi}^{BC}}_{||C}) 
\sin\theta\rd \theta\rd \phi \] But on the other hand
\[ 16\pi \int_V {\pi}^{\mu\nu} {\cal L}_X {A}^{0}_{\mu\nu}=
\int_V {\pi}^{\mu\nu} {A}^{0}_{\mu\nu,\phi}= \int_V
\Pi_{AB}\psi^{AB}{_{,\phi}} \]
 and
\[16\pi \partial_0 \int_C {\pi}^{\mu\nu} {A}^{0}_{\mu\nu,\phi}=   
\int_C \dot
\Pi_{AB}\psi^{AB}{_{,\phi}}-\Pi_{AB}{_{,\phi}}\dot\psi^{AB}=
\frac 12 \int_{\partial C}\sin\theta r^2\dot{\gamma}_{AB}
\gamma^{AB}{_{,\phi}} \] 
 We will show in the next section that non-conservation law for
 angular momentum agrees in terms of the asymptotics:
\be\label{dotJonC} 16\pi \dot J_z=- \int_{\partial C}
\frac 12 \kolo{\chi}{_{AB,u}}{\kolo{\chi}^{AB}}_{,\phi}
\sin\theta\rd \theta\rd \phi \ee

\subsection{Symplectic structure on scri}
 Let us observe that we can use previous results (\ref{piA0}) and
 (\ref{piA3}) to reduce the form
\[ \pi^{\mu\nu}\delta{A}^{v}_{\mu\nu} =\pi^{\mu\nu}\delta({A}^{0}_{\mu\nu}
 +2{A}^{3}_{\mu\nu}) \] Let us also remind coordinate system which can
 be used to describe the situation in a similar way as in section 2.3
 for scalar field and 3.1 for electrodynamics.  $(u,r) \rightarrow
 (v,\overline u)$, $\overline u =-2r$ $v=u+2r$,
 $\partial_u=\partial_v$, $\partial_r=-2\partial_{\overline u}
 +\partial_v$, $\rd u\wedge\rd r=\frac 12 \rd \overline u \wedge \rd
 v$ and finally $\pi^{\mu\nu}\delta{A}^{v}_{\mu\nu}\rd
 r\rd\theta\rd\phi=\frac 12
\pi^{\mu\nu}\delta{A}^{v}_{\mu\nu} \rd \overline u\rd\theta\rd\phi$.

If we put \be 16\pi{\pi}^{\mu\nu} \delta A^{0}_{\mu\nu} = -\frac12
\sin\theta \left( r \gamma_{AB}\right)_{\! ,3} \delta
\left( r
\gamma^{AB}\right)   + \delta\left[ 2r^4\sin\theta
\left(\frac{\beta}{r^2}\right)_{,3}\right]  \label{piA0bis}
\ee
 and
\[ 16\pi \pi^{\mu\nu} \delta  A^3{_{\mu\nu}} 
 = 2\sin\theta\left(2V- r^2 U^B{_{||B}}\right)\delta\beta
 +\sin\theta\gamma^{AB}\delta U_{A||B} -\frac 12 \sin\theta
\left(
\dot{g}_{AB} -\frac Vr {g_{AB}}{_{,3}}  \right)\delta \gamma^{AB} + \] \be
 +r^2\sin\theta \E{-2\beta}U^B{_{,3}}g_{BA}\delta U^A -\delta
\left[ 2
 r^2\sin\theta \left(\frac V{r^2}+ \left(\frac{V}{2r}\right)_{,3}
 +\frac Vr \beta _{,3} -\dot{\beta} -U^B\beta_{,B} \right)
\right] \label{piA3bis} \ee
 assuming asymptotic behaviour on $\scri^+$ we obtain the following
 formula on the future null infinity:
\be\label{ns} 
\left. 16\pi \pi^{\mu\nu} \delta  A^v{_{\mu\nu}} \right|_{\scri^+}=
 -\sin\theta r\dot\gamma_{AB}\delta r\gamma^{AB}+4\delta(\sin\theta M)
\ee Let $N=[u_i,u_f]\times
 S^2 \subset \scri^+$ is a ``finite piece'' of $\scri^+$. The relation
 with the TB mass is based on the following observations. First of all
 from (\ref{ns}) we obtain
\[ 
 16\pi\int_N \frac 12 \pi^{\mu\nu}A^v{_{\mu\nu,0}} \rd \overline u
\rd\theta\rd\phi= -\frac 12 \int_N
 r^2\sin\theta\dot\gamma_{AB}\dot\gamma^{AB} \rd \overline u
\rd\theta\rd\phi  +2\int_{\partial N} M \sin\theta\rd\theta\rd\phi \]
 and secondly
\[ 16\pi\int_N \pi^{\mu\nu}{\cal L}_X A^v{_{\mu\nu}}=\int_{\partial N}
\sqrt{-g} (\nabla^{\overline u}X^v-\nabla^v X^{\overline u}) =
 -\int_{\partial N} 2M\sin\theta\rd\theta\rd\phi \] where
 $X=\partial_0$, so finally

\[ -4\int_{\partial N}M\sin\theta\rd\theta\rd\phi = -\frac 12 \int_N
 r^2\sin\theta\dot\gamma_{AB}\dot\gamma^{AB} \rd \overline u
\rd\theta\rd\phi \]
 The left-hand side of the above formula represents the change of
 Bondi mass from initial state $u_i$ to final state $u_f$ ($\partial
 N=
\{ u_f\}\times S^2 \cup \{ u_i\}\times S^2 $) but the right-hand side is a
 flux of the energy through $N$ which is a piece of $\scri^+$ between
 initial and final state.

Similarily for angular momentum we have
\[  16\pi\int_N \pi^{\mu\nu}{\cal L}_X A^v{_{\mu\nu}}
 = 16\pi\int_N \frac 12 \pi^{\mu\nu} A^v{_{\mu\nu,\phi}} \rd
\overline u \rd\theta\rd\phi= -\frac 12 \int_N
 r^2\sin\theta\dot\gamma_{AB}\gamma^{AB}_{,\phi} \rd \overline u
\rd\theta\rd\phi \]
 where now $X=\partial_\phi$.

\section{Multipole structure of Bondi--van der Burg--Metzner--Sachs equations}
 Let $v=u+2r$ than the metric (\ref{gB}) takes the following form:
\begin{eqnarray} 
 g_{\mu\nu}\rd x^\mu\rd x^\nu =& & \left( -\frac Vr +
 r^2\gamma_{AB}U^AU^B + {\rm e}^{2\beta} \right) \rd u^2 - {\rm
 e}^{2\beta} \rd u \rd v -2r^2\gamma_{AB}U^B\rd u \rd x^A+
\nonumber \\ 
 & & +r^2\gamma_{AB}\rd x^A \rd x^B \end{eqnarray} We shall rewrite
 formulae from van der Burg paper \cite{Burg} in a ``spherically
 covariant'' way. More precisely we denote:\\ 0.  $M$ i $V$ are
 scalars\\ 1.pairs of functions $U,W$ and $N,P$ can be combined in two
 vectors $U^A$ and $N^A$ respectively: \[ U_\theta=U^\theta=U \] \[
 U_\phi =\sin^2\theta U^\phi=W\sin\theta \] \[ N_\theta=N^\theta=N \]
\[ N_\phi
 =\sin^2\theta N^\phi=P\sin\theta \] 2.pairs of functions $c,d$, $C,H$
 and $D,K$ correspond to the symmetric traceless tensors
 $\kolo\chi_{AB}$, $C_{AB}$ i $D_{AB}$: \[
\kolo\chi^{\theta}{_\theta}=-\kolo\chi^{\phi}{_\phi}=2c \]
\[
\kolo\chi^{\theta}{_{\phi}}=\sin^2\theta\kolo\chi_{\theta}{^{\phi}}
 =2d\sin\theta \] Similarily $C^{\theta}{_\theta}=C$,
 $D^{\theta}{_\theta}=D$ etc.  The reason of this notation arises in a
 natural way if we change the parameterization of the 2-dimensional
 metric $\gamma_{AB}$.  Let us remind that van der Burg in \cite{Burg}
 (p. 112) proposed the following parameterization:
\be
\gamma_{AB}{\rd x}^A{\rd x}^B=
 e^{2\gamma} \cosh (2\delta){\rd \theta}^2 + 2 \sinh (2\delta)
\sin \theta
\rd\theta{\rd \phi} 
 + e^{-2\gamma} \cosh (2\delta) \sin^2 \theta {\rd \phi}^2
\end{equation}
\noindent which differs from original Sachs formulation by linear
 transformation of functions $\gamma$ and $\delta$ (see
\cite{Sachs} p. 107).
 Next the used functions $\gamma$ and $\delta$ are expanded as follows
\[ \gamma = c/r +(C-\frac 16c^3-\frac 32 cd^2)r^{-3} +D
 r^{-4}+ O(r^{-5})\] \[ \delta=d/r +(H-\frac 16d^3+\frac 12
 c^2d)r^{-3} +K r^{-4}+ O(r^{-5}) \] Let us notice that there is no
 $r^{-2}$ term which was analyzed in \cite{XIV} and vanishing of this
 term is called ``outgoing radiation condition''.\\ We propose to
 change this parameterization in such a way that for original Bondi
 axially symmetric metric both formulations are the same and the main
 advantage of our change is that the expansion terms take a nice
 geometric form (mainly the term of order $r^{-3}$ takes a nice form).
 
Let us fix the frame $\rd\theta$, $\sin\theta\rd\phi$ which is
 orthonormal with respect to the background metric $\kolo\gamma_{AB}$.
 The symmetric matrix (close to unity)
\begin{equation}
\label{gamma}
\left( \begin{array}{cc}
 e^{2\gamma} \cosh (2\delta) & \sinh (2\delta) \\
\sinh (2\delta)  & e^{-2\gamma} \cosh (2\delta) 
\end{array} \right)
\end{equation}
 with determinant equal 1 can be also parameterized in a natural way
 by exponential mapping \[ \exp(a\sigma_x+b\sigma_z) \] where
 $\sigma_x$ and $\sigma_z$ are Pauli matrices:
\[ \sigma_x=\left( \begin{array}{cc}
 1 & 0 \\ 0 & -1 \end{array} \right) \quad \sigma_z=\left(
\begin{array}{cc}
 0 & 1 \\ 1 & 0 \end{array} \right) \] The solution of the matrix
 equation \begin{equation}
\label{matrix}
\left( \begin{array}{cc}
 e^{2\gamma} \cosh (2\delta) & \sinh (2\delta) \\
\sinh (2\delta)  & e^{-2\gamma} \cosh (2\delta) 
\end{array} \right) = \exp(a\sigma_x+b\sigma_z) 
\end{equation}
 leads to the relation between $a,b$ and $\gamma,\delta$ in the
 following form: \[ a=\sinh(2\gamma)\cosh(2\delta){{\rm arccosh}
 (\cosh(2\delta)\cosh(2\gamma)) \over
\sqrt{ \sinh^2(2\delta)+\cosh^2(2\delta)\sinh^2(2\gamma)}}
\]
\[ b=\sinh(2\delta){{\rm arccosh}(\cosh(2\delta)\cosh(2\gamma))
\over
\sqrt{ \sinh^2(2\delta)+\cosh^2(2\delta)\sinh^2(2\gamma)}}
\]
 but the asymptotic relation for small $\gamma,\delta$ is simpler: \[
 a=2\gamma +\frac83\gamma\delta^2 + O(\gamma,\delta)^5 \] \[ b=2\delta
 +\frac43\gamma^2\delta + O(\gamma,\delta)^5 \] and it gives only
 correction in $r^{-3}$ in our expansion.  More precisely
\[ \frac12 a = c/r +(C-\frac 16(c^2+d^2)c)r^{-3} +D r^{-4}+
 O(r^{-5})\] \[ \frac12 b =d/r +(H-\frac 16(d^2+ c^2)d)r^{-3} +K
 r^{-4}+O(r^{-5}) \] but now we can write the expansion in a matrix
 form:
\be\label{lg}
\log\gamma_{AB}= \kolo\chi_{AB}/r +
\left( 2C_{AB}-\frac1{48}\kolo\chi_{CD}\kolo\chi^{CD}
\kolo\chi_{AB}\right)r^{-3}+2D_{AB}r^{-4}+O(r^{-5})
\ee
 and each term of the expansion is a traceless symmetric tensor on a
 sphere.  The indices are raised with respect to the inverse
 $\kolo\gamma{^{AB}}$ of the background metric (which is a standard
 unit sphere). It is diagonal in our coordinates
 $\kolo\gamma_{\theta\theta} =1$ and $\kolo\gamma_{\phi\phi}
 =\sin^2\theta$.  Metric connection of the $\kolo\gamma_{AB}$ has the
 following non-vanishing components \[
\Gamma^\theta{_{\phi\phi}}=-\sin\theta\cos\theta \; , \quad
\Gamma^\phi{_{\phi\theta}}= \cot\theta \] We are ready to show
 the asymptotic expansions for the rest of the quantities which appear
 in Bondi--Sachs type metric (\ref{gB}).  They were introduced in
\cite{Burg} (p.114) but now we can rewrite them in
 a covariant way on $S^2$:
\[ U^A=-{1\over 2r^2}{\kolo{\chi}^{AB}}_{||B}+{2N^A\over r^3}+{1\over
 r^3}\left[
\frac12{\kolo{\chi}^A}_B{\kolo{\chi}^{BC}}_{||C}+{1\over
 16}\left(\kolo{\chi}_{CD} \kolo{\chi}^{CD}\right)^{||A}\right]
\]

\[ U_A:=r^2\gamma_{AB}U^B=-{1\over 2}{\kolo{\chi}^{B}}_{A||B}+{2N_A\over r}
 +{1\over 16r}\left(\kolo{\chi}_{CD}
\kolo{\chi}^{CD}\right)_{||A} \]

\[ 1-{V\over r}={2M\over r}+{{N^A}_{||A}\over r^2}-{1\over
 r^2}\left[\frac14 {\kolo{\chi}^{AB}}_{||B} {{\kolo{\chi}_A}^C}_{||C}+
 {1\over 16}\kolo{\chi}^{CD}\kolo{\chi}_{CD}\right] \]

\[ \beta =-{1\over 32}\cdot{1\over r^2}\kolo{\chi}_{AB}\kolo{\chi}^{AB} \]

Basic equations, (eq. 13--15 in \cite{Burg}):
\be\label{Mdot}
 M_{,u}=-\frac18 \kolo{\chi}_{AB,u}{\kolo{\chi}^{AB}}_{,u} +\frac14
 {\kolo{\chi}^{AB}}_{||AB,u} \ee

\be\label{Ndot}
 3{N^A}_{,u}=-M^{||A}-\frac14 \kolo{\ve}^{AB}
\left({{\kolo{\chi}_C}^D}_{||DE} \kolo{\ve}^{EC}\right)_{||B}
 -\frac34 \kolo{\chi}^A{_B} \kolo{\chi}^{BC}{_{||C,u}} - \frac14
\kolo{\chi}^{CD}{_{,u}} \kolo{\chi}^A{_{C||D}}  \ee
 (eq. 8--9 i 11--12 in \cite{Burg}): \[ -4 {\dot C}_{AB} -\frac
 18\kolo\chi^{CD}\kolo\chi_{CD}\kolo\chi_{AB,u} + \frac 14
\kolo\chi_{AB}\kolo\chi^{CD}\kolo\chi_{CD,u}
 = N_{A||B}+N_{B||A}-\kolo\gamma_{AB}N^C{_{||C}} + \]
\be\label{Cdot}  
 - M \kolo\chi_{AB}-\frac14
\kolo\varepsilon_{AC}\kolo\chi^{C}{_B}
\kolo\chi_E{^F}{_{||FG}}\kolo\varepsilon^{EG} \ee
\be\label{Ddot}
 -4 {\dot D}_{AB}= ({\bf a} +4)C_{AB} -(\kolo\chi_A{^C}N_B)_{||C}
 -(\kolo\chi_B{^C}N_A)_{||C}
 +\kolo\gamma_{AB}(\kolo\chi^{CD}N_D)_{||C} \ee Let us observe that
 mono-dipole but also {\em quadrupole} part of the right-hand side of
 the equation (\ref{Ddot}) vanishes. More precisely, from the relation
\[ C^{AB||C}{_{CAB}} = C^{AB}{_{ABC}}{^C} + 2 C^{AB}{_{||AB}} \]
 we obtain that
\[  \left[ ({\bf a} +4)C^{AB} \right]_{||AB} = ({\bf a} +6)\left(
 C^{AB}{_{||AB}} \right) \] and similarily for $\hat
 C_{AB}:=\varepsilon_{AD} C^D{_B}$.  Let us rewrite equation
 (\ref{Ddot}) in the following way:
\be\label{DdotH}
 -4 {\dot D}_{AB}= ({\bf a} +4)C_{AB} - S_A{^C}{_{B||C}} \ee where
\[ 
 S_{ABC}:= \kolo\chi_{AC}N_B +\kolo\chi_{BC}N_A -\kolo\gamma_{AB}
\kolo\chi_{CD}N^D
\]
 It is easy to check that $S_{ABC}$ is traceless symmetric tensor (in
 each pair of indices) and the same holds for $\hat
 S_{ABC}:=\varepsilon_{AD} S^D{_{BC}}$. One can prove that
 $S^{ABC}{_{||CAB}}$ and $\hat S^{ABC}{_{||CAB}}$ are orthogonal to
 the first three eigenvalue spherical harmonics (with $l=0,1,2$). This
 way we get 10-dimensional space of quadrupole Newman--Penrose charges
 in $D_{AB}$ which are conserved (\cite{NP}, \cite{Burg}).  More
 precisely, quadrupole (and also mono-dipole) part of $\partial_u
 D^{AB}{_{||AB}}$ and $\partial_u {\hat D}^{AB}{_{||AB}}$ have to
 vanish. However for the polyhomogeneous asymptotics it may be not
 conserved (see
\cite{XIV}).

The linearized metric of Bondi-Sachs type
\[ H\cong \frac12\cdot{1\over r^2}\cdot \kolo{\chi}_{AB}\kolo{\chi}^{AB}
\cong 0 \]

\[ h_{uu}\cong 1-{V\over r}+2\beta +r^2\kolo{\gamma}_{AB}U^AU^B 
\cong \frac{2M}r +\frac{N^A{_{||A}}}{r^2} \]

\[ h_{uv}\cong -\beta \cong 0 \]

\[  h_{uA}\cong -r^2\kolo{\gamma}_{AB}U^B-r\kolo{\chi}_{AB}U^B 
\cong -\frac12 \kolo{\chi}^B{_{A||B}} +\frac{N_A}{r} \]

\[ \chi_{AB}\cong r\kolo{\chi}_{AB},
\quad h_{AB}\cong r\kolo{\chi}_{AB}+\frac12
 r^2\kolo{\gamma}_{AB} H \cong r\kolo{\chi}_{AB} \]

and linearized asymptotics of invariants
\[ {\bf x}\cong {4M\over r}+{6{N^A}_{||A}\over r^2} \quad
\left[ +{1\over 2r}
\left(\kolo{\chi}_{AB}\kolo{\chi}^{AB}\right)_{,u}  
 +{3\over 8r^2}\kolo{\chi}_{AB}\kolo{\chi}^{AB} -{1\over
 8r^2}a\left(\kolo{\chi}_{CD}\kolo{\chi}^{CD}\right)\right] \]

\[ {\bf y}\cong -{1\over r}{{\kolo{\chi}_A}^C}_{||CB} \kolo{\ve}^{AB} 
 + {6\over r^2}N_{A||B} \kolo{\ve}^{AB} \]

\[ \Psi_x=r{\bf x} \cong 4M \, , \quad 
\dot\Psi_x=r{\bf x}_{,u} \cong {\kolo{\chi}^{AB}}_{||AB,u} \]

\[ \Psi_y= r{\bf y} \cong -{{\kolo{\chi}_A}^C}_{||CB}\kolo{\ve}^{AB}
\, , \quad \dot\Psi_y=r{\bf y}_{,u}\cong -
\left({{\kolo{\chi}_A}^C}_{||CB}
\kolo{\ve}^{AB}\right)_{,u}  \]
 give an indication how to relate linearized theory with van der Burg
 asymptotics. This observation will be used in the sequel.

\subsection{Supertranslations}
 Let us consider $\scri^+$ as a cartesian product $S^2\times R^1$ or
 rather trivial affine bundle over $S^2$ with typical fiber $R^1$ then
 the supertranslation corresponds to the null section of this affine
 bundle. On the other hand the boost transformation leads to the
 nontrivial scaling factor in a fiber and a conformal transformation
 on a base manifold $S^2$ (see
\cite{Sachs} p. 111).

Prolongation of the supertranslation from scri ``to the center'' in
 Bondi coordinates (metric (\ref{gB})) leads to the following
 asymptotic relations (see also \cite{Sachs} p. 119): \[ {\overline
 x}^A=x^A +\frac 1r \alpha^{||A} -\frac 1{2r^2} \left(
 {\kolo\chi}^{AB}\alpha_{||B}
 -2\alpha^{||AB}\alpha_{||B}+\Gamma^A{_{BC}}
\alpha^{||B}\alpha^{||C}\right) +\ldots \]
\[ {\overline u}= u-\alpha -\frac 1{2r}\alpha^{||A}\alpha_{||A}
 + \frac 1{4r^2} \left[ {\kolo\chi}^{AB}\alpha_{||A}\alpha_{||B} -
\alpha^{||A}\left(\alpha_{||B}
\alpha^{||B}\right)_{||A}\right]+\ldots
\] \[ {\overline r}=r-\frac 12 {\bf a}\alpha + \frac 1{2r}
\left[ {\kolo\chi}^{AB}_{||B}\alpha_{||A}+\frac12
 {\kolo\chi}^{AB}\alpha_{||AB} +\frac12
 {\kolo\chi}^{AB}{_{,u}}\alpha_{||A}\alpha_{||B} -\frac 12
\alpha^{||AB} \alpha_{||AB}
 - \alpha_{||A} \alpha^{||A}+ \right. \] \[ \left.  +\frac 14 ({\bf
 a}\alpha)^2 - \left( {\bf a}\alpha\right)^{||A}\alpha_{||A}
\right]+\ldots \] Now we can
 check the transformation law for $\kolo\chi$ and $M$: \[
\overline M = M + \frac 12 {\kolo\chi}^{AB}_{||B,u}\alpha_{||A}+\frac 14
\kolo\chi^{AB}_{,u}\alpha_{||AB} + \frac14{\kolo\chi}^{AB}_{,uu}
\alpha_{||A} \alpha_{||B}\]

\[ \overline{\kolo\chi}_{AB}={\kolo\chi}_{AB} - 2\alpha_{||AB}+
 {\kolo\gamma}_{AB}{\bf a}\alpha \] \[
\overline{\partial_A}={\partial_A}+\alpha_{,A}{\partial_0} \]
 and finally we obtain that certain combination \[ \overline{4M -
 {\kolo\chi}^{AB}{_{||AB}}}= 4M - {\kolo\chi}^{AB}{_{||AB}}+ {\bf
 a}({\bf a}+2)\alpha \] has a simple transformation law with respect
 to the supertranslations.  Moreover, mono-dipole part of $M$ is
 invariant with respect to the supertranslations. It corresponds to
 the mass and linear momentum at null infinity.  This way we have
 proved the following:\\[1ex]
\underline{Theorem 1.} The energy-momentum 4-vector at null infinity is
 invariant with respect to the supertranslations.\\[1ex] On the other
 hand angular momentum is not invariant with respect to the
 supertranslations but it transforms probably as follows:
\[ 16\pi \overline J_z =16\pi J_z + \int_{S^2} 4M\alpha_{,\phi} \]

\underline{Remark} The supertranslation gauge freedom also
 exists in the linearized theory. The linearized part of the
 supertranslation corresponds to the gauge condition which preserves
 five components of the linearized metric: $h_{ur}$, $H$, $h_{rr}$,
 $h_{rA}$. More precisely, it is a solution of the gauge conditions:
\[ \xi^u{_{,r}}=0 \, , \quad \xi^{u||A}{_A}=(\xi^A{_{||A}})_{,r} \, ,
\quad (\xi_{A||B}\varepsilon^{AB})_{,r}=0 \]
\[ \xi^A{_{||A}}+\frac 2r \xi^r=0 \, , \quad \xi^r{_{,r}}+\xi^u{_{,u}}=0 \]
 We use here Minkowski background metric in the form:
\[ \eta_{\mu\nu}\rd x^\mu \rd x^\nu = -\rd u^2 - 2 \rd u \, \rd r
 +r^2 \kolo\gamma_{AB} \rd x^A \rd x^B \] and the solution of the
 gauge equations is the following:
\[ \xi^A=\frac 1r \kolo\gamma^{AB}\alpha_{||B} \, , \quad \xi^u=-\alpha \,
 , \quad \xi^r=-\frac 12 {\bf a}\alpha \] where $\alpha$ is any real
 mapping $\alpha : S^2 \mapsto R$ and mono-dipole part of $\alpha$
 corresponds to the usual translations in Minkowski space. The gauge
 transformation for traceless symmmetric tensor $\chi_{AB}$
\[ {\chi}_{AB} \longrightarrow {\chi}_{AB} - 2r\alpha_{||AB}+
 r{\kolo\gamma}_{AB}{\bf a}\alpha \] is similar to the nonlinear case.

How to remove supertranslation gauge freedom?\\ Assume at time $u_0$
 that $4\underline M-\kolo\chi^{AB}{_{||AB}}=0$ then stationary
 solution becomes simple stationary solution. \\ This procedure allows
 to treat $\kolo\chi^{AB}$ as invariant asymptotic degrees of
 freedom.\\[1ex]
\underline{Remark} The Kerr-Newman metric in Bondi-Sachs coordinates can
 be asymptotically represented in such a way that $\underline M= 0
 =\kolo\chi_{AB}$.

\subsection{Hierarchy of asymptotic solution on scri for scalar wave
 equation} Let us rewrite wave equation in null coordinates $(u,v)$
\be\label{wave} \rho^{-1}(\rho^{-1}\varphi)^{,a}{_a}
 +{\bf a}\varphi=0 \ee and suppose we are looking for a solution of
 the wave equation (\ref{wave}) as a series
\be\label{szereg} \varphi = \varphi_1 \rho + \varphi_2 \rho^2 + \varphi_3
\rho^3 +\ldots \ee 
 where each $\varphi_n$ is a function on scri, $\partial_v
\varphi_n =0$.

If we put the series (\ref{szereg}) into the wave equation
 (\ref{wave}) we obtain the following recursion:
\be\label{fir}
\partial_u \varphi_{n+1}=-\frac1{2n}[{\bf a}+(n-1)n]\varphi_n
\ee
 Compare with equations 2, 3, 4 in \cite{BVM}. \\[1ex]
\underline{Remark}
 The kernel of the operator $[{\bf a} + l(l+1)]$ corresponds to the
 $l$-th spherical harmonics. The right-hand side of (\ref{fir})
 vanishes on the $n-1$ spherical harmonics subspace.  This means that
 the corresponding multipole in $\varphi_{n+1}$ does not depend on
 $u$. In particular for $n=3$ we have quadrupole charge in the fourth
 order. The nonlinear counterpart of this object is called {\em
 Newman-Penrose charge}. We discuss some features related with NP
 charges in section 8.5.
 
\subsection{Linear theory, asymptotic hierarchy, ``charges''}
 Let us first check that linearized theory can be obtained if we
 reject nonlinear terms in asymptotic hierarchy
 (\ref{Ndot}--\ref{Ddot}):
\[ 4\dot M=\kolo\chi^{AB}{_{||AB,u}} \]
\[ 3 \dot N^A{_{||A}}=-{\bf a}M \quad 3\dot N_{A||B}{\kolo\ve}^{AB}=-\frac
 14 {\bf a} \hat\chi^{AB}{_{||AB}} \]
\[ -4 \dot C^{AB}{_{||AB}}=({\bf a}+2)N^A{_{||A}} \quad
 -4 \hat C^{AB}{_{||AB,u}}=({\bf a}+2) N_{A||B}{\kolo\ve}^{AB} \]
\[ -4 \dot D^{AB}{_{||AB}}=({\bf a}+6)C^{AB}{_{||AB}} \quad
 -4 \hat D^{AB}{_{||AB,u}}=({\bf a}+6)\hat C^{AB}{_{||AB}} \]
 
\[ {\bf x}=4M\rho+6N^A{_{||A}}\rho^2 + 6C^{AB}{_{||AB}} \rho^3 +
 4D^{AB}{_{||AB}} \rho^4 + O(\rho^5) \] \[ {\bf
 y}=\hat\chi^{AB}{_{||AB}} \rho +6N_{A||B}{\kolo\ve}^{AB} \rho^2 +
 6\hat C^{AB}{_{||AB}} \rho^3 + 4\hat D^{AB}{_{||AB}} \rho^4 +
 O(\rho^5) \] Full agreement with (\ref{fir}) up to the 4-th order.
\[ M=m+3p+\underline M \, , \quad 4\underline{\dot M}
 =\kolo\chi^{AB}{_{||AB,u}} \, , \quad \dot m=\dot p =0 \]
\[ N^A=-p^{||A}u-k_0^{||A} -{\kolo\ve}^{AB}s{_{||B}}+\underline N^A \]

Let us notice that mono-dipole part of invariants:
\[ {\bf x}=4m\rho+12j^{l0}x_l\cdot\rho^3 +12p^lx_l\cdot \rho^3\cdot
\left(u+{2\over \rho}\right) \]

\[  {\bf y}=12s^lx_l\cdot\rho^3 \]
 or
\[ {\bf x}=4(m+3p)\cdot \rho+12(k_{0}+p\cdot u)\cdot\rho^2
\quad {\bf y}=12s\cdot\rho^2 \] is the same as (\ref{cx}) and
 (\ref{cy}).

\subsubsection{nonradiating solutions}
 Suppose $\int_{S^2} \dot M=0$ then from basic equation (\ref{Mdot})
 we know
\[ 0= \int_{S^2} \dot M=-\frac18 \int_{S^2}
\kolo{\chi}_{AB,u}{\kolo{\chi}^{AB}}_{,u}  \] 
 and we get $\kolo{\chi}_{AB,u}=0$ and finally also $\dot M=0$.
 Moreover equation (\ref{Ndot}) gives the following relation: \[ 3
\dot N_{A||B}\kolo\varepsilon^{AB}=\frac 14
\hat\chi^{CD}{_{||CD}} \]
 so the dipole part $\mbox{dip} (\dot N_{A||B}\kolo\varepsilon^{AB})$
 vanishes and this means that the angular momentum is conserved. This
 way we have proved the theorem formulated at the end of subsection
 2.2, namely:\\[1ex]
\underline{Theorem 2.} If the TB mass is conserved than angular momentum is
 conserved too.\\[1ex] The general solution of this type (namely $\dot
 M=0=\kolo\chi_{AB,u}$) will be called {\em nonradiating solution} and
 it has the following form:
\[ M=m+3p+\underline M \]
\[ N^A=-p^{||A}u-k_0^{||A}-\kolo\varepsilon^{AB}s_{||B}+ \tilde
 N^A +\frac u3 \left( \frac 14\kolo\varepsilon^{AB}
\hat\chi^{CD}{_{||CDB}} -\underline M^{||A} \right)
\] \[4\dot C_{AB}= -N_{A||B} - N_{B||A} +\kolo\gamma_{AB}
 N^C{_{||C}} +(m+3p+\underline M)\kolo\chi_{AB} +\frac14
\hat\chi_{AB}\hat\chi^{CD}{_{||CD}} 
\] \[ 4 {\dot D}_{AB}= -({\bf a} +4)C_{AB} + S_A{^C}{_{B||C}} \]
\[
 S_{ABC}= \kolo\chi_{AC}N_B +\kolo\chi_{BC}N_A -\kolo\gamma_{AB}
\kolo\chi_{CD}N^D \]

\subsection{How to relate linearized theory with van der Burg equations,
 stationary solutions} The ``first order'' asymptotics of the
 Bondi-Sachs type metric on $\scri^+$ is described by three functions
 $M,\kolo\chi_{AB}$. We shall try now to relate these data with
 boundary value on $\scri^+$ of our invariants $\Psi_\Upsilon$ in the
 linearized theory in such a way that non-conservation laws for the
 mass and angular momentum are similar in both cases. Suppose we know
 $M,\kolo\chi_{AB}$ at the moment $u=u_0$ and $\kolo\chi_{AB,u}$ on
 $\scri^+$ (we need only in the neighbourhood of $u_0$). We propose to
 perform supertranslation which is related with data at $u_0$ in such
 a way that
\be\label{sg} (4\underline M
 -\kolo\chi^{AB}_{||AB})(u_0)=0 \ee Let us call the condition
 (\ref{sg}) supertranslation gauge at the moment $u_0$.  This way we
 have removed quasi-locally supertranslation ambiguity at the moment
 $u_0$. We stress that the relation $4\underline
 M-\kolo\chi^{AB}_{||AB}=0$ holds {\em only} at $u_0$ because
\underline{QF}$(\kolo\chi_{AB,u}):= 
 4\underline M_{,u}-\kolo\chi^{AB}_{||AB,u}$ is not vanishing in
 general however for the nonradiating solutions it may be fullfilled
 globally. This is the main difference between linearized theory where
 the condition $4\underline M-\kolo\chi^{AB}_{||AB}=0$ can be
 fullfilled globally and nonlinear data where we can only demand this
 condition to be fullfilled at one moment $u_0$. Nevertheless these
 procedure allows us to relate nonlinear data at $u_0$ with linearized
 theory namely
\[ 4\underline M=\kolo\chi^{AB}_{||AB} \rightarrow \Psi_x \, ,
\quad \kolo\chi^{AB}_{||AB,u} \rightarrow \dot\Psi_x \]
\[ \hat\chi^{AB}_{||AB} \rightarrow \Psi_y \, ,
\quad \hat\chi^{AB}_{||AB,u} \rightarrow \dot\Psi_y \]
 Now it is easy to verify analogy.  The calculations from the previous
 sections devoted to the linearized gravity should convince the reader
 that in linear theory we can believe in the following equations:

\be\label{lmdot} -16\pi \partial_0 m_{\mbox{\tiny TB}} =
\int_{S(s,0)} \sin\theta \rd\theta\rd\phi
 [ \dot\Psi_x {\bf a}^{-1}({\bf a}+2)^{-1} \dot\Psi_x +
\dot\Psi_y {\bf a}^{-1}({\bf a}+2)^{-1} \dot\Psi_y] 
\ee

\be\label{lJdot}
 -16\pi\partial_0 J_z =\int_{S(s,0)} \sin\theta \rd\theta\rd\phi
\left[  \dot\Psi_x {\bf a}^{-1}({\bf a}+2)^{-1} {\Psi_x}{_{,\phi}} +
\dot\Psi_y {\bf a}^{-1}({\bf a}+2)^{-1} {\Psi_y}{_{,\phi}}
\right]
\ee
 On the other hand the TB energy can be defined in terms of the
 asymptotics on $\scri^+$
\[ 16\pi m_{\mbox{\tiny TB}}= \int_{S^2} 4 M\sin\theta\rd\theta\rd\phi \]
 From (\ref{Mdot}) we get non-conservation law for the TB mass:
\[ -16\pi \partial_0 m_{\mbox{\tiny TB}} =
 -\int_{S^2} 4\dot M\sin\theta \rd\theta\rd\phi=\frac12
\int_{S^2} \kolo\chi_{AB,u}\kolo\chi^{AB}{_{,u}}\sin\theta \rd\theta\rd\phi
 = \] \be\label{nlm0} = \int_{S^2} \sin\theta \rd\theta\rd\phi
\left[
 (\kolo\chi^{AB}_{||AB})_{,u} {\bf a}^{-1}({\bf a}+2)^{-1}
 (\kolo\chi^{AB}_{||AB})_{,u} + (\hat\chi^{AB}_{||AB})_{,u} {\bf
 a}^{-1}({\bf a}+2)^{-1} (\hat\chi^{AB}_{||AB})_{,u} \right] \ee
 Similarity between (\ref{lmdot}) and (\ref{nlm0}) is obvious provided
 $\kolo\chi^{AB}_{||AB,u} \rightarrow \dot\Psi_x$,
 $\hat\chi^{AB}_{||AB,u} \rightarrow \dot\Psi_y$.

Similarily angular momentum (around $z$-axis) can be defined as
 follows
\[ 8\pi J_z= 3\int_{S^2} {\breve N}_{A||B}\kolo{\ve}^{AB}\cos\theta
\sin\theta\rd\theta \rd\phi \]
 where
\[ {\breve N}_A:= N_A +\frac1{12} \kolo{\chi}_{AB}{\kolo\chi}^{BC}{_{||C}}
\]

We have promised at the end of the previous section to show the
 relation (\ref{dotJonC}) for angular momentum.  The following
 sequence of equalities holds:
\[ 16\pi\dot J_z= -\int_{S^2} 6\dot N_{\phi} +\frac 12 \partial_u
 (\kolo\chi_{\phi B}\kolo\chi^{BC}{_{||C}})=\int_{S^2}
\kolo\chi_{\phi B}\kolo\chi^{BC}{_{||C,u}} +\frac 12
\kolo\chi^{AB}{_{,u}}\kolo\chi_{\phi A||B}-\frac 12\kolo\chi_{\phi
 B,u}\kolo\chi^{BC}{_{||C}} =\] \[ = -\frac 12 \int_{S^2}
\kolo\chi^{AB}{_{,u}}\kolo\chi_{AB,\phi} \]
 where in the middle we have used equation (\ref{Ndot}). The last
 equality (in the above sequence) is a nontrivial identity and can be
 generally denoted:
\[
\int_{S^2} X^A\chi_{AB}\dot\chi^{BC}{_{||C}} +\frac 12
\dot\chi^{BC}\chi_{AB||C}X^A -\frac 12 X^A \dot\chi_{AB}\chi^{BC}{_{||C}}=
\]
\be\label{idX}
 =-\frac 12\int_{S^2} \dot\chi^{AB} (
 X^C\chi_{AB||C}+X^C{_{||B}}\chi_{CA} +X^C{_{||A}}\chi_{CB} )
\ee
 where now $\chi_{AB}$ and $\dot\chi_{AB}$ are any symmetric traceless
 tensors on a unit sphere, $X^A\partial_A:=\partial_{\phi}$ and
\[ \chi_{AB,\phi}= X^C\chi_{AB||C}+X^C{_{||B}}\chi_{CA}
 +X^C{_{||A}}\chi_{CB} \] Another form of (\ref{idX}) can be
 transformed as follows:
\[ \frac 12\int_{S^2} \dot\chi^{BC}\left( X^A\chi_{AB||C}
 +X_C\chi_B{^A}{_{||A}} -X^A\chi_{BC||A} \right) =0 \] is equivalent
 to
\[ \int_{S^2} \dot\chi^{BC}X^A\left(\chi_{BC||A}-\chi_{BA||C}\right) =
\int_{S^2} \dot\chi^{BC}X_C\chi_B{^A}{_{||A}} \] and last
 equality holds for integrands \[
\dot\chi^{BC}X^A\left(\chi_{BC||A}-\chi_{BA||C}\right)=
\dot\chi^{BC}X^A\varepsilon_{AC}\hat\chi_B{^D}{_{||D}}=
 X^A\dot\chi^{BC}\varepsilon_{AC}\varepsilon_{BF}\chi^{FD}{_{||D}}=
 X^A\dot\chi_{AF}\chi^{FD}{_{||D}} \] This way we have proved
 (\ref{idX}) and finally (\ref{dotJonC}) which can be rewritten as
 follows
\[ -16\pi\dot J_z = \frac 12 \int_{S^2}
\kolo\chi^{AB}_{,u}\kolo\chi_{AB,\phi} =\]
\[ =
\int_{S^2} \sin\theta \rd\theta\rd\phi \left[
 (\kolo\chi^{AB}_{||AB})_{,u} {\bf a}^{-1}({\bf a}+2)^{-1}
 (\kolo\chi^{AB}_{||AB})_{,\phi} + (\hat\chi^{AB}_{||AB})_{,u} {\bf
 a}^{-1}({\bf a}+2)^{-1} (\hat\chi^{AB}_{||AB})_{,\phi}
\right] \]
 The similarity with (\ref{lJdot}) is obvious provided that
 supertranslation ambiguity is removed.

Let us introduce the following objects in the full nonlinear
 asymptotics:
\[ 4M=4m+12p+Q+B_x \]
\[ \chi^{AB}{_{||AB}}=B_x+Q_0  \quad \hat\chi^{AB}{_{||AB}}=B_y \]
 where $m$ -- monopole (mass), $p$ -- dipole (linear momentum), $Q$,
 $Q_0$, $B_x$, $B_y$ -- mono-dipole free.  $Q_0$ represents
 supertranslation ambiguity: \[ \overline Q_0=Q_0- {\bf a}({\bf
 a}+2)\alpha \, , \quad \dot Q_0=0 \] and equation (\ref{Mdot}) is
 equivalent to \[ 4\dot m +12 \dot p +\dot Q =-\frac 12
\dot\chi_{AB}\dot\chi^{AB}
 ={\rm QF} (\dot B_x,\dot B_y) \] where QF is a quadratic quasi-local
 functional.  Supertranslation gauge $Q(u_0)=Q_0$ allows to relate
 $\Psi_\Upsilon$ at $\scri^+$ with $B_\Upsilon$.  The corresponding
 names $m$ -- mass, $p$ -- linear momentum, $Q$ -- ``supertranslation
 charge'' are obvious.  This decomposition is chosen in a convenient
 manner for the situation of the so-called ``sandwich-wave''. \\
 Suppose $B_x$ and $B_y$ have compact support on $\scri^+$
 (supp$B_\Upsilon \subset [u_i,u_f]
\times S^2 \subset \scri^+$). Let us also suppose that below $u_i$ and upper
 $u_f$ our gravitating system is stationary. These two assumptions
 define ``sandwich-wave''. \\ When nonradiating solution becomes
 stationary?\\ From $\dot N^A=0$ we obtain $p=\underline
 M=\hat\chi^{AB}{_{||AB}}=0$, $m,k_0,s$ are not restricted but also
 $\kolo\chi^{AB}{_{||AB}}$ does not vanish.  From $\dot D_{AB}=0$ we
 get $C_{AB}=({\bf a}+4)^{-1} S_A{^C}{_{B||C}}$. Similarily $\dot
 C_{AB}=0$ gives $m\kolo\chi_{AB}=\tilde N_{A||B}+\tilde N_{B||A}
 -\kolo\gamma_{AB}
\tilde N^C{_{||C}}$ or $\tilde N_{A||B}\kolo\varepsilon^{AB}=0$ and 
 $\kolo\chi^{AB}{_{||AB}}=({\bf a}+2)\tilde N^A{_{||A}}$.  Let us call
 {\em simple stationary} solution the situation when
 $\kolo\chi_{AB}=0$ and $M=m=\mbox{const.}$ described by van der Burg
 in static case (sec.5 in \cite{Burg}).\\[1ex]
\underline{Remark} The equations related with Newman-Penrose charge in
 static situation presented in \cite{Burg} at the end of page 119 can
 be denoted in our notation as follows:
\[ ({\bf a}+10)D_{AB}=15(MC_{AB}-N_A N_B +\frac12 \kolo\gamma_{AB} N^C
 N_C) \] We have defined three categories of special solutions:
\[ \mbox{simple stationary solutions} \subset \mbox{stationary solutions}
\subset\mbox{nonradiating solutions} \] 
 and let us observe that the supertranslation gauge leads to the
 conclusion that every stationary solution in supertranslation gauge
 (\ref{sg}) is simple.
\[ (\mbox{nonradiating sol.in supertr. gauge})\cup (\mbox{stationary
 sol.}) = \mbox{simple stationary sol.} \] On the other hand in the
 case of the ``sandwich wave'' the supertranslation gauge at $u_i$ and
 at $u_f$ is not the same in general.  The difference depends on
\[ \int_{u_i}^{u_f} \dot Q \rd u = \int_{u_i}^{u_f} \mbox{\underline{QF}}
\rd u \]
 so in general the initial and final states cannot be simple in the
 same Bondi coordinates.

\subsection{Special solutions of asymptotic hierarchy, Newman--Penrose
 charges} Equations (\ref{Mdot} -- \ref{Ddot}) represent nonlinear
 analogue (up to the fourth order) of the hierarchy (\ref{fir}) for
 usual wave equation. \\ We could define as a generalized NP charge
 any solution which starts in the $n+1$-th order from ``multipole
 constant''. More precisely, if $\varphi
\in \ker [{\bf a}+(n-1)n]$ then from (\ref{fir})
 $\varphi_{n+1,u}=0$. Let us observe that if this charge vanishes we
 can derive ``finite'' Janis solution \cite{Janis}, which is obtained
 by ``cutting the series'' and derive hierarchy ``upward''.  More
 precisely,
\[ \varphi=\varphi_1\rho+\varphi_2\rho^2 +\ldots +\varphi_n\rho^n \]
\[ \varphi_n \in  \ker [{\bf a}+n(n-1)] \Longrightarrow
\dot\varphi_{n+1}=0 \, , \quad \varphi_n=C(u)Y_{n-1}(\theta,\phi) \]
\[ \varphi_{k-1}=\frac{2k-2}{n(n-1)-(k-1)(k-2)}\dot\varphi_k \;\; k\leq n \]
 and $Y_l$ is a spherical harmonics ($[{\bf a}+l(l+1)]Y_l=0$).  In
 particular when $\dot\varphi_1=0$ then $C(u)$ is a polynomial of
 degree $n-1$.\\ On the other hand if NP charge is not vanishing the
 solution $\varphi$ can not be stationary.  Moreover, monopole and
 dipole examples show that these solutions are singular (but on
 $\scri^-$).  The monopole example is the following \[ \varphi=
 2\varphi_2 \rho v^{-1}=\varphi_2
\frac{2\rho^2}{2+\rho u} \quad
 {\bf a}\varphi_2=0 \quad \partial_u \varphi_2=0
\]
 similarily the dipole one \[ \varphi= \frac 43 \varphi_3 \rho
\frac{3v-u}{v^2(v-u)}
 = -\varphi_3 \frac{4}{3} \frac{\rho^3}{(2+\rho u)^2} \quad ({\bf
 a}+2)\varphi_3=0 \quad \partial_u \varphi_3=0
\]
 and generaly
\[ \varphi_{n+1} \in  \ker [{\bf a}+n(n-1)] \Longrightarrow
\dot\varphi_{n+1}=0 \, , \quad \varphi_{n+1}=C Y_{n-1}(\theta,\phi) \]
 but now $\dot C=0$ and $\dot\varphi_{n+2}=-\frac n{n+1}
\varphi_{n+1}\neq 0$.

For gravity we have:\\ 1. $D_{AB}=0$ $\Longrightarrow$ Janis
 solution\\ 2. $D_{AB}$ -- pure quadrupole $\Longrightarrow$ NP charge
 solution.

In Janis paper there are only linearized solutions. We shall try now
 to construct asymptotic quadrupole solution of nonlinear hierarchy
 (\ref{Mdot}--\ref{Ddot}). Let us assume that $\underline M$ and
 $\hat\chi^{CD}{_{||CD}}$ are given quadrupoles ($0=({\bf
 a}+6)\underline M=({\bf a}+6)\hat\chi^{CD}{_{||CD}}$).
\[ M_{,u}=\kolo\chi_{AB,u}=0 \]
\[ 4\underline M= \tilde x(\theta,\phi) \, \quad
 {{\kolo{\chi}_C}^D}_{||DE} \kolo{\ve}^{CE}=\tilde y(\theta,\phi)
\]
\[ D_{AB}=0 \] \[ S_{ABC}=\kolo\chi_{AC}N_B +\kolo\chi_{BC}N_A
 -\kolo\gamma_{AB} \kolo\chi_{CD}N^D \] \[ C_{AB}=({\bf
 a}+4)^{-1}(S_A{^C}{_{B||C}}) -\frac1{24} u^2 \left( n_{A||B}+n_{B||A}
 -\kolo\gamma_{AB} n^C{_{||C}}\right) \] \[
 N^A=-p^{||A}u-k_0^{||A}-\kolo\varepsilon^{AB}s_{||B}+ \tilde N^A
 +\frac u3 n^A \] \[ M=m+3p+\underline{M} \] \[ n^A:= \frac
 14\kolo\varepsilon^{AB} \hat\chi^{CD}{_{||CDB}} -\underline M^{||A}
\] \[ \tilde N_{A||B} +\tilde N_{B||A} -\kolo\gamma_{AB}
\tilde N^C{_{||C}} :=
 (m+3p+\underline M)\kolo\chi_{AB} +\frac14
\hat\chi_{AB}\hat\chi^{CD}{_{||CD}} -4({\bf a}+4)^{-1}(S_A{^C}{_{B||C,u}})
\] This a special example of general nonradiating asymptotic
 solution defined by the condition that the TB mass is conserved.

\small
\section*{List of symbols}
\begin{itemize}
\item[$V$] -- three-dimensional volume or function in Bondi-Sachs type metric
\item[$L $] -- lagrangian density
\item[$\Sigma $] -- hyperboloid
\item[$\varphi $] -- scalar field
\item[$\psi $] -- rescaled scalar field
\item[$\delta $] -- ``variational'' derivative
\item[$\partial_{\mu} $] -- partial derivative
\item[$T^\mu{_\nu} $] -- symmetric energy-momentum tensor
\item[$\eta_{\mu\nu} $] -- flat Minkowski metric
\item[$\eta $] -- $\det\eta_{\mu\nu}$
\item[${\cal T}^\mu{_\nu} $] -- canonical energy-momentum density
\item[$\delta^\mu{_\nu} $] -- Kronecker's delta
\item[$p^\mu $] -- canonical field momenta
\item[$\pi $] -- canonical momenta
\item[$\cal H $] -- hamiltonian, energy generator
\item[$H $] -- density of a hamiltonian or two-dim. trace of $h_{AB}$
\item[$x^\mu, y^\nu $] -- coordinates on $M$
\item[$t $] --  time coordinate on $M_2$
\item[$r$] --  radial coordinate on $M_2$
\item[$\omega $] --  related radial coordinate on $M_2$, $r=\sinh\omega$
\item[$\rho$] --  ``inverse'' radial coordinate on $M_2$, $r=\rho^{-1}$
\item[$s$] -- ``hyperboloidal time'' coordinate on $M_2$, $s=t-\sqrt{1+r^2}$
\item[$u,v $] -- null coordinates on $M_2$, $u=t-r$, $v=t+r$
\item[$\overline u $] --  coordinate on $M_2$, $\overline u =-2r$
\item[$\theta,\phi $] -- spherical coordinates on $S^2$
\item[$\rd $] -- exterior derivative
\item[$\mu,\nu,\ldots $] -- four-dimensional indices running $0,\ldots,3$
\item[$k,l,\ldots $] -- three-dimensional indices running $1,\ldots,3$
\item[$A,B,\ldots $] -- two-dimensional indices on a sphere
\item[$a,b,\ldots $] -- two-dimensional ``null'' indices on $M_2$
\item[$\mbox{\msa\symbol{3}}$] -- d'Alambertian, wave operator
\item[$\overline{\mbox{\msa\symbol{3}}}$] -- conformally related wave operator
\item[$\overline\eta_{\mu\nu}$] -- conformally related metric
\item[$R$] -- scalar curvature
\item[$X$] -- vector field
\item[$i^0$] -- spatial infinity
\item[$\scri$] -- null infinity
\item[$\scri^+$] -- future null infinity
\item[$\scri^-$] -- past null infinity
\item[$N$] -- null surface ``parallel'' to $\scri^+$ or a piece of $\scri^+$
\item[$m_{\rm ADM}$] -- ADM mass
\item[$S^2$] -- sphere parameterized by $\theta,\phi$
\item[$S(s,\rho)$] -- sphere in $M$ corresponding to coordinates $s,\rho$
\item[$S_s(\omega)$] -- sphere in $M$ corresponding to coordinates $s,\omega$
\item[$S(s,0)$] -- sphere on $\scri^+$
\item[$S(1)$] -- unit sphere
\item[$\kolo\gamma_{AB}$] -- metric on a unit sphere
\item[$\bf a $] -- two-dimensional laplacian on a unit sphere
\item[$\kolo\varepsilon^{AB}$] -- skew-symmetric tensor on a unit sphere,
 $\sin\theta\varepsilon^{\theta\phi}=1$
\item[$\varepsilon^{AB}$] -- two-dimensional skew-symmetric tensor,
 $r^2\sin\theta\varepsilon^{\theta\phi}=1$
\item[$||$] -- two-dimensional covariant derivative on a sphere
\item[$\hat\partial_A $] -- dual of $\partial_A$, 
 $\hat\partial_A=\varepsilon_A{^B}\partial_B$
\item[${\cal F}^{\mu\nu}$] -- electromagnetic induction density
\item[$f_{\mu\nu}$] -- electromagnetic field
\item[$A_\mu$] -- electromagnetic potential
\item[$\psi$,$*\!\psi$] -- gauge-invariant positions for electromagnetism
\item[$\pi$,$*\!\pi$] -- gauge-invariant momenta for electromagnetism
\item[$\tilde J_z$] -- angular momentum
\item[$g_{kl} $] -- three-dimensional riemannian metric
\item[$P^{kl} $] -- ADM momentum
\item[$K^{kl} $] -- extrinsic curvature
\item[$R^\mu{_{\nu\lambda\sigma}} $] -- curvature tensor 
\item[$R_{\mu\nu} $] -- Ricci tensor
\item[$\Gamma^\lambda{_{\mu\nu}} $] -- Christoffel symbol
\item[$n^\mu$] -- normal unit future directed vector
\item[$\cal R$] -- three-dimensional scalar curvature
\item[$h_{kl}$] -- linearized metric
\item[$\varpi^{kl}$] -- linearized momentum
\item[$p^{kl}$] -- ``new'' linearized momentum
\item[$g $] -- $\det g_{kl}$
\item[$\Lambda $] --  volume element, $\Lambda=r^2\sin\theta$
\item[$\xi_\mu $] -- gauge in linearized gravity
\item[$\kappa $] -- $\kappa=\coth\omega$ 
\item[${\bf x},{\bf X},{\bf y},{\bf Y}$] -- invariants
\item[$\chi_{AB}$] -- traceless part of $h_{AB}$
\item[$S_{AB}$] -- traceless part of $p_{AB}$
\item[$S$] -- trace of $p_{AB}$
\item[$H$] -- trace of $h_{AB}$
\item[$\triangle_\Sigma$] -- laplacian on a hyperboloid 
\item[$J_z$] -- angular momentum generator
\item[$P_z$] -- linear momentum generator
\item[$\Psi_x, \Psi_y$] -- ``asymptotic position'' on a hyperboloid
\item[$\Pi_x, \Pi_y$] -- ``asymptotic momenta'' on a hyperboloid
\item[$\Upsilon$] --  abstract index, $\Upsilon = x,y$
\item[$E_{ab}$] -- $\frac 12 E_{ab}\rd x^a\wedge \rd x^b =\rd u\wedge \rd v$
\item[${\bf y}_a$] -- invariant in null coordinates
\item[${\bf x}_{ab}$] -- invariant in null coordinates
\item[$\beta, V, U^A, \gamma_{AB}$] -- parameters describing Bondi-Sachs
 type metric
\item[$C$] -- null cone or van der Burg asymptotics
\item[$m_{\rm TB}$] -- the Trautman-Bondi mass
\item[$\Psi^{AB}$] -- nonlinear asymptotic position on a null cone
\item[$\Pi_{AB}$] -- nonlinear asymptotic momenta on a null cone
\item[${\cal L}_X$] -- Lie derivative with respect to vector field $X$
\item[$s^z, s^l, {\bf s}$] -- spin charge
\item[$\bf m$] -- mass charge
\item[$p^z, p^l, {\bf p}$] -- linear momentum charge
\item[$j^{l0}, {\bf k}_0$] -- static momentum charge (center of mass)
\item[$M$] -- Minkowski space or asymptotics of function $V$ in van der
 Burg notation
\item[$\gamma,\delta$] -- van der Burg parameterization of $\gamma_{AB}$
\item[$U,W$] -- van der Burg parameterization of $U^A$
\item[$N,P$] -- van der Burg parameterization of $N^A$ 
\item[$N^A$] -- asymptotics of $U^A$
\item[$c,C,D$] -- van der Burg notation for the asymptotics of $\gamma$
\item[$d,H,K$] -- van der Burg notation for the asymptotics of $\delta$
\item[$\kolo\chi_{AB}, C_{AB}, D_{AB}$] -- asymptotics of $\gamma_{AB}$
\item[$\sigma_x, \sigma_z$] -- Pauli matrices
\item[$S_{ABC}$] -- traceless symmetric tensor appearing in eq. (\ref{Ddot})
\item[$\hat S_{ABC}$] -- ``dual'' of $S_{ABC}$, $\hat S_{ABC}=
\kolo\ve_A{^D} S_{DBC}$ 
\item[$\hat C_{AB}$] -- ``dual'' of $C_{AB}$, 
 $\hat C_{AB}=\kolo\ve_A{^D} C_{DB}$
\item[$\hat\chi_{AB}$] -- ``dual'' of $\kolo\chi_{AB}$, $\hat
\chi_{AB}=\kolo\ve_A{^D} \kolo\chi_{DB}$ 
\item[$\hat D_{AB}$] -- ``dual'' of $D_{AB}$, $\hat D_{AB}=\kolo\ve_A{^C}
 D_{CB}$
\item[${\rm mon}(F)$] -- monopole part of $F$
\item[${\rm dip}(F)$] -- dipole part of $F$
\item[$\underline F$] -- mono-dipole free part of $F$
\item[$\overline F$] -- supertranslation of $F$
\item[$Y_l$] -- spherical harmonics with eigenvalue $-l(l+1)$ of the
 laplacian $\bf a$
\end{itemize}

\end{document}